\documentclass[a4paper,11pt]{article}
\pdfoutput=1
\usepackage{jheppub} 
\usepackage[mathscr]{euscript}

\let\emptyset\varnothing

\usepackage{amsfonts}
\usepackage{amscd}
\usepackage{amssymb}
\usepackage{amsmath,bbm}
\usepackage{graphicx}
\usepackage{epsfig}
\usepackage{latexsym}
\usepackage{mathtools}
\usepackage{hyperref}
\usepackage[vcentermath]{youngtab}
\hypersetup{
    colorlinks=true,
   linkcolor=black,
    citecolor=black,
    filecolor=black,
    urlcolor=black,}
    
\def \be  {\begin{equation}}
\def \ee  {\end{equation}}
\def \bea {\begin{equation}\begin{aligned}}
\def \eea {\end{aligned}\end{equation}}
\def \ba  {\begin{eqnarray}}
\def \ea  {\end{eqnarray}}
\def \bb  {\begin{thebibliography}}
\def \eb  {\end{thebibliography}}
\def \lab #1 {\label{#1}}

\newcommand\cB{\mathcal{B}}

\newcommand\cD{\mathcal{D}}

\newcommand\cI{\mathcal{I}}

\newcommand\cL{\mathcal{L}}
\newcommand\cM{\mathcal{M}}
\newcommand\cN{\mathcal{N}}
\newcommand\cO{\mathcal{O}}

\newcommand\cE{\mathcal{E}}
\newcommand\cW{\mathcal{W}}

\newcommand\cZ{\mathcal{Z}}
\newcommand\al{\alpha}

\newcommand\bC{\mathbb{C }}

\newcommand\bR{\mathbb{R}}

\newcommand\bZ{\mathbb{Z}}

\newcommand\lb{\lambda}

\newcommand\la{\langle}
\newcommand\ra{\rangle}
\newcommand\del{\partial}

\newcommand{\ep}{\epsilon}

\renewcommand{\Im}{\mathop{\mathrm{Im}}}

\newcommand{\CN}{\mathcal{N}}

\newcommand{\CR}{\mathcal{R}}

\newcommand{\CZ}{\mathcal{Z}}

\makeatletter
\newcommand{\ellSN}{\mathop{\operator@font sn}\nolimits}
\newcommand{\ellCN}{\mathop{\operator@font cn}\nolimits}
\newcommand{\ellDN}{\mathop{\operator@font dn}\nolimits}
\newcommand{\ellAM}{\mathop{\operator@font am}\nolimits}
\newcommand{\ellK}{\mathop{\smash{\operator@font K}\vphantom{a}}\nolimits}
\newcommand{\ellE}{\mathop{\smash{\operator@font E}\vphantom{a}}\nolimits}
\makeatother

\ifx\genfrac\sdflkaj

\else

\fi

\newcommand{\beq}{\begin{equation}}
\newcommand{\eeq}{\end{equation}}

\makeatletter
\def\mr@ignsp#1 {\ifx\:#1\@empty\else #1\expandafter\mr@ignsp\fi}%
\newcommand{\multiref}[1]{\begingroup
\xdef\mr@no@sparg{\expandafter\mr@ignsp#1 \: }%
\def\mr@comma{}%
\@for\mr@refs:=\mr@no@sparg\do{\mr@comma\def\mr@comma{,}\ref{\mr@refs}}%
\endgroup}
\makeatother

\newcommand{\hypref}[2]{\ifx\href\asklfhas #2\else\href{#1}{#2}\fi}
\newcommand{\Secref}[1]{Section~\multiref{#1}}
\newcommand{\secref}[1]{Sec.~\multiref{#1}}

\newcommand{\appref}[1]{App.~\multiref{#1}}
\newcommand{\Tabref}[1]{Table~\multiref{#1}}

\newcommand{\figref}[1]{Fig.~\multiref{#1}}
\renewcommand{\eqref}[1]{(\multiref{#1})}

\newtheorem{theorem}{Theorem}[section]

\newtheorem{proposition}[theorem]{Proposition}
\newtheorem{corollary}[theorem]{Corollary}

\def\[{\begin{equation}}
\def\]{\end{equation}}
\def\<{\begin{eqnarray}}
\def\>{\end{eqnarray}}

\ifx\href\asklfhas\newcommand{\href}[2]{#2}\fi

\title{Defects and Quantum Seiberg-Witten Geometry}

\author[a]{Mathew Bullimore,}
\author[b]{Hee-Cheol Kim,}
\author[b]{Peter Koroteev\,}

\affiliation[a]{Institute for Advanced Study \\ Einstein Dr., Princeton\\ NJ 08540, USA}
\affiliation[b]{Perimeter Institute for Theoretical Physics \\ 31 Caroline Street North, Waterloo\\ Ontario N2L2Y5, Canada}

\emailAdd{mbullimore@ias.edu}
\emailAdd{hkim@perimeterinstitute.ca}
\emailAdd{pkoroteev@perimeterinstitute.ca}

\abstract{We study the Nekrasov partition function of the five dimensional $U(N)$ gauge theory with maximal supersymmetry on $\mathbb{R}^4\times S^1$ in the presence of codimension two defects. The codimension two defects can be described either as monodromy defects, or by coupling to a certain class of three dimensional quiver gauge theories on $\mathbb{R}^2\times S^1$. We explain how these computations are connected with both classical and quantum integrable systems. We check, as an expansion in the instanton number, that the aforementioned partition functions are eigenfunctions of an elliptic integrable many-body system, which quantizes the Seiberg-Witten geometry of the five-dimensional gauge theory.}

\begin{document}
\maketitle

\pagebreak
\section{Introduction}\label{Sec:Intro}
Supersymmetric gauge theories provide a rich source of inspiration for various branches of mathematics. From a practical viewpoint, they can also provide a powerful set of techniques to solve challenging mathematical problems using physics. The interplay between supersymmetric gauge theories and mathematics is enhanced by introducing defects that preserve some amount of supersymmetry. 

In this work, we study 5d $U(N)$ $\cN=2$ supersymmetric gauge theories with codimension two and codimension four defects and how they are connected to the quantization of the integrable system associated to its Seiberg-Witten geometry~\cite{Seiberg:1994rs,Seiberg:1994aj}. We focus on a class of codimension two defects preserving $\cN=4$ supersymmetry in three dimensions. They can be described either as  Gukov-Witten monodromy defects~\cite{Gukov:2006jk,Gukov:2008sn} or by coupling to a class of 3d $\cN=4$ supersymmetric gauge theories. We will concentrate on the surface defect obtained by introducing the most generic monodromy for the gauge field, or alternatively by coupling to the 3d $\cN=4$ theory $T[U(N)]$~\cite{Gaiotto:2008ak}. The codimension four defects are described by supersymmetric Wilson loops.

As a preliminary step towards understanding and computing with surface defects, we will first find a reformulation of the twisted chiral ring of a canonical deformation of the $T[U(N)]$ theory on $S^1 \times \mathbb{R}^2$, building on the work of~\cite{Gaiotto:2013bwa}. We will show that the twisted chiral ring relations are equivalent to the spectral curve of an associated classical $N$-body integrable system, known as the complex trigonometric Ruijsenaars-Schneider (RS) system. In addition, this provides a reformulation of the equivariant quantum K-theory of the cotangent bundle to a complete flag variety, via the Nekrasov-Shatashvili correspondence \cite{Nekrasov:2009ui, Nekrasov:2009uh}. We will also explore the connection with quantum K-theory for more general linear quiver gauge theories.

This classical integrable system can be quantized by turning on an equivariant parameter $\ep$ for rotations in $\mathbb{R}^2$, otherwise known as the three-dimensional Omega background $S^1 \times \mathbb{R}^2_\ep$. As it is technically simpler, we will first consider the squashed $S^3$ partition function of $T[U(N)]$ theory using results from supersymmetric localization~\cite{Kapustin:2009kz,Hama:2011ea}. The partition functions on $S^1\times \mathbb{R}^2_\ep$ can then be obtained by factorization of the $S^3$ partition function~\cite{Pasquetti:2011fj,Beem:2012mb}. We show that these supersymmetric partition functions are eigenfunctions of the quantized trigonometric Ruijsenaars-Schneider system with the Planck constant proportional to $\ep$. The corresponding eigenvalues are given by supersymmetric Wilson loops for background $U(N)$ flavor symmetries. The twisted chiral ring relations, or equivalently the spectral curve of the classical integrable system, are reproduced in the semi-classical limit $\ep \to 0$. 

In coupling the three-dimensional theory $T[U(N)]$ theory as a surface defect in 5d $U(N)$ $\cN=2$ gauge theory, the twisted chiral ring relations are deformed by an additional complex parameter $Q$, which is related to the 5d holomorphic gauge coupling. According to \cite{Gaiotto:2013sma} it is expected that this deformation provides a presentation of the Seiberg-Witten curve of the 5d theory on $S^1 \times \mathbb{R}^4$. The Seiberg-Witten curve of the 5d $U(N)$ $\cN=2$ supersymmetric gauge theory is known to correspond to the spectral curve of the $N$-body elliptic Ruijsenaars-Schneider system~\cite{Nekrasov:1996cz}. This is indeed a deformation of the trigonometric RS system by an additional complex parameter $Q$.

In order to test this relationship, we will compute the 5d Nekrasov partition function of $\cN=2$ $U(N)$ supersymmetric gauge theory on $S^1 \times \mathbb{R}_{\ep_1,\ep_2}^4$ in the presence of surface defects wrapping one of the two-planes $S^1 \times \mathbb{R}_{\ep_1}^2$. This computation is performed by treating the surface defect as a monodromy defect and applying the orbifolding procedure introduced in~\cite{Alday:2010vg,Kanno:2011fw}. In order to check that the Gukov-Witten monodromy defect is reproducing the same surface defect as coupling to $T[U(N)]$ theory, we check that this computation reproduces the $S^1\times \mathbb{R}^2_{\ep_1}$ partition function of $T[U(N)]$ theory in the limit $Q\to0$ where the coupling to the 5d degrees of freedom is turned off. In particular, we note that the Gukov-Witten monodromy parameters are identified with the Fayet-Iliopoulos parameters of the 3d gauge theory supported on the defect.

After performing this preliminary check, we study the full Nekrasov partition function on $S^1 \times \mathbb{R}^4_{\ep_1,\ep_2}$ as an expansion in the parameter $Q$. In the the Nekrasov-Shatashvili limit $\ep_2 \to 0$, we will show that the expectation value of the most generic surface defect is formally an eigenfunction of the elliptic RS system. Furthermore, we find that the corresponding eigenvalues are given by the expectation values of supersymmetric Wilson loops wrapping $S^1$ in the 5d gauge theory. This computation provides a quantization of the Seiberg-Witten geometry. 

We will also study another type of codimension two defect by coupling directly to 3d hypermultiplets~\cite{Lamy-Poirier:2014sea,GK2014}. We focus on the simplest example where two free hypermultiplets of $U(2)$ flavor symmetry are coupled to the bulk gauge field of the 5d $U(2)$ $\mathcal{N}=2$ gauge theory. We will show that the partition function of this coupled system solves an eigenfunction equation of the so-called two-body {\it dual} elliptic Ruijsenaars-Schneider system. Indeed, the S-transformation of the 3d theory in~\cite{Witten:2003ya} relates this partition function to the partition function of the $U(2)$ gauge theory with a monodromy defect.

The paper is organized as follows. In \Secref{Sec:TwistChiralRings} we will relate the twisted chiral ring of $T[U(N)]$ to the spectral curve of the classical trigonometric Ruijsenaars-Schneider system and discuss connections to equivariant quantum K-theory. In \Secref{Sec:3dpf} we show that partition functions on squashed $S^3$ and $S^1 \times \mathbb{R}^2$ are eigenfunctions of the corresponding quantized integrable system. Then in \Secref{Sec:5dRamification} we explain how to compute the expectation values of surface defects in $\cN=2$ $U(N)$ gauge theory on $S^1 \times \mathbb{R}^4$ and demonstrate that they are eigenfunctions of the elliptic RS system. Later in \Secref{Sec:Toda} we present a cursory discussion of various limits and degenerations of the results presented in this paper. Finally, in section \Secref{Sec:IntApplications} we summarize the connections of this work to integrable systems and discuss areas for further research.

 
\section{Twisted Chiral Rings}\label{Sec:TwistChiralRings}
In this section, we will study 3d $\CN=4$ linear quiver gauge theories on $S^1 \times \mathbb{R}^2$, deformed by hypermultiplet masses, FI parameters, and with the canonical $\CN=2^*$ mass deformation. 
It was observed by Nekrasov and Shatashvili \cite{Nekrasov:2009ui, Nekrasov:2009uh} that the equations determining the supersymmetric massive vacua on $S^1 \times \mathbb{R}^2$, or the twisted chiral ring relations, can be identified with Bethe ansatz equations for a quantum XXZ integrable spin chain.

We will focus for the most part on the triangular quiver gauge theory: $T[U(N)]$. We will reformulate the statement of its twisted chiral ring in terms of the spectral curve of a classical $N$-body integrable system known as the complexified trigonometric Ruijsenaars-Schneider system. Alternatively, it can be viewed as a lagrangian correspondence that diagonalizes this classical integrable system. This reformulation will be important when we come to couple this theory as a codimension two defect in five-dimensions.

We will explain how the corresponding statements for more general linear quivers can be obtained by a combination of Higgsing and mirror symmetry, and demonstrate this in a simple example. We will also briefly discuss connections to results in the mathematical literature on the equivariant quantum K-theory of the cotangent bundles to partial flag manifolds.

\subsection{The Nekrasov-Shatashvili Correspondence}
Three-dimensional theories with $\cN=4$ supersymmetry have $SU(2)_H \times SU(2)_C$ R-symmetry and flavor symmetries $G_H \times G_C$ acting on the fields parametrizing the Higgs and Coulomb branches respectively. For the purpose of this paper, it is important to turn on a canonical deformation preserving only $\cN=2$ supersymmetry. The corresponding $U(1)_R$ is the diagonal combination of Cartan generators of $SU(2)_H \times SU(2)_C$. The anti-diagonal combination becomes an additional flavor symmetry $U(1)_\ep$ with real mass parameter $\ep$. Furthermore, we turn on real deformation parameters by coupling to $\cN=2$ vectormultiplets for $G_H \times G_C$ and giving a vacuum expectation value to the real scalar. In a UV description, these deformation parameters enter as real hypermultiplet masses denoted typically by $m$ and FI parameters denoted by $t$. We refer to this setup as $\cN=2^*$ supersymmetry. We refer the reader to \cite{Gaiotto:2013bwa} for a more complete description of this setup.

\begin{figure}[!h]
\begin{center}
\includegraphics[scale=0.4]{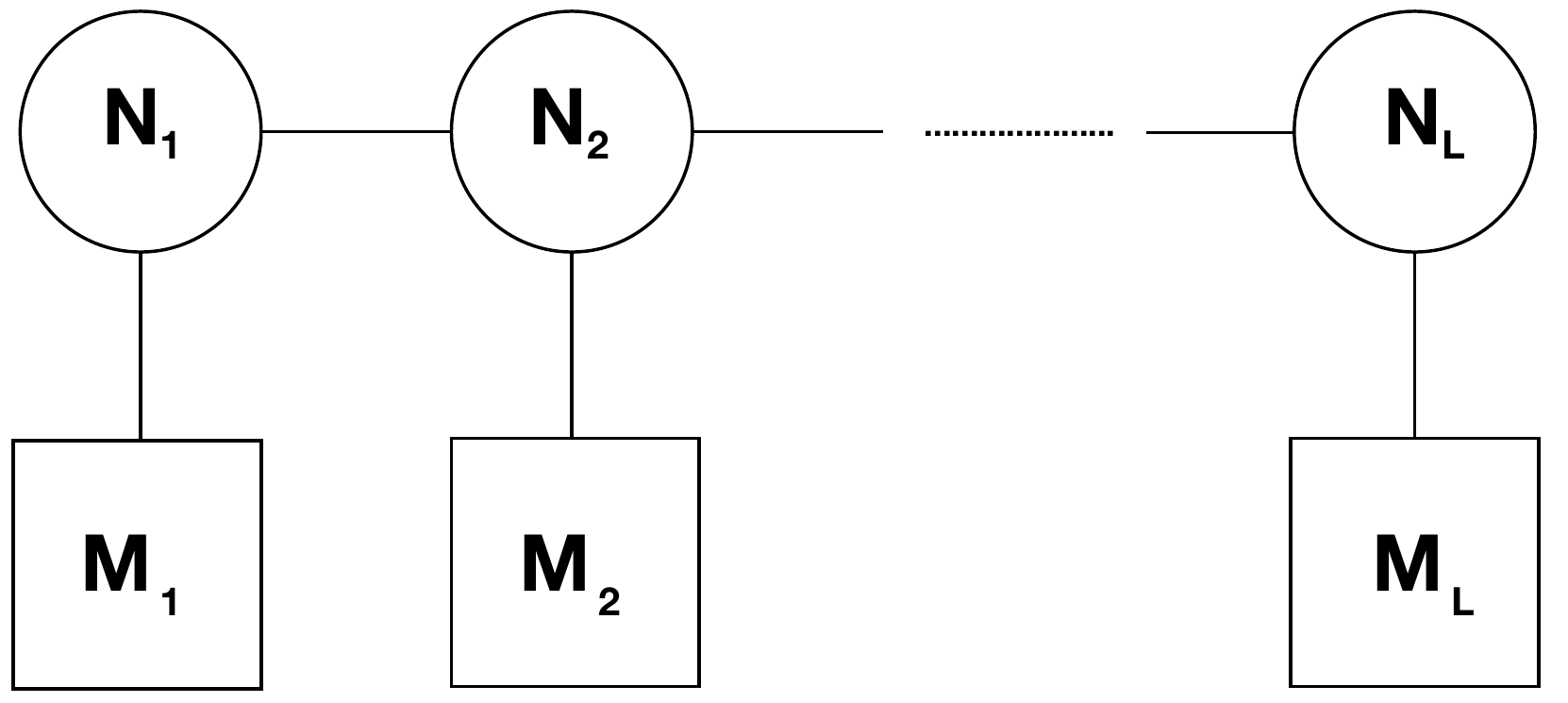}
\caption{3d $\CN=4$ quiver gauge theory of type $A_{L}$ with gauge group $U(N_1)\times\dots\times U(N_L)$ and $M_i$ fundamental hypermultiplets at $i$-th gauge node.}
\label{Fig:quivergauge}
\end{center}
\end{figure}
Here, we focus on theories with a UV description as a linear quiver with unitary gauge groups. Our notation is summarized in figure~\ref{Fig:quivergauge}. It is convenient to introduce a $U(1)$ symmetry acting trivially such that the Higgs branch symmetry is given by $U(M_1) \times \cdots \times U(M_L)$. The corresponding mass parameters are denoted by $\{ m^{(j)}_1,\ldots,m^{(j)}_{M_j} \}$. Similarly, we introduce an additional topological $U(1)$ so that the Coulomb branch symmetry manifest in the UV description is $U(1)^{L+1}$ with corresponding parameters $\{t_1,\ldots,t_{L+1}\}$. The physical FI parameter at the $j$-th gauge node is $t_{j+1} - t_{j}$. This symmetry can be enhanced by monopole operators up to a maximum of $U(L+1)$ in the IR.

We will focus on he twisted chiral ring of the effective 2d $\cN=(2,2)$ theory obtained by compactifying on a circle of radius $R$. In this case, the real deformation parameters are complexified by background Wilson lines wrapping the circle and behave as twisted masses in the language of $\cN=(2,2)$ supersymmetry. 

For generic deformation parameters there is a discrete set of massive supersymmetric vacua on $S^1 \times \mathbb{R}^2$ each with an associated effective twisted superpotential $\cW^{(i)}(m,t,\ep)$. This is a holomorphic function that is independent of superpotential and gauge couplings. In the UV theory one can integrate out three-dimensional chiral multiplets to find an effective twisted superpotential $\cW(m,t,\ep,s)$ for the dynamical vectormultiplets. The supersymmetric vacua are then solutions to
\be
\exp{\left[ 2\pi R \frac{\del\cW}{\del s_i} \right]}= 1\,,
\label{eq:BetheEqs}
\ee
which can be identified with the twisted chiral ring relations of the effective two-dimensional theory.

The effective twisted superpotential of the generic linear quiver shown in figure~\ref{Fig:quivergauge} is given by
\bea
\cW(s,m,t,\epsilon) = \sum_{i=1}^{L} \left(t_{i}-t_{i+1}+\frac{i\delta_j}{2R} \right)\sum_{\al=1}^{N_i} s_\al^{(i)} + \cW_{\text{1-loop}}(s,m,\epsilon) \\
+\sum_{j=1}^{L}\sum_{i=1}^{j}t_{i+1}\sum_{a=1}^{M_i}m_a\,.
\label{eq:Wtwgen}
\eea
The first term includes contributions from the FI parameters at each node together with phase $\delta_i =M_i+N_{i-1}+N_i+N_{i+1}-1$~\footnote{Sign conventions are slightly altered compared to \cite{Gaiotto:2013bwa}.}. The second term includes the 1-loop contributions from the KK tower of chiral multiplets. The basic building block of the 1-loop contributions is the contribution $\ell(m)$ from a three-dimensional chiral multiplet of mass $m$, which is a solution of the differential equation $2\pi R \, \del_m \ell(m) =\log( 2\sinh \pi R m)$. We refer the reader to~\cite{Gaiotto:2013bwa} for an explicit expression. The final term is included to ensure that mirror symmetry acts straightforwardly in the presence of the spurious $U(1)$ symmetries.

As the imaginary components of the twisted mass parameters are periodic, it is convenient to introduce exponentiated parameters
\be
\mu^{(i)}_a = e^{2\pi R m^{(i)}_a}\,, \qquad \tau_i =e^{2\pi R t_i}\,, \qquad \sigma^{(i)}_\al = e^{2\pi R s^{(i)}_\al}\,, \qquad  \eta = e^{\pi R \ep} \, .
\label{eq:TrigLabels}
\ee
With this notation, the equations for the supersymmetric vacua are
\begin{equation}
\frac{\tau_{i}}{\tau_{i+1}}\prod_{\beta=1}^{N_{i-1}}\frac{\eta \sigma^{(i)}_\al- \sigma^{(i-1)}_{\beta}}{\eta \sigma^{(i-1)}_{\beta}-\sigma^{(i)}_\al} \cdot \prod_{ \beta \neq \al}^{N_{i}}\frac{\eta^{-1} \sigma^{(i)}_\al - \eta \sigma^{(i)}_\beta}{\eta^{-1}\sigma^{(i)}_{\beta}-\eta \sigma^{(i)}_\al} \cdot \prod_{\beta=1}^{N_{i+1}}\frac{\eta \sigma^{(i)}_\al - \sigma^{(i+1)}_{\beta}}{\eta \sigma^{(i+1)}_{\beta}-\sigma^{(i)}_\al}\cdot\prod_{a=1}^{M_i}\frac{\eta \sigma^{(i)}_\al-\mu^{(i)}_a}{\eta \mu^{(i)}_a-\sigma^{(i)}_\al}=(-1)^{\delta_i}\,, 
\label{eq:XXZGen}
\end{equation}
for all $i=1,\ldots,L$. It was observed by Nekrasov and Shatashvili~\cite{Nekrasov:2009ui, Nekrasov:2009uh} that these equations can be identified with the Bethe equations of a quantum integrable XXZ spin chain. A complete dictionary for linear quivers can be found in \cite{Gaiotto:2013bwa}.

In order to write down the twisted chiral ring it is necessary to introduce a generating function for the gauge invariant combinations of the $\sigma^{(i)}_\al$'s. For this purpose, we introduce an auxiliary parameter $u$ and monic polynomials
\begin{equation}
Q_i(u) = \prod_{\al=1}^{N_i}(u - \sigma^{(i)}_\al)\,, \qquad P_i(u) = \prod_{a=1}^{M_i}(u - \mu^{(i)}_a) \, .
\end{equation}
The equations for supersymmetric vacua~\eqref{eq:XXZGen} can be expressed in terms of these polynomials as
\begin{equation}
\eta^{-\Delta_i} \frac{\tau_{i}P_i^+ Q_{i-1}^+ Q_{i}^{--} Q_{i+1}^+}{\tau_{i+1}P_i^- Q_{i-1}^- Q_{i}^{++} Q_{i+1}^-} = -1\,,
\label{eq:TQBaxter}
\end{equation}
where the polynomials are understood to be evaluated at $ u = \sigma^{(i)}_\al$ for $\al = 1,\ldots,N_i$. We defined $\Delta_i = M_i+N_{i+1}+N_{i-1}-2N_i$ to be the 1-loop contribution to the scaling dimensions of monopole operators charged under $i$-th gauge group. The superscripts on the polynomials are shorthand for multiplicative shifts of the arguments by $\eta$, for example $Q^+_i(u) = Q_i(\eta u)$, $Q^-_i(u) = Q_i(\eta^{-1}u)$. The twisted chiral ring relations for gauge invariant combinations of $\sigma^{(i)}_\al$'s are given by expanding equations~\eqref{eq:TQBaxter} in $u$.  

In what follows, it will be useful to introduce another slightly less familiar reformulation of the twisted chiral ring~\footnote{We thank Davide Gaiotto for exhibiting us this calculation in a sample example.}. We first rescale the FI parameters by $\widetilde \tau_i = \tau_i \, \eta^{\sum_{j=1}^{i-1}\Delta_j}$ to absorb the dependence on $\Delta_j$. Then we introduce the polynomial equations
\begin{equation}
\widetilde \tau_{i+1} Q^+_i \widetilde Q^-_i - \widetilde \tau_{i} Q^-_i \widetilde Q^+_i = (\widetilde \tau_{i+1}-\widetilde \tau_i)P_iQ_{i-1}Q_{i+1}\, ,
\label{eq:QQrelations}
\end{equation}
where $\widetilde Q_i(u)$ are auxiliary polynomials of rank $M_i+N_{i-1}-N_i+N_{i+1}$. To recover equations~\eqref{eq:TQBaxter} for the supersymmetric vacua we shift the argument this polynomial equation by $\eta^\pm$ and evaluate both at roots $\sigma^{(i)}_\al$ of $Q_i(u)$. Dividing one equation by the other, the combination $(\widetilde \tau_{i+1}-\widetilde \tau_i)$ and the auxiliary polynomials $\widetilde Q_i(u)$ cancel out and we reproduce~\eqref{eq:XXZGen}. The twisted chiral ring relations are given by expanding equations~\eqref{eq:QQrelations} in $u$. 

Finally, 3d $\cN=4$ theories have a remarkable duality known as mirror symmetry~\cite{Intriligator:1996ex,deBoer1997148}. This can be understood from brane constructions in Type IIB String theory \cite{Hanany:1996ie}, where it is realized as the S-duality. Mirror symmetry acts in quite a non-trivial manner on the data $(N_i,M_i)$ of the linear quiver, which is spelled out in reference~\cite{Gaiotto:2013bwa}. Mirror symmetry interchanges mass parameters and FI parameters of the quivers and also acts non-trivially on the $\cN=2^*$ mass deformation. Schematically, we have~\footnote{Here the mirror map for $\eta$ has a different sign compared to \cite{Gaiotto:2013bwa}, where $\eta$ was mapped onto $-\eta^{-1}$.}.
\begin{equation}
\mu_a \leftrightarrow \tau_a \qquad \eta \leftrightarrow \eta^{-1}
\label{eq:MirrorSymmetryMap}
\end{equation}
where the check symbol designates parameters of the dual theory.



\subsection{Quantum/Classical Duality}\label{Sec:Trig}
As mentioned above, the equations for supersymmetric vacua can be identified with the Bethe equations for a quantum integrable spin chain. Remarkably, the same system of equations are related to a second, classical integrable system of interacting relativistic particles in one dimension -- the complexified trigonometric Ruijsenaars-Schneider system \cite{Gaiotto:2013bwa}. 

\begin{figure}[h]
\begin{center}
\includegraphics[scale=0.5]{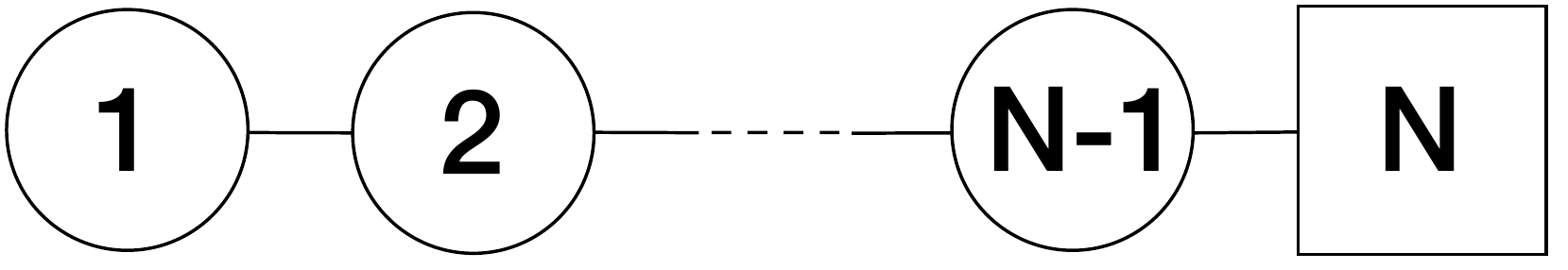}
\caption{A Lagrangian description of the $T[U(N)]$ theory consists of a sequence of gauge groups $U(1)\times\dots \times U(N-1)$ with bifundamental matter and $N$ hypermultiplets at the final node.}
\label{Fig:NakajimaComplete}
\end{center}
\end{figure}

This correspondence is most straightforward to understand in the case of the triangular quiver $T[U(N)]$ (see \figref{Fig:NakajimaComplete}) -- this will be our main example throughout this paper. 
\vspace{0.5cm}

The effective twisted superpotential \eqref{eq:Wtwgen} of this theory is given by
\bea
\cW & = \sum_{j=1}^{N} \left(t_{j}-t_{j+1}+\frac{i\delta_j}{2R} \right)\sum_{\al=1}^j s^{(j)}_\al \\
& + \sum_{j=1}^{N-1} \sum_{\al=1}^j \sum_{\al'=1}^{j+1} \left( \ell\left( s_\al^{(j)} - s_{\al'}^{(j+1)}+\frac{\ep}{2} \right)+\ell\left( - s_\al^{(j)} + s_{\al'}^{(j+1)}+\frac{\ep}{2} \right) \right) \\
& + \sum_{j=1}^{N-1} \sum_{\al \neq \al'} \ell\left(s^{(j)}_\al - s_{\al'}^{(j)} - \ep\right)\,,
\label{eq:T[U(N)]w}
\eea
where $\delta_j = j-1$\footnote{In the notations of \eqref{eq:XXZGen} $N_j=j$, so the this definition of $\delta_j$ is consistent with the above conventions modulo two.} and we define $s^{(N)}_\al = m_\al$ and $t_{N+1}=0$ to simplify notation. For this theory, the Higgs and Coulomb branch symmetries are both $U(N)$ and we have corresponding exponentiated mass $\mu_j$ and FI parameters $\tau_j$ with $j=1,\ldots,N$. This theory is invariant under mirror symmetry with the transformation $\mu_j \leftrightarrow \tau_j, \eta \leftrightarrow \eta^{-1}$.

By introducing the conjugate momenta to $\mu_j$ and $\tau_j$
\begin{equation}
p_\mu^j = \exp{\left[ 2\pi R \frac{\del\cW}{\del m_j} \right]}\,,
\qquad p_\tau^j = \exp{\left[ 2\pi R \frac{\del\cW}{\del t_j} \right]}\,,
\label{eq:ConjMomenta}
\end{equation}
we provide canonical coordinates on two copies $\cM_\mu$ and $\cM_\tau$ of the cotangent bundle to $(\mathbb{C}^*)^N$ with the following holomorphic symplectic forms
\be
\Omega_\mu= \sum_{j=1}^N \frac{ d \mu_j}{\mu_j} \wedge \frac{d p_\mu^j}{p_\mu^j}\,,  \qquad
\Omega_\tau= \sum_{j=1}^N  \frac{d \tau_j}{\tau_j} \wedge \frac{d p_\tau^j}{p_\tau^j}  \,.
\ee
This is the phase space of our complex classical integrable system. The defining equations for the conjugate momenta~\eqref{eq:ConjMomenta} sweep out a complex Lagrangian in the product $\cL \subset \cM_\mu \times \cM_\tau$ with holomorphic symplectic form $\Omega_\mu - \Omega_\tau$ and generating function given by the on-shell twisted effective superpotential $\cW(m_i,t_j,\ep)$. 

It is straightforward to find an explicit description of the Lagrangian $\cL$ for $T[U(2)]$ theory. The supersymmetric vacua equations read
as follows
\be
\frac{\tau _1 \left(\mu _1-\eta  \sigma _1\right) \left(\mu _2-\eta  \sigma _1\right)}{\tau _2 \left(\eta  \mu _1-\sigma _1\right) \left(\eta  \mu _2-\sigma _1\right)} = 1\,.
\label{eq:TU2BetheEq1}
\ee
The conjugate momenta for the FI terms are as follows
\begin{equation}
p_\tau^1  = \sigma_1\,, \qquad p_\tau^2 = \frac{\mu_1 \mu_2}{\sigma_1}\,, \\
\label{eq:momentarelationsTU2}
\end{equation}
and for the masses
\begin{equation}
p_\mu^1  = \tau _2 \frac{\eta  \mu _1-\sigma _1}{\mu _1-\eta  \sigma _1}\,, 
\qquad p_\mu^2 = \tau_2\frac{\eta  \mu _2-\sigma _1}{\mu _2-\eta  \sigma _1} \,.
\label{eq:momentarelationsTU2Masses}
\end{equation}
Given the above definitions of conjugate momenta vacua equation \eqref{eq:TU2BetheEq1} can be presented in two equivalent ways. First, 
as
\begin{equation}
\frac{\tau_1 \eta - \tau_2 \eta^{-1}}{\tau_1-\tau_2} p_\tau^1+\frac{\tau_2 \eta - \tau_1 \eta^{-1}}{\tau_2-\tau_1} p_\tau^2  = \mu_1+\mu_2 \,, \qquad p_\tau^1 p_\tau^2 = \mu_1\mu_2\,,
\label{eq:tRSrelationsTU2}
\end{equation}
and, second, as
\begin{equation}
\frac{\mu_1 \eta^{-1} - \mu_2 \eta}{\mu_1-\mu_2} p_\mu^1+\frac{\mu_2 \eta^{-1} - \mu_1 \eta}{\mu_2-\mu_1} p_\mu^2  = \tau_1+\tau_2\,, \qquad p_\mu^1p_\mu^2 = \tau_1\tau_2\,.
\label{eq:tRSrelationsTU2Mag}
\end{equation}
In \eqref{eq:tRSrelationsTU2} the combinations appearing on the left are the Hamiltonians of the complex trigonometric RS system for two particles with positions $\tau_1,\tau_2$ and momenta $p_\tau^1, p_\tau^2$. The right hand sides are independent of the momenta $p_\mu^1, p_\mu^2$ so the Lagrangian correspondence $\cL$ diagonalizes the system. From this perspective, $\cW(m_a,t_a,\ep,s_\al)$ evaluated on supersymmetric vacua are solutions of the relativistic Hamilton-Jacobi equation. The evident symmetry under $\mu_j \leftrightarrow \tau_j$ and $\eta \leftrightarrow \eta^{-1}$ means this Lagrangian also diagonalizes the same system with the coupling inverted $\eta \to \eta^{-1}$.
 
To eliminate dynamical vectormultiplet scalars $\sigma_\al$ from the supersymmetric vacuum equations in favor of the conjugate momenta $p_\mu^j$ or $p_\tau^j$ in the case $N>2$ is rather non-trivial task. Below we demonstrated that one can do this in two equivalent ways. First as
\begin{equation}
\det \left(u - L(\tau,p_{\tau},\eta)  \right)  = \prod_{j=1}^N(u-\mu_j)\,,
\label{eq:tRSW}
\end{equation}
and, second, as
\begin{equation}
\det \left(u - L(\mu,p_{\mu},\eta^{-1})  \right) = \prod_{j=1}^N(u-\tau_j)\,.
\label{eq:tRST}
\end{equation}
In both relations above
\be
L_{ij}(\alpha,p_{\alpha},\beta) = \frac{\prod\limits_{k \neq j}^N(\alpha_i \, \beta- \alpha_k \, \beta^{-1})}{\prod\limits_{k \neq i}^N( \alpha_i-\alpha_k) } p_{\alpha}^j\,,
\ee
is the Lax matrix for the $N$-body complex trigonometric RS system. One can clearly see that \eqref{eq:tRSW} and \eqref{eq:tRST} are related to each other by mirror symmetry map \eqref{eq:MirrorSymmetryMap}. Therefore, somewhat artificially, we can refer to the former relation as written in the `electric' frame, where eigenvalues of $L$ are related to masses $\mu_j$, and to the latter relation as presented in the `magnetic' frame, in which, using mirror frame variables, the eigenvalues of $L$ are identified with FI parameters $\tau_j$.

As we mentioned above, in order to understand why \eqref{eq:tRSW} and \eqref{eq:tRST} are true it is convenient to re-formulate the supersymmetric vacuum equations arising from this twisted superpotential. We introduce monic degree-$j$ polynomials $Q_j(u)$, $\widetilde Q_j(u)$ for each node $j=1,\ldots,N$ of the quiver. Note that we treat the matter polynomial in a uniform manner, that is we define $m_a = \sigma^{(N)}_a$ and hence $P(u) = Q_N(u)$. With this definition, we have $Q\widetilde Q$ equations \eqref{eq:QQrelations}
\be
\tau_{j+1} Q_j^+ \widetilde Q_j^- - \tau_j Q_j^- \widetilde Q_j^+ = (\tau_{j+1} - \tau_j) Q_{j-1} Q_{j+1}
\label{eq:QQTUN} \, .
\ee
Note that in this case $\Delta_j=0$ and there is no need to redefine $\tau_j$. The original supersymmetric vacuum equations are obtained by shifting arguments in the above by $\eta^{\pm}$, evaluating at the roots $\sigma_{\al}^{(j)}$ of $Q_j(u)$ and eliminating the auxiliary polynomials $\widetilde Q_j(u)$. They can be expressed uniformly as
\be
\frac{\tau_{j}}{\tau_{j+1}} \frac{Q^{+}_{j-1} Q^{--}_j Q^+_{j+1}}{Q^{-}_{j-1} Q^{++}_j Q^-_{j+1}}  = - 1\,,
\label{eq:BetheEqQ}
\ee
evaluated on the roots $u = \sigma^{(j)}_\al$.

\subsubsection{Electric Frame}
Firstly, we explain how to eliminate the $\sigma^{(j)}_\al$ in favor of the momenta conjugate to the FI parameters \eqref{eq:tRSW}. For this purpose, we will set up an inductive procedure to solve the $Q\widetilde Q$ equations recursively node by node. We first note that by the definition \eqref{eq:ConjMomenta}
\be
p_{\tau}^j = - \frac{Q_j(0) }{Q_{j-1}(0) }
\ee
and hence by evaluating the $Q\widetilde Q$ equation at $u=0$ we can immediately solve for the constant terms in the polynomials $Q_j(u)$ and $\tilde Q_j(u)$ as follows
\bea
Q_j(0) & = (-1)^j p_\tau^1 \ldots p_{\tau}^j
\qquad \\
 \widetilde Q_j(0) & = (-1)^j p_\tau^1 \ldots p_{\tau}^{j-1} p_{\tau}^{j+1} \, .
 \label{lowest}
\eea
Now, given the polynomials $Q_i(u)$ and $\widetilde Q_i(u)$ for $i \leq j$ we can determine $Q_{j+1}(u)$ from the $Q\tilde Q$ equation~\eqref{eq:QQTUN}. Then, by shifting $j \to j+1$ in the same equation and evaluating it on the $j$ roots $\sigma^{(j)}_\al$ of $Q_j(u)$ we have just enough data to determine the $j$ unknown coefficients in $\widetilde Q_{j+1}(u)$. 

To illustrate this process, let us perform the first interation explicitly. It is convenient to introduce the monic degree one polynomials $q_i=u-p_\tau^i$. Then from equation~\eqref{lowest} we have $Q_1(u)=q_1(u)$ and $\widetilde Q_1(u)=q_2(u)$. Then, from equation from \eqref{eq:QQTUN} with $j=1$, we find
\begin{equation}
Q_2 = \frac{\tau_2 q^+_1  q^-_2 - \tau_1 q^-_1q^+_2}{\tau_2-\tau_1}\,.
\end{equation}
Now, evaluating equation~\eqref{eq:QQTUN} with $j=2$ on the root of the polynomial $Q_1(u)$ it is straightfoward to compute the coefficient of the linear term in $\widetilde Q_2(u)$ and hence find
\be
\widetilde Q_2 = \frac{\tau_3 q_1^+ q^-_3 - \tau_1 q^-_1 q^+_3}{\tau_3-\tau_1}\,.
\ee
We can now immediately compute the polynomial $Q_3(u)$ using equation~\eqref{eq:QQTUN} with $j=2$

We have implemented this procedure to many orders in $j$ and found experimentally that the solution can be expressed as follows. We introduce the following $j \times j$ matrices
\be
M_{i_1,\ldots,i_j} = \begin{bmatrix} \,  q_{i_1}^{j-1} & \tau_{i_1} q_{i_1}^{j-3} & \cdots & \tau_{i_1}^{j-1} q_{i_1}^{1-j} \\ \vdots & \vdots & \ddots & \vdots \\  q_{i_j}^{j-1}  & \tau_{i_j} q_{i_j}^{j-3} & \cdots & \tau_{i_j}^{j-1} q_{i_j}^{1-j} \, \end{bmatrix}\,,
\qquad
M_{i_1,\ldots,i_j}^{(0)} = \begin{bmatrix} \, 1 & \tau_{i_1} & \cdots & \tau_{i_1}^{j-1} \\ \vdots & \vdots & \ddots & \vdots \\ 1 & \tau_{i_j} & \cdots & \tau_{i_j}^{j-1} \,  \end{bmatrix}\,,
\ee
where we define $q_i=u-p_\tau^i$ and we remind the reader that superscripts are not exponentials but shifts of the argument by $\eta$. Then the solution is given by a ratio of Vandermonde-like determinants
\be
Q_j(u)= \frac{\text{det}\Big( M_{1,\ldots,j} \Big)}{\text{det}\Big( M^{(0)}_{1,\ldots,j} \Big)}\,, 
\qquad
 \widetilde Q_j(u)= \frac{\text{det}\Big( M_{1,\ldots,j-1,j+1} \Big)}{\text{det}\Big( M^{(0)}_{1,\ldots,j-1,j+1} \Big)}\,.
\ee
Solutions of this form for similar functional equations have appeared in the integrability literature. We expect that these techniques could be used to prove the solution we have found (see e.g. \cite{Kazakov:2007na}).

Since all polynomials $q_i$ are monic of degree one the above ratios can be simplified and by inverting the matrix $M^{(0)}_{1,\ldots,N}$, the ratio of determinants can be reexpressed as a single spectral determinant
\be
Q_N(u) = \text{det}\Big(u - L \Big)\,,
\ee
where
\be
L_{ij} = \frac{\prod\limits_{k \neq j}^N \left( \tau_i \, \eta - \tau_k \, \eta^{-1} \right) }{\prod\limits^N_{k \neq i} \left( \tau_i-\tau_k\right) }   p_\tau^j\,,
\label{eq:tRSLAX}
\ee
is the Lax matrix of the $N$-body complex trigonometric RS system. At the final stage of the recursion, the polynomial $Q_N(u)$ becomes the matter polynomial $P(u) = \prod\limits_{j=1}^N(u-\mu_j)$, providing us with the required relation \eqref{eq:tRSW}. By expanding both sides of \eqref{eq:tRSW} in $u$ we find explicitly the Hamiltonians 
\begin{equation}
 \text{det}\left(u - L(\tau_i, p_\tau^i,\eta) \right) = \sum_{r=0}^N T_r(\tau_i, p_\tau^i,\eta) u^r\,,
\label{eq:tRSLaxDecomp}
\end{equation}
and their eigenvalues 
\begin{equation}
\prod\limits_{j=1}^N(u-\mu_j) = \sum_{r=0}^N \chi_r(\mu_i) u^r\,,
\end{equation}
Thus we can explicitly write the full set of conserved charges for trigonometric RS system $T_r(\tau_i, p_\tau^i,\eta) = \chi_r(\mu_i)$, or, more explicitly as
\begin{equation}
\sum_{\substack{\mathcal{I}\subset\{1,\dots, N\} \\ |\mathcal{I}|=r}}\prod_{\substack{i\in\mathcal{I} \\ j\notin\mathcal{I}}}\frac{\tau_i\, \eta-\tau_j \, \eta^{-1}}{\tau_i-\tau_j}\prod\limits_{k\in\mathcal{I}}p_\tau^k = \sum_{\substack{\mathcal{I}\subset\{1,\dots, N\} \\ |\mathcal{I}|=r}} \prod_{k\in\mathcal{I}}\mu_k\,,
\label{eq:tRSRelationsEl}
\end{equation}
where $r=0,1,\ldots,N$. 

\subsubsection{Magnetic Frame}
Let us now look at the other presentation of twisted chiral ring \eqref{eq:tRST}. Now we want to eliminate $\sigma_\al^{(j)}$ in favor of the momentum conjugate to the masses, $p_{\mu_j}$. In this case, it will not be possible to provide an argument that lands directly on the Lax matrix formulation of the complex trigonometric RS model. Instead, we attempt to verify the mirror equations
\begin{equation}
\sum_{\substack{\mathcal{I}\subset\{1,\dots, N\} \\ |\mathcal{I}|=r}}\prod_{\substack{i\in\mathcal{I} \\ j\notin\mathcal{I}}} \frac{\mu_i\, \eta^{-1} -\mu_j \, \eta}{\mu_i-\mu_j}\prod\limits_{k\in\mathcal{I}}p_\mu^k = \sum_{\substack{\mathcal{I}\subset\{1,\dots, N\} \\ |\mathcal{I}|=r}} \prod\limits_{k\in\mathcal{I}}\tau_k\,,
\label{eq:tRSRelationsM}
\end{equation}
related to those above by $\tau_j \leftrightarrow \mu_j$ and $\eta \leftrightarrow \eta^{-1}$.

Let us first consider the first independent Hamiltonian with $r=1$. The momentum conjugate to the masses \eqref{eq:ConjMomenta} can be expressed in terms of the polynomials $Q_j(u)$ as follows
\be
p_{\mu}^\alpha 
= \tau_N (\eta^{-1})^{N-1} \frac{Q_{N-1}^+(\sigma_\al^{(N)} )}{Q_{N-1}^-(\sigma_\al^{(N)} )} \,,
\ee
where we remind the reader that $\sigma^{(N)}_\al = \mu_\al$. It is now straightforward to see that the first Hamiltonian can be expressed as a contour integral
\be
\sum_{\al=1}^N \Bigg[ \, \prod_{\beta \neq \al} \frac{\mu_\al\, \eta^{-1} -\mu_\beta \, \eta}{\mu_\al-\mu_\beta} \, \Bigg] p_\mu^\al 
= \tau_N \frac{\eta^2 }{1-\eta^2} \oint_{C_N} \frac{du}{u} \frac{Q^{--}_N(u)}{Q_N(u)}\frac{Q^{+}_{N-1}(u)}{Q^{-}_{N-1}(u)}\,,
\label{eq:1dpartitionftn-r=1}
\ee
where the contour $C_N$ surrounds the roots of $Q_N(u)$ i.e. the masses $\mu_\al$. Our proposition is that this contour integral evaluates to $\tau_1+\cdots +\tau_N$.

We prove this proposition by induction. To perform the inductive step, we contract the contour $C_N$ such that it surrounds the roots $\sigma^{N-1}_\al$ of $Q^{-}_{N-1}(u)$, $u=0$ and $u=\infty$, then eliminate the dependence on $Q_N(\sigma_\al^{(N-1)})$ using the supersymmetric vacuum equations \eqref{eq:BetheEqQ}, and then express the result once again as contour integral. Performing these steps, we find
\bea
\tau_N \frac{\eta^2 }{1-\eta^2} \oint_{C_N} \frac{du}{u} \frac{Q^{--}_N(u)}{Q_N(u)}\frac{Q^{+}_{N-1}(u)}{Q^{-}_{N-1}(u)} 
& = \tau_N - \tau_N \frac{\eta^2 }{1-\eta^2} \oint_{C_{N-1}} \frac{du}{u} \frac{Q^{-}_N(u)}{Q^{+}_N(u)}\frac{Q^{++}_{N-1}(u)}{Q_{N-1}(u)} \\
& =\tau_N + \tau_{N-1} \frac{\eta^2 }{1-\eta^2} \oint_{C_{N-1}} \frac{du}{u} \frac{Q^{--}_{N-1}(u)}{Q_{N-1}(u)} \frac{Q^{+}_{N-2}(u)}{Q^{-}_{N-2}(u)} \\
& =\tau_N + \tau_{N-1} + \ldots + \tau_1\,,
\eea
as required. In the second line above contour $C_{N-1}$ surrounds only roots of $Q_{N-1}(u)$.

The argument for the Hamiltonian appearing at order $u^r$ proceeds in a similar manner. We first express the Hamiltonian as a contour integral
\bea
\sum_{\substack{\mathcal{I}\subset\{1,\dots, N\} \\ |\mathcal{I}|=r}}\prod_{\substack{i\in\mathcal{I} \\ j\notin\mathcal{I}}} \frac{\mu_i\, \eta^{-1} -\mu_j \, \eta}{\mu_i-\mu_j}\prod\limits_{k\in\mathcal{I}}p_\mu^k & = 
\oint_{C_N} \frac{du_1}{u_1} \ldots \frac{du_r}{u_r} \frac{\prod\limits_{m \neq n} (u_m-u_n)}{\prod\limits_{m,n=1}^r(\eta^{-2}u_m-u_n)} \\
& \times \tau_N^r \prod_{m=1}^r \frac{Q^{--}_N(u_m)}{Q_N(u_m)}\frac{Q^{+}_{N-1}(u_m)}{Q^{-}_{N-1}(u_m)}\, .
\eea
where the contour $C_N$ surrounds the poles arising from the denominators $Q_N(u_m)$. Note the critical role of the numerator factor $\prod\limits_{m \neq n}(u_m-u_n)$ in ensuring that the non-zero residues are labelled by sets $i_1<\cdots <i_r$. This contour integral is the path integral of a supersymmetric gauged quantum mechanics on the circle, an observation we explain further below. Our claim is that this contour integral evaluates to
\be
\chi_r(\tau) = \sum_{i_1<\ldots i_r} \tau_{i_1} \cdots \tau_{i_r} \, .
\ee
To prove this statement, we again proceed by induction. The inductive step depends on the following contour integral identity
\bea
& \oint_{C_N} \frac{du_1}{u_1} \ldots \frac{du_r}{u_r} \frac{\prod\limits_{m \neq n} (u_m-u_n)}{\prod\limits_{m,n=1}^r(\eta^{-2}u_m-u_n)}  \tau_N^r \prod_{m=1}^r \frac{Q^{--}_N(u_m)}{Q_N(u_m)}\frac{Q^{+}_{N-1}(u_m)}{Q^{-}_{N-1}(u_m)} \\
&  = \tau_N \oint_{C_{N-1}} \frac{du_1}{u_1} \ldots \frac{du_{r-1}}{u_{r-1}} \frac{\prod\limits_{m \neq n} (u_m-u_n)}{\prod\limits_{m,n=1}^{r-1}(\eta^{-2}u_m-u_n)}  \tau_{N-1}^{r-1} \prod_{m=1}^{r-1} \frac{Q^{--}_{N-1}(u_m)}{Q_{N-1}(u_m)}\frac{Q^{+}_{N-2}(u_m)}{Q^{-}_{N-2}(u_m)} \\
& +  \oint_{C_{N-1}} \frac{du_1}{u_1} \ldots \frac{du_r}{u_r} \frac{\prod\limits_{m \neq n} (u_m-u_n)}{\prod\limits_{m,n=1}^r(\eta^{-2}u_m-u_n)}  \tau_{N-1}^r \prod_{m=1}^r \frac{Q^{--}_{N-1}(u_m)}{Q_{N-1}(u_m)}\frac{Q^{+}_{N-2}(u_m)}{Q^{-}_{N-2}(u_m)} \\
& = \tau_N \chi_{r-1}(\tau_1,\ldots,\tau_{N-1}) + \chi_{r}(\tau_1,\ldots,\tau_{N-1}) \\
& = \chi_r(\tau_1,\ldots,\tau_N) \,,
\label{eq:1dcontour}
\eea
which we have checked in numerous examples. 

We expect that this expression as well as \eqref{eq:1dpartitionftn-r=1} can be interpreted as  partition functions of quantum mechanics on the 1d supersymmetric defect on $S^1$ coupled to the 3d gauge theory on $S^1\times \mathbb{R}^2$. The integral relation~\eqref{eq:1dcontour} can be interpreted as an identity between the partition functions of 1d defects coupled to neighboring nodes of the quiver. This observation requires further study.

\subsection{Line Operators and Interfaces in $\cN=4$ SYM}

Many of the computations presented in the preceding section, in particular the connections to classical integrable systems, have a useful interpretation in terms of interfaces between copies of four-dimensional $U(N)$ $\cN=2^*$ theory. 

The starting point for this construction is the moduli space of vacua $\cM$ of $U(N)$ $\cN=2^*$ theory on $S^1\times \mathbb{R}^3$. There is always a region at infinity in the moduli space where the gauge group is broken to the maximal abelian subgroup $U(1)^N$ and complex coordinates $(\al_i, p_{\al_i})$ valued in $(\mathbb{C}^*)^{2N}$ corresponding to complexified electric and magnetic Wilson lines in each abelian factor, complexified by vectormultiplet scalars. Classically, we would have $\cM^{\mathrm{cl}} = (\mathbb{C}^*)^{2N}$, which is the phase space of the complex trigonometric RS system \cite{Gaiotto:2013bwa}.

The quantum corrected moduli space $\cM$ in the appropriate complex structure is given by the space of $GL(N)$ flat connections on a torus with puncture. This can be described by the holonomies $A$ and $B$ around the two cycles of the torus, which must obey $A B A^{-1} B^{-1} = E$ where $E$ has eigenvalues $\eta^{-2},\ldots,\eta^{-2},\eta^{2N-2}$. Remarkably, these equations can be solved in terms of coordinates $(\al_i, p_{\al_i})$ such that
\bea
\mathrm{Tr}_{\Lambda^r}(A) & = \sum_{\substack{\mathcal{I}\subset\{1,\dots, N\} \\ |\mathcal{I}|=r}} \prod\limits_{i\in\mathcal{I}}\al_i \\
\mathrm{Tr}_{\Lambda^r}(B) & = \sum_{\substack{\mathcal{I}\subset\{1,\dots, N\} \\ |\mathcal{I}|=r}}\prod_{\substack{i\in\mathcal{I} \\ j\notin\mathcal{I}}} \frac{\al_i\, \eta^{-1} -\al_j \, \eta}{\al_i-\al_j}\prod\limits_{i\in\mathcal{I}}p_\al^i \, .
\eea
In particular, there is a choice of gauge where $B$ becomes the Lax matrix of the complexified trigonometric RS system. It can be confirmed by localization computations that $\mathrm{Tr}_{\Lambda^r}(A)$ and $\mathrm{Tr}_{\Lambda^r}(B)$ correspond respectively to BPS Wilson and 't Hooft loops in the anti-fundamental representations of $U(N)$ wrapping the $S^1$.

Interfaces between two theories with moduli spaces $\cM_L$ and $\cM_R$ correspond to Lagrangian submanifolds $\cL \subset \cM_L \times \cM_R$. Let us recall that the three-dimensional $T[U(N)]$ theory can be identified with the S-duality interface for the $U(N)$ $\cN=2^*$ theory. In this context the complex parameters of the three-dimensional theory $(\mu_i, p_{\mu}^i)$ and $(\tau_i , p_{\tau}^i)$ are identified with the Darboux coordinates for the moduli space on either side of the interface. The corresponding Lagrangian submanifold $\cL$ is then described precisely by the equations \eqref{eq:tRSW} and \eqref{eq:tRST}. These relations are interpreted as Ward identities for line operators at the interface: a 't Hooft loop approaching from one side is equivalent to a Wilson loop approaching from the other. This is expected since Wilson loops and 't Hooft loops are interchanged under S-duality. 

Similarly, the contour integral relations in equation \eqref{eq:1dcontour} can be interpreted as boundary Ward identities for 't Hooft loops at an interface between $\cN=2^*$ theories with gauge groups $U(N)$ and $U(N-1)$. This interpretation are discussed further in \secref{Sec:Interfaces}. 

\subsection{Quantum Equivariant K-theory}\label{Sec:EqQuntKThNak}
From the work of Nekrasov and Shatashvili \cite{Nekrasov:2009ui, Nekrasov:2009uh} it is known that the twisted chiral ring of 3d $\cN=4$ quiver gauge theories on $S^1 \times \mathbb{R}^2$ should provide a representation of the equivariant quantum K-theory ring of the Higgs branch. For generic linear quivers the Higgs branch is a quiver variety introduced by Nakajima~\cite{nakajima1994, nakajima19944, nakajima1998}. For our main example $T[U(N)]$ this is the cotangent bundle to the $N$-dimensional complete flag variety.  

It is also known from the work Givental and collaborators \cite{givental1995,1996alg.geom7001K,2001math8105G} that there is a deep connection between equivariant quantum cohomology / quantum K-theory and classical many-body integrable systems. The results of this sections can be interpreted in this light: the equivariant quantum K-theory of the cotangent bundle to a $N$-dimensional complete flag is determined by the $N$-body trigonometric RS integrable system. Later in \cite{Braverman:2005kq} the results of Givental and Lee were proved using different methods which utilize the action of quantum groups. This method is also applicable to affine quiver varieties of type-A.

We can therefore formulate the following
\begin{proposition}
The $T$-equivariant quantum K-ring of the cotangent bundle to the complete N-dimensional complex flag variety is given by
\begin{equation}
QK^\bullet_T(T^*\mathbb{F}_N)\simeq \mathbb{C}\left[(p_\tau^i)^{\pm1},\tau_i^{\pm},\eta^{\pm},\mu_i^{\pm}\right]/\mathcal{I}\,,\quad i=1,\dots, N\,,
\end{equation}
where ideal $\mathcal{I}$ is given by relations \eqref{eq:tRSRelationsEl} and $T$ is the maximal torus of $U(N)\times U(1)$ with equivariant parameters $\mu_1,\dots \mu_N$ for $U(N)$ and equivariant parameter $\eta$ for $U(1)$. The correspondence between physical and geometrical parameters is summarized in \Tabref{tab:PhysGeomCorr}.
\label{Th:NakajimaComplete}
\end{proposition}
\begin{table}[!h]
\begin{center}
\begin{tabular}{|c|c|}
\hline
Twisted chiral ring of 3d theory  & Quantum Equivariant K-theory \\
\hline \hline
FI parameters $\tau_i$ & Quantum deformation parameters \\
\hline
$\CN=2^*$ mass $ \eta $ & Equivariant parameter for $U(1)_\ep$ \\
\hline
Hypermultiplet masses $\mu_i$ & Equivariant parameters for $U(N)$ \\
\hline
\end{tabular}
\caption{Identification of mass parameters of 3d quiver theories and equivariant quantum K-rings. \label{tab:PhysGeomCorr}}
\end{center}
\end{table}
Note that we have kept equivariant parameters appropriate for $U(N)$ (rather than $SU(N)$) global symmetry. For $SU(N)$ symmetry one needs to add additionally impose the conditions $p_\tau^1 \cdot\dots\cdot p_\tau^N = \mu_1\cdot\dots\cdot\mu_N=1$ so that one equivariant parameter and one generator are excluded.

In the limit $\eta\to 1$ twisted chiral ring relations \eqref{eq:tRSRelationsEl} have the simple solutions $p_\tau^i=\mu_{\sigma(i)}$ for any permutation $\sigma$. In terms of trigonometric RS model, this limit describes $N$ non-interacting relativistic particles. In terms of 3d gauge theory, this describes the point in the parameter space where the supersymmetry is enchanted from $\CN=2^*$ to $\CN=4$. 

Recently we became aware of reference \cite{2014arXiv1411.0478R} where quantum K-rings of flag varieties were discussed. The conjecture the authors make about the quantum K-ring is in agreement with Proposition \ref{Th:NakajimaComplete}.

\subsection{More Genetic Quiver Varieties}

So far we have considered only the triangular quiver $T[U(N)]$ which has the property of being self-dual under three-dimensional mirror symmetry. We now briefly discuss how aspects of more general linear quivers can be studied through a combination of Higgsing and mirror symmetry.

For generic hypermultiplet masses and FI parameters, there are only discrete massive supersymmetric vacua on $S^1 \times \mathbb{R}^2$. However, by tuning the mass parameters $m_a$ (or FI parameters $t_a$ in the mirror frame) one can open up a Higgs branch. Moving out onto this branch and flowing to the infrared we can partially Higgs the theory to obtain a new quiver.

Using Type IIB brane constructions one can develop an algorithm for constructing a generic quiver from $T[U(N)]$ by repeating this procedure together with mirror symmetry. In terms of the parameter space of hypermultiplet mass deformations, Higgsing corresponds to restriction to a certain subvariety specified by the Higgs branch locus of the original quiver theory (see  \cite{Gaiotto:2013bwa} for examples). Below we shall illustrate the idea using couple of simple examples.

\subsubsection{Higgsing $T[U(2)]$}
The twisted chiral ring relation of $T[U(2)]$ theory is equivalent to the spectral curve of the 2-body trigonometric RS model as presented in equation \eqref{eq:tRSrelationsTU2}. These equations reproduce the Bethe equations upon inserting the on-shell values of the momenta \eqref{eq:momentarelationsTU2}. In order to Higgs theory we impose we impose the condition
\begin{equation}
\eta^{-1}\mu_1 = \eta\mu_2=\sigma_1:=\mu\,
\label{eq:Higgsingmu}
\end{equation}
from which we find $p_\tau^1=p_\tau^2=\mu$ from the definition of the momenta \eqref{eq:momentarelationsTU2}. Equivalently, inserting the condition \eqref{eq:Higgsingmu} into the spectral curve \eqref{eq:tRSrelationsTU2} we find
\begin{equation}
\left(\eta +\eta^{-1}\right) \left(\tau _1-\tau _2\right)=\tau _1
   \left(\frac{\mu \eta}{p_\tau^1}+\frac{p_\tau^1}{\eta\mu}\right)-\tau _2 \left(\frac{\mu }{\eta  p_\tau^1}+\frac{p_\tau^1\eta}{\mu }\right)\,,
\end{equation}
which determines $p_\tau^1=\mu$. After Higgsing $T[U(2)]$ theory becomes free, hence its chiral ring relations are trivial. 

\subsubsection{Higgsing $T[U(3)]$}
The twisted chiral ring of $T[U(3)]$ theory computes the equivariant quantum K-ring of $T^*\mathbb{F}_3$. In what follows we shall study vacua of $T[U(3)]$ theory subject to a Higgsing condition similar to \eqref{eq:Higgsingmu} and compute quantum K-ring for the Nakajima quiver variety corresponding to the Higgsed theory. From 3d mirror symmetry it can be shown that the quantum K-ring of the latter variety is canonically isomorphic to the quantum K-ring $T^*\mathbb{P}^2$.

\begin{figure}[!h]
\begin{center}
\includegraphics[scale=1.0]{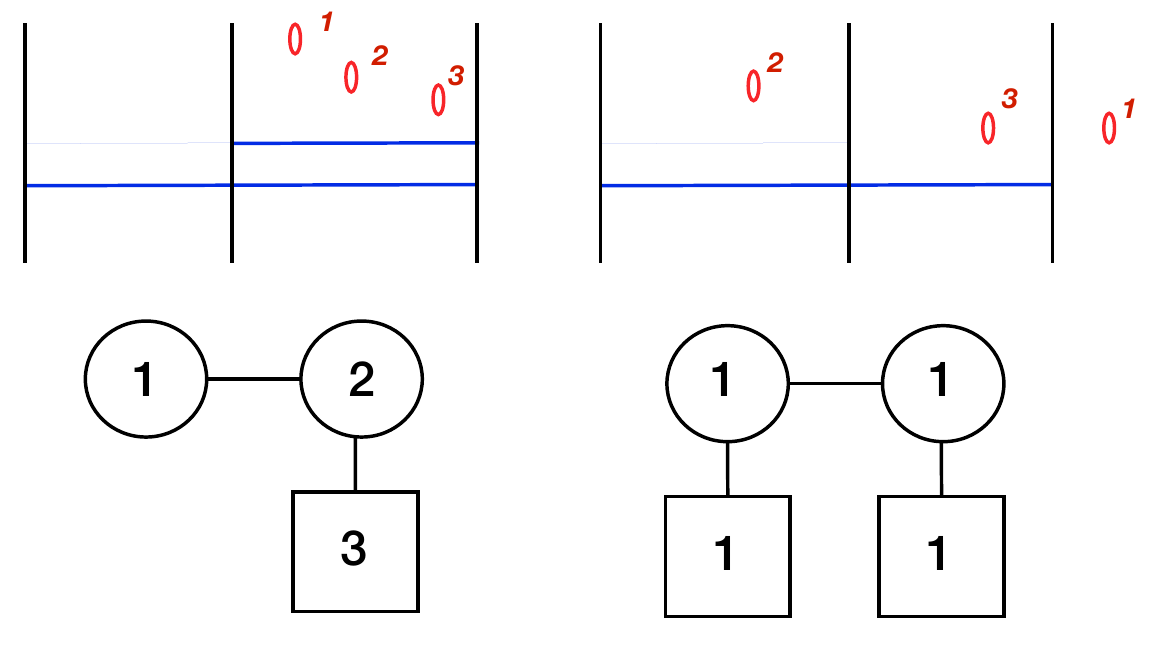}
\caption{Higgsing $T[U(3)]$ theory using Type IIB brane construction. Vertical black lines denote NS5 branes along directions 012789, horizontal blue lines denote D3 branes along 0123 directions, and oval red circles stand for D5 branes stretched along 012456 directions. The left side of the figure describes $T[U(3)]$ theory, whereas the right side shows how to obtain $A_2$ quiver with labels $(1,1)(1,1)$ by Higgsing flavor branes 1 and 2. After applying the S-duality to the right figure D5 and NS5 branes switch roles, and D3 branes are self dual. The newly obtained structure of two NS5 branes and three D5 branes describes $U(1)$ theory with three flavors.}
\label{Fig:TU31Higgsing}
\end{center}
\end{figure}

The $T[U(3)]$ theory corresponds to the three-particle trigonometric RS model which is described by the Hamiltonians given in \eqref{eq:tRSRelationsEl}. As we Higgs the 3d theory we impose (see \figref{Fig:TU31Higgsing})
\begin{equation}
\eta^{-1}\mu_1 = \eta\mu_2=\sigma^{(2)}_1:=\mu\,,
\label{eq:HiggsingTU3}
\end{equation}
where $\sigma^{(2)}_1$ is one of the Coulomb branch parameters for the $U(2)$ gauge group (second node of the quiver).
The theory is therefore reduced to the $A_2$ quiver with labels $(1,1)(1,1)$; its mirror is $A_1$ quiver with labels $(1,3)$, or in other words (which lead to $T^*\mathbb{P}^2$), the $U(1)$ theory with three flavors. 

Let us first describe the chiral ring relations for the $A_2$ quiver. Constraint \eqref{eq:HiggsingTU3} affects the r.h.s. of \eqref{eq:tRSRelationsEl} by imposing the relationship between the masses and the l.h.s by restricting some of the conjugate momenta $p_\tau$
\begin{equation}
p_\tau^1 = \sigma^{(1)}\,,\quad p_\tau^2 = \frac{\sigma^{(2)}_1\sigma^{(2)}_2}{\sigma^{(1)}}\,,\quad p_\tau^3 = \frac{\mu_1\mu_2\mu_3}{\sigma^{(2)}_1\sigma^{(2)}_2}\,,
\end{equation}
as follows
\begin{equation}
p_\tau^1 = \sigma^{(1)}\,,\quad p_\tau^2 = \frac{\mu\sigma^{(2)}_2}{\sigma^{(1)}}\,,\quad p_\tau^3 = \frac{\mu\mu_3}{\sigma^{(2)}_2}\,.
\end{equation}
which are consistent with the definition of the conjugate momenta for $(1,1)(1,1)$ quiver, where $\mu=\eta\mu_2$ is the mass of the hypermultiplet on the first node of the quiver.
Therefore the only difference between chiral rings on the $T[U(3)]$ (or $(3,2)(1,0)$) theory and the $(1,1)(1,1)$ theory -- one simply imposes \eqref{eq:HiggsingTU3} on the mass parameters
\begin{align}
T_1 = (\eta^{-1}+\eta)\mu + \mu_3\,,\notag\\
T_2 = (\eta^{-1}+\eta)\mu\mu_3+\mu^2\,,\notag\\
T_3 = \mu^2\mu_3\,,
\label{eq:TU3HamsGenHiggsedm}
\end{align}
where $T_r$ are defined in \eqref{eq:tRSLaxDecomp}. One can also explicitly check that the Bethe equations for $(1,1)(1,1)$ theory are equivalent to \eqref{eq:TU3HamsGenHiggsedm}. 

It is instructive to see how Higgsing applies in the magnetic frame, namely while using $\mu_i$ and $p_\mu^i$ in tRS Hamiltonians \eqref{eq:tRSRelationsM}. Thus we want to derive vacua equations for the Higgsed $T[U(3)]$ together with the conjugate momenta $p_\mu^i$ and compare them with the data of the $(1,1)(1,1)$ quiver.

Let us evaluate the effective twisted superpotential \eqref{eq:T[U(N)]w} for $T[U(3)]$ theory at the locus of the Higgs branch \eqref{eq:HiggsingTU3}, which can be written in the original variables as 
\begin{equation}
m_1 = m_2 + \epsilon\,,\quad s^{(2)}_1=m_2+\frac{\epsilon}{2}\,.
\end{equation}
Using the twisted effective superpotential we can compute conjugate momenta to mass parameters
\begin{equation}
p_\mu^2=\tau _2 \tau _3\frac{ \left(\eta  \mu _2-\mu _3\right)
   \left(\eta  \mu _2-\sigma _1\right)}{\left(\mu _2-\eta  \mu
   _3\right) \left(\eta  \sigma _1-\mu _2\right)}\,,\quad p_\mu^3=\tau _3\frac{
   \left(\mu _2-\eta  \mu _3\right) \left(\eta  \mu _3-\sigma
   _2\right)}{\left(\eta  \mu _2-\mu _3\right) \left(\eta 
   \sigma _2-\mu _3\right)}\,,
\end{equation}
where we have substituted $\sigma_1^{(1)}=\sigma_1$ and $\sigma^{(2)}_2=\sigma_2$. We can see that the above expressions reproduce the corresponding conjugate momenta for the $A_2$ quiver with labels $(1,1)(1,1)$
\begin{equation}
p_\mu^{(1)}=\tau _2 \tau _3\frac{ 
   \eta  \mu _2-\sigma _1}{\eta  \sigma _1-\mu_2}\,, \quad p_\mu^{(2)}=\tau _3\frac{\eta  \mu _3-\sigma
   _2}{\eta\sigma _2-\mu _3}\,,
\label{eq:Momenta1111}
\end{equation}
up to rational function
\begin{equation}
A = \frac{\eta  \mu _2-\mu _3}{\mu _2-\eta  \mu_3}\,,
\label{eq:ContactTermTU3}
\end{equation}
namely, $p_\mu^2 = A p_\mu^{(1)}$ and $p_\mu^3 = A^{-1} p_\mu^{(2)}$. Note that $-A$ provides a contribution of the free fundamental hypermultiplet of mass $m_3-m_2$ to the effective twisted superpotential. We can interpret contact term \eqref{eq:ContactTermTU3} as coming from the global symmetry of the Higgs branch of $T[U(3)]$ theory. After Higgsing the dependence on $\mu_1$ (the position of particle 1 on the right of \figref{Fig:TU31Higgsing}) and its momentum disappears. 

Let us now explicitly describe the quantum K-ring of $T^*\mathbb{P}^2$ using vacua equations. Since $A_2$ quiver with labels $(1,1)(1,1)$ is mirror dual to $U(1)$ theory with three flavors we can use vacua equations of the latter theory to derive the desired chiral ring relations. In the mirror variables one gets
\begin{equation}
\mu_2 \prod_{j=1}^3 (\eta\sigma -\tau_j) + \mu_3\prod_{j=1}^3(\eta \tau_j -\sigma) = 0\,,
\label{eq:BetheEq1w3}
\end{equation}
together with momenta conjugate to the FI parameters 
\[
 p_2=\sigma\,, \qquad p_3 = \frac{\tau_1\tau_2\tau_3}{\sigma} \,,
\]
and hence $p_2 p_3 = \tau_1\tau_2\tau_3$. We can expand the vacuum equations in powers of $\sigma$ and then eliminate $\sigma$ using $p_2:=p$. By doing so \eqref{eq:BetheEq1w3} becomes 
\begin{equation}
[\mu_{23}]_{-1}\,p^3 -\chi_1(\tau)[\mu_{23}]_0\, p^2 +\chi_2(\tau)[\mu_{23}]_1 \,p - \chi_3(\tau)[\mu_{23}]_2  =0\,,
\label{eq:U13p3eq}
\end{equation}
where we defined
\begin{equation}
[\mu_{23}]_{a}=\frac{\eta^a\mu_2-\eta^{-a}\mu_3}{\mu_2-\mu_3}\,,
\end{equation}
so $[\mu_{23}]_0=1$. Equivalently we could have started with vacua equations for $(1,1)(1,1)$ quiver and using momenta defined in \eqref{eq:Momenta1111} derive \eqref{eq:U13p3eq}.

With more experimentation we can find the general formula for the $A_1$ quiver with $N$ fundamental hypermultiplets
\[
\sum_{a=0}^N (-1)^a [\mu]_{a-1} \chi_{a}(\tau_i) \,p^a = 0\,,
\label{eq:HypergeomEqClass}
\]
where $[\mu]_{a}=\frac{\eta^a\mu-\eta^{-a}}{\mu-1}$ and $\chi_a$ stand for characters of the $a$-th antisymmetric tensor representation of $\mathfrak{sl}_{N}$ (for the trivial representation $s_0=1$). Interestingly, the above formula provides the classical limit of an $N$-th order difference equation for the $q$-hypergeometric series ${}_{N}F_{N-1}$. Indeed, if we promote $p$ to a shift operator \eqref{eq:HypergeomEqClass} is nothing but the functional equation for quantum ${}_{N}F_{N-1}$ function.  It is also well known that holomorphic blocks of the $U(1)$ theory with $N$ flavors obey exactly the same equation\footnote{It can presumably be interpreted as Ward identity for line operators.}.

We can therefore make the following
\begin{corollary}
$T$-equivariant quantum K-ring of $T^*\mathbb{P}^N$ is isomorphic to
\begin{equation}
QK^\bullet_T(T^*\mathbb{P}^N)\simeq \mathbb{C}\left[p^{\pm 1},(\tau_i)^{\pm 1},\eta^{\pm 1},\mu^{\pm 1}\right]/\mathcal{I}\,, \quad i=1,\dots,N\,,
\end{equation}
where ideal $\mathcal{I}$ is given by relation \eqref{eq:HypergeomEqClass}.
\end{corollary}

Note that in the limit when K\"{a}hler parameter vanishes $\mu\to 0$ we have $[\mu]_{a-1}\to \eta^{a-1}$. Therefore, by redefining $p_a\to \eta^{1-a} p_a$ we obtain that the ideal given in \eqref{eq:HypergeomEqClass} is reduced to 
\begin{equation}
(p-\tau_1)\cdot\dots\cdot(p-\tau_N)\,,
\label{eq:PNKringRel}
\end{equation}
which is in the agreement with the known result about classical K-rings and cohomology rings of complex projective spaces. Note also that the dependence on $\eta$ drops out suggesting that the action of $U(1)_\eta$ is trivial in this case; which confirms the known fact that equivariantly $T^*\mathbb{P}^N$ retracts to $\mathbb{P}^N$.

\pagebreak
\section{3d Partition Functions}\label{Sec:3dpf}
In \Secref{Sec:TwistChiralRings}, we explained how twisted chiral ring of $T[U(N)]$ theory is encoded in the spectral curve of a classical integrable system of $N$ interacting relativistic particles called the trigonometric RS system.

In this section, we will quantize this classical integrable system by studying $T[U(N)]$ theories on a curved three-dimensional background. For most of this section, we concentrate on squashed $S^3$, where the quantization parameter is identified with squashing parameter traditionally denoted by $b$ (and, as we explain below, also $b^{-1}$). In this background, the conjugate momenta to $\tau_i$ and $\mu_i$ are promoted to difference operators $p_\tau^i$ and $p_\mu^i$ and the spectral curve becomes an operator equation, which annihilates the partition function. From the squashed $S^3$ partition function, we will obtain the partition function of $T[U(N)]$ on $S^1 \times \mathbb{R}^2$ by factorization. 

\subsection{$S^3$ Partition function}
Let us begin by summarizing the supersymmetric partition function of 3d $\cN=4$ gauge theories on squashed $S^3$, as computed using supersymmetric localization in \cite{Kapustin:2009kz,Hama:2010av, Hama:2011ea}. This partition function depends on a squashing parameter $b >0$ and real mass parameters valued in the Cartan subalgebra of the global symmetry group $G$. The latter are introduced by coupling to a background $\cN=2$ vectormultiplet for $G$ and giving a vacuum expectation value to the real scalar. For the most part we consider $U(N)$ flavor symmetries and use shorthand notation $\vec m = (m_1,\ldots,m_N)$. 

As in the previous section, it is convenient to introduce exponentiated mass parameters. The conjugate momenta now become elementary difference operators. For a $U(N)$ global symmetry we define in this section 
\be
\mu_j = e^{2\pi b m_j} \qquad p_{\mu}^j = e^{i b \del_{m_j}} 
\ee
They obey the following relation
\be
p_{\mu}^i \mu_j = q^{\delta_{ij}}\mu_j p_{\mu}^i \,,
\ee
where $q = e^{2\pi i b^2}$. We will also introduce a second set of exponentiated mass parameters whose variables denoted by $\widetilde\mu_j$, $p_{\widetilde\mu}^j$ and $\widetilde q$. They are defined in exactly the same fashion except we replace $b \to b^{-1}$. 

The partition function can also depend on real FI parameters for $U(N)$ gauge groups. The real FI parameter $t$ associated to a $U(N)$ gauge group is a real mass parameter for the topological $U(1)$. The corresponding exponentiated FI parameters $\tau$, $\widetilde\tau$ and momenta $p_{\tau}$, $p_{\widetilde\tau}$ are defined as above.

As in \secref{Sec:TwistChiralRings}, it will be important to turn on a real mass parameter $m$ for the diagonal combination $U(1)_\varepsilon \subset U(1)_H \times U(1)_C$ breaking to $\cN=2^*$ supersymmetry in three dimensions. It is convenient to introduce the notation 
\be
\varepsilon = b_+ + im\,, \qquad \varepsilon^* = b_+ -im \,,
\ee
together with the exponentiated parameters 
\begin{equation}
\eta^2 = q\, e^{-2 \pi i b \varepsilon}=e^{2\pi i b(b_--im)}\,,
\label{eq:etaqdef}
\end{equation}
where we define $b_\pm=(b\pm b^{-1})/2$. We will not need to consider the conjugate momentum for this symmetry.

Let us now summarize the building blocks of the squashed $S^3$ partition function of 3d $\cN=2^*$ quiver gauge theories. They are built from bifundamental hypermultiplets, vectormultiplets and FI parameters,
\begin{enumerate}
\item $U(N_1) \times U(N_2)$ bifundamental hypermultiplet:
\be
Q_{N_1,N_2}(m^{(1)},m^{(2)}) = \prod_{i=1}^{N_1} \prod_{j=1}^{N_2} S\left( \frac{\varepsilon^*}{2} + im^{(1)}_{i} - im^{(2)}_{j} \right)S\left( \frac{\varepsilon^*}{2} - im^{(1)}_i + im^{(2)}_j \right)
\ee
\item $U(N)$ vectormultiplet: 
\bea
\nu_N(m) & = \prod\limits_{\substack{ i,j=1 \\ i\neq j}}^N S(im_i-im_j)^{-1} \prod\limits_{i,j=1}^N S\left(\varepsilon+im_i - im_j\right)  \\
& = (-1)^{\frac{N(N-1)}{2}}\prod\limits_{ i < j }^N  \, 2 \sinh  \pi b (m_i - m_j ) \, 2 \sinh \pi b^{-1} (m_i - m_j)  \\  
& \qquad \times \prod\limits_{i,j=1}^N S\left(\varepsilon+im_i - im_j\right)
\eea
\item FI parameter for $U(N)$ symmetry
\be
e^{-2\pi i t (m_1 + \cdots + m_N)}
\ee
\end{enumerate}
Turning off the $\cN=2^*$ mass deformation $\varepsilon \to b_+$, the contributions from the adjoint $\cN=2$ chiral multiplet cancel in pairs and we recover the familiar $\cN=2$ vectormultiplet measure involving only the hyperbolic sine function. It is important to include these contribution when $m\neq 0$.

Here we are using the double sine function $S(z) = S_2(z|b,b^{-1})^{-1}$. This is meromorphic in $z$ with simple poles at $z =m b+ n b^{-1}$ for $n,m \leq 0$ and simple zeroes for $n,m \geq 1$. The most important properties we will need are
\bea 
S(z+b^{\pm}) & = 2 \sin(\pi b^{\pm} z ) S(z)\,, \\
S(x)S(b_+-x) & = 1\,,
\eea
where $b_+\equiv b+b^{-1}$. Further properties are summarized in~\appref{appendix:doublesine}. 

The $S^3$ partition function is related to a quantization of the Lagrangian submanifold $\cL$ of section~\ref{Sec:Trig}. The relevant quantization parameter $\hbar$ is related to the squashing parameter by $\hbar = 2\pi b^2$ (for the parameters $\mu$, $p_{\mu}$,\ldots) and $\hbar = 2\pi b^{-2}$ (for the parameters $\tilde \mu$, $\tilde p_{\mu}$,\ldots). In the semi-classical limit $b \to 0$, the $S^3$ partition function schematically has asymptotic behavior 
\be
\CZ_{S^3} = \int d\sigma  \, e^{ \, - 2\pi i \,\cW(\sigma) \, +\, \ldots}
\ee 
where $\cW(\sigma)$ is the effective twisted superpotential with the radius $R$ replaced by the squashing parameter $b$. In this limit, the conjugate momenta become $p_{\mu} = e^{2\pi b\del_m\cW}$ and hence matching their semi-classical counterparts.

\subsection{'t Hooft Operators}
The building blocks for 3d $\cN=4$ quiver gauge theories have an interesting interplay with a system of commuting difference operators, which quantize the classical Hamiltonians studied in section. We will consider two sets of operators acting on the mass parameters $m = (m_1,\ldots,m_N)$ of a $U(N)$ symmetry. The first set is defined by
\bea
T_r(m) & = \sum_{\substack{\cI \subset \{1,\ldots,n\}  \\ | \, \cI \,  |=r} } \, \prod_{\substack{i \in \cI \\ j \notin \cI}} \; \frac{\sin \pi b\left( \epsilon -i m_{ij} \right)}{\sin\pi b\left( -im_{ij} \right)} \, \prod_{j \in \cI} \, e^{ib \del_{m_j}} \\
& = \sum_{\substack{\cI \subset \{1,\ldots,n\}  \\ | \, \cI \,  |=r} } \, \prod_{\substack{i \in \cI \\ j \notin \cI}} \; \frac{q^{1/2} \eta^{-1} \mu_i - \eta q^{-1/2} \mu_j}{\mu_i -\mu_j} \, \prod_{j \in \cI} \, p_{\mu}^j\,,
\eea
where $r=1,\ldots,N$. The second set, which we denote by $\widetilde{T}_r(m)$, is obtained by replacing $b\to b^{-1}$ in the first line or replacing the exponentiated parameters by their tilded counterparts in the second line. It can be shown that the two sets of operators commute among themselves. 

Note that $T_r(m)$ is expressed as a sum over the states of the anti-symmetric tensor representation of $U(N)$ of rank $r$ where the mass parameters $ (m_1,\ldots, m_N)$ are shifted by an amount proportional to the corresponding weight. Similarly, the operators with reversed masses $T_r(-m)$ are associated to the conjugate anti-symmetric tensor representation of $U(N)$ of rank $r$. 

This is no coincidence. The same difference operators appear in the computation of the expectation value of supersymmetric 't Hooft loops in $\cN=4$ SYM theory with gauge group $U(N)$ on squashed $S^4$~\cite{Gomis:2011pf}~\footnote{The expectation value of 't Hooft loops on a round four-sphere were computed in~\cite{Gomis:2011pf}. In the case of a squashed four-sphere the expectation value can be computed under the assumption of the AGT correspondence from Verlinde loop operators in Toda CFT~\cite{Gomis:2010kv}}. For general squashing, the supersymmetric 't Hooft loops can be supported on two Hopf-linked circles corresponding to the operators $T_r(m)$ and $\widetilde{T}_r(m)$ respectively. As shown in references~\cite{Bullimore:2014nla,Bullimore:2013xsa}, the expectation value of a supersymmetric 't Hooft loop in the $r$-th anti-symmetric representation of $U(N)$ can be massaged into the form
\be
\int d^Nm \, \nu_N(m) \, \overline{G(m,\ep,\tau)} \, \left[ \, T_r(m) \cdot G(m,\ep,\tau) \, \right]
\ee
with a similar equation for $\widetilde{T}_r(m)$. 

The wavefunction $G(m,\ep,\tau)$ is expected to be the partition function of $U(N)$ $\cN=4$ SYM on a hemisphere of squashed $S^4$ with Dirichlet boundary conditions at the boundary. There is a global $U(N)$ symmetry at the boundary whose real mass parameter is $m$. The boundary is identified with the squashed $S^3$ geometry we consider in this section. The supersymmetric partition function on $S^4$ is computed by including contributions from each of the hemispheres, reintroducing a 3d $\cN=4$ vectormultiplets with partition function $\nu_N(m)$ and gauging.

The operators $T_r(m)$ obey some remarkable properties in the way they interact with the $S^3$ partition function of 3d $\cN=4$ vectormultiplets and bifundamental hypermultiplets. Each of these properties can be understood in terms of the action of 't Hooft loops on interfaces in 4d $\cN=4$ SYM.

\subsection{Interfaces}\label{Sec:Interfaces}
The first important property of the difference operators is that they are self-adjoint with respect to the measure $\nu_N(m)$ on the Cartan subalgebra of $U(N)$ defined by the $\cN=4$ vectormultiplet contribution. As a preliminary step we show that the vectormultiplet measure obeys the difference equation
\bea
(p_{\mu}^i)^{-1} \, \nu_N(m) 
& = \Bigg[ \, \prod_{j \neq i} \frac{\sin\pi b(-im_{ij}-b)}{\sin\pi b(im_{ij})} \frac{\sin\pi b(\varepsilon+im_{ij})}{\sin\pi b(\varepsilon-im_{ij}-b)} \, \Bigg] \, \nu_N(m) \\
& = \Bigg[ \, \prod_{j \neq i} \frac{\mu_i q^{-1/2}-q^{1/2}\mu_j}{\mu_j-\mu_i} \frac{q^{1/2} \eta^{-1}\mu_j-\eta q^{-1/2}\mu_i}{\eta^{-1} \mu_i - \eta q^{-1/2}\mu_j } \, \Bigg] \, \nu_N(m)
\eea
together with the same equation for the tilded variables. Let us now pick two functions $f(x)$ and $g(x)$ and assume that they have no poles in the region $-b< \Im(x) < b$. Then by deforming the contour integration $x_i \to x_i - b$ for each mass parameter in the set $i \in \cI$ and using the above result, a short calculation shows that
\be
\int d \nu(m)  \, f(m) \, \left[ \, T_r(m) \cdot g(x) \, \right]  = \int d\nu(m) \, \left[\, T_{r}(-m) \cdot f(m) \right] \, g(m) \, .\\
\ee
This relation can be used to `integrate by parts' inside the contour integral expressions for partition functions, provided no poles are crossed in deforming the contour.

This can be interpreted in terms of a trivial interface. Namely, the partition function of $U(N)$ $\cN=4$ theory on a four-sphere can be built from two copies of the hemisphere partition function with Dirichlet boundary conditions by introducing a 3d $\cN=4$ vectormultiplet on the equator. The partition function is constructed by taking $f(m) = \overline{G(m,\ep,\tau)}$ and $g(m) = G(m,\ep,\tau)$. The equation says that the 't Hooft loop is unchanged on passing through the boundary. 

The second property states that the partition function of a $U(N) \times U(N-M)$ bifundamental hypermultiplet intertwines a difference operator with its decomposition under the symmetry breaking pattern $U(N) \to U(N-M)$. In this case we proceed by example before stating the general result.

Let us begin by considering a bifundamental $U(N) \times U(N)$ hypermultiplet. The partition function obeys
\be
\left( T_r(m^{(1)}) - T_r(-m^{(2)}) \right) \, Q_{N,N}(m^{(1)},m^{(2)}) = 0 
\label{conjugaterep}
\ee
together with an isomorphic equation involving $\widetilde T_r(m)$. Firstly, acting on the partition function of bifundamental hypermultiplets with the momenta $p_{\mu^{(1)}}^i$ and $p_{\mu^{(2)}}^i$ and using the difference equation obeyed by the double sine function we have
\be
p^i_{\mu^{(1)}} Q  = \prod_{j=1}^n \frac{\mu^{(1)}_i -  \eta^{-1} \mu_j^{(2)}}{q^{1/2} \eta^{-1} \mu_i^{(1)} -  q^{-1/2}\mu_j^{(2)}} \, Q\,, 
\qquad\qquad
(p^i_{\mu^{(2)}})^{-1} Q  = \prod_{j=1}^n \frac{\mu^{(1)}_j - \eta^{-1} \mu_i^{(2)}}{q^{1/2} \eta^{-1} \mu_j^{(1)} - q^{-1/2} \mu_i^{(2)}} \, Q  \, .
\ee
Using these results the first line of equation~\eqref{conjugaterep} is a consequence of the rational function identity 
\bea
\sum_{\substack{\cI \subset \{1,\ldots,N\} \\ | \, \cI \, | = r }} \, \prod_{\substack{ i \in \cI \\ j \notin \cI }} \, 
\frac{q^{1/2} \eta^{-1} \mu^{(1)}_i - \eta q^{-1/2}\mu^{(1)}_j}{\mu^{(1)}_i -\mu^{(1)}_j }
\prod_{\substack{ i \in \cI \\ j =1,\ldots,N }} \frac{\mu^{(1)}_i - \eta^{-1} \mu_j^{(2)}}{q^{1/2} \eta^{-1} \mu_i^{(1)} - q^{-1/2} \mu_j^{(2)}}\\
= \sum_{\substack{\cI \subset \{1,\ldots,N\} \\ | \, \cI \, | = r }} \, \prod_{\substack{ i \in \cI \\ j \notin \cI }} \, 
\frac{q^{1/2} \eta^{-1} \mu^{(2)}_j - \eta q^{-1/2}\mu^{(2)}_i}{\mu^{(1)}_j -\mu^{(1)}_i }
\prod_{\substack{ i \in \cI \\ j =1,\ldots,N }} \frac{\mu^{(1)}_j -  \eta^{-1} \mu_i^{(2)}}{q^{1/2} \eta^{-1} \mu_j^{(1)} - q^{-1/2} \mu_i^{(2)}}\, .
\label{identity1}
\eea
This has been proven for precisely the same purpose in reference~\cite{2012JMP....53l3512H}. An isomorphic argument applies for the equation involving the operators $\widetilde T_r(m)$. Thus we see that the partition function of $U(N) \times U(N)$ hypermultiplets intertwines the quantum hamiltonians associated with the fundamental and anti-fundamental representations of $U(N)$.

We now consider the more interesting case of $U(N) \times U(N-1)$ bifundamental hypermultiplets. The partition function obeys
\be
\left[ \, T_r(m^{(1)})  - T_r(-m^{(2)})  - T_{r-1}(-m^{(2)})  \, \right] \, Q_{N,N-1}(m^{(1)},m^{(2)}) = 0
\ee
together with an isomorphic equation involving $\widetilde T_r(m)$. This equation is a consequence of the identity
\bea
\sum_{\substack{\cI \subset \{1,\ldots,n\} \\ | \, \cI \, | = r }} \, \prod_{\substack{ i \in \cI \\ j \in \{1,\ldots,n\} / \cI }} \, 
\frac{q^{1/2}\eta^{-1} \mu^{(1)}_i - \eta q^{-1/2} \mu^{(1)}_j }{\mu^{(1)}_i - \mu^{(1)}_j }
\prod_{\substack{ i \in \cI \\ j =1,\ldots,n -1}}  \frac{\mu^{(1)}_i - \eta^{-1} \mu_j^{(2)}}{q^{1/2} \eta^{-1} \mu_i^{(1)} - q^{-1/2} \mu_j^{(2)}} \\
= \sum_{\substack{\cI \subset \{1,\ldots,n - 1\} \\ | \, \cI \, | = r }} \, \prod_{\substack{ i \in \cI \\ j \in \{1,\ldots,n - 1\} / \cI }} \, 
\frac{q^{1/2} \eta^{-1} \mu^{(2)}_j - \eta q^{-1/2}\mu^{(2)}_i}{\mu^{(1)}_j -\mu^{(1)}_i }
\prod_{\substack{ i \in \cI \\ j =1,\ldots,n }}\frac{\mu^{(1)}_j - \eta^{-1/2} \mu_i^{(2)}}{q^{1/2} \eta^{-1} \mu_j^{(1)} - q^{-1/2} \mu_i^{(2)}}\\
+ \sum_{\substack{\cI \subset \{1,\ldots,n - 1\} \\ | \, \cI \, | = r - 1 }} \, \prod_{\substack{ i \in \cI \\ j \in \{1,\ldots,n - 1\} / \cI }} \, 
\frac{q^{1/2} \eta^{-1} \mu^{(2)}_j - \eta q^{-1/2}\mu^{(2)}_i}{\mu^{(1)}_j -\mu^{(1)}_i }
\prod_{\substack{ i \in \cI \\ j =1,\ldots,n }} \frac{ \mu^{(1)}_j - \eta^{-1} \mu_i^{(2)}}{q^{1/2} \eta^{-1} \mu_j^{(1)} - q^{-1/2} \mu_i^{(2)}}\, .
\label{identity2}
\eea
which can be obtained from equation~\eqref{identity1} by replacing $\mu^{(2)}_N \to \mu_N^{(2)} \gamma$ and take the limit $\gamma \to \infty^+$. This corresponds to making $N$ of the hypermultiplets very heavy and integrating them out to recover the bifundamental of $U(N) \times U(N-1)$. The partition function now intertwines a hamiltonian with the conjugate of its decomposition under $U(N) \to U(N-1)$. In terms of representations of $U(N)$ we have
\bea
\Lambda^{(N)}_r \, & \to \, \bar\Lambda_r^{(N-1)} + \bar\Lambda_{r-1}^{(N-1)}  \\
\bar\Lambda^{(N)}_r \, & \to \, \Lambda_r^{(N-1)} + \Lambda_{r-1}^{(N-1)} \, .
\eea
where we have denoted $r$-th anti-symmetric tensor power of the fundamental representation by $\Lambda_r^{(N)}$ and the conjugate representation by $\bar\Lambda_r^{(N)}$. 

Although we will not need it for the triangular quiver $T(U(N)$ it is interesting to study the partition function of $U(N) \times U(N-M)$ bifundamental hypermultiplets. The argument proceeds by induction and the details can be found in. The result of the computation is that
\be
\left( T_r(m^{(1)})  - \sum_{s=0}^{\mathrm{min}(r,M)} \mathrm{Dim}(\eta,\Lambda^{(M)}_s) \, T_{(r-s)}(-m^{(2)}) \right) \, Q_{N,N-M}(m^{(1)},m^{(2)}) = 0 \, ,
\label{intgen}
\ee
where
\be
\mathrm{Dim}(\eta,\Lambda^{(M)}_s)  
=\chi_{\Lambda^{(M)}_s} \left( (q/\eta^2)^{M-1},(q/\eta^2)^{M-3},\ldots, (q/\eta^2)^{1-M} \right) 
\ee
is the quantum dimension of the representation $\Lambda^{(M)}_s$ of $U(M)$ with quantum parameter $\eta$ and in the formula we have defined $\chi_{\Lambda}(x_1,\ldots,x_M)$ to be the Schur polynomial for representation $\Lambda$ of $U(M)$. In the limit $\eta \to1$ the quantum dimension becomes the ordinary dimension. An example is the hamiltonian associated to the fundamental representation of $U(N)$,
\be
\left[ T_1(m^{(1)}) -  T_1(-m^{(2)}) - \left( (q/\eta)^{M-1} + (q/\eta)^{M-3} + \ldots + (q/\eta)^{1-M} \right) \, \right] \, Q_{N,N-M}(m^{(1)},m^{(2)})  =0 
\ee
where we have used $T_0(-m^{(2)})=1$. 

Summarizing, we have found that the partition function of $U(N) \times U(N-M)$ bifundamental hypermultiplets intertwines a difference operator associated to a antisymmetric tensor representation of $U(N)$ with the conjugate of its decomposition under $U(N) \to U(N-M)$. The coefficients in the expansion are the quantum dimensions associated to the number of times that representation appears in the decomposition with quantum parameter $\eta$.

This equation has an interpretation in terms of an interface between two copies of the $\cN=2^*$ theory with gauge groups $U(N)$ and $U(N-j)$ respectively. The hemisphere partition functions are coupled by adding three-dimensional bifundamental hypermultplets of $U(N) \times U(N-M)$ and gauging these symmetries on the interface. We can then interpret the intertwining property as the mathematical statement of how a 't Hooft loop of $U(N)$ decomposes into 't Hooft loops of $U(N-j)$ on passing through the interface. 

That the 't Hooft loops should decompose in the same way that characters of the corresponding representations of $U(N)$ decompose under $U(N-1)$ is perhaps clearer in an S-dual picture where the theories are coupled by a reduction of symmetry due to a Nahm pole boundary condition. The S-dual of a 't Hooft loop is a Wilson loop whose contribution is a character. From this perspective, it is clear that a Wilson loop will decompose according to the decomposition of the characters of representations under $U(N) \to U(N-M)$ when brought to the interface~\footnote{We thank Davide Gaiotto for suggesting this interpretation.}. 

\subsection{$T[U(N)]$}
Let us now consider the partition function of $T[U(N)]$, which we denote by $\cZ_{U(N)}(m,t)$. It depends on mass parameters $m =(m_1,\ldots,m_N)$ for the Higgs branch symmetry $U(N)_H$ and FI parameters $t = (t_1,\ldots,t_N)$ for the Coulomb branch symmetry $U(N)_C$. As before, the FI parameter associated to the $U(j)$ gauge group is $t_j-t_{j+1}$.

This theory arises on an interface between two copies of the $\cN=2^*$ theory in four dimensions with gauge groups $U(N)$ and holomorphic couplings related by the S-duality transformation $\tau \to -1/\tau$. The coupling between the bulk and interface degrees of freedom is performed by gauging the $U(N)_H$ symmetry on one side of the interface and the $U(N)_C$ symmetry on the other. Therefore the mass parameters $m$ and FI parameters $t$ are identified with the vacuum expectation value of the vectormultiplet scalar of the bulk theory on either side of the interface.

It is expected that Wilson loops and 't Hooft loops are interchanged by the S-duality transformation $\tau \to -1/\tau$. Therefore, we expect that acting with a 't Hooft loop one one side of the interface is equivalent to acting with a Wilson loop on the other. Concentrating on the minuscule representations, this can be turned into the mathematical statement
\be
\left( T_r(m) - W_r(t) \right) Z_{U(N)}(m,t) = 0
\ee
where
\be
W_r(t) = \chi_r(\tau_1,\ldots,\tau_N) = \sum_{i_1 < \ldots < i_r} e^{2\pi i b (t_{i_1}+\ldots+t_{i_r})}\, .
\ee
In words, the partition function should be an eigenfunction of the 't Hooft loop difference operators whose eigenvalue if the expectation value of a Wilson loop. This is the quantized counterpart of the `gauge invariant' formulation of the Bethe equations for supersymmetric vacua which we have found earlier. As always, in the context of the $S^3$ partition function there is an isomorphic statement for the tilde parameters.

We shall prove this conjecture by induction on the rank of $U(N)$ following an argument presented in. The initial condition is the three-sphere partition function of $T(U(1))$. This is simply an FI parameter $t$ for a background $U(1)$ vectormultiplet with mass parameter $m$,
\be
\CZ_{U(1)}(m , t) = e^{2\pi i m t} \, .
\ee
It is immediate that this partition function obeys the difference equations $p_{\mu} = \tau $ and $p_{\widetilde\mu} = \widetilde\tau$ as required. The partition function of $T[U(N)]$ is now defined recursively by
\bea
\CZ_{U(N)}(m,t) = e^{2 \pi i t_N \sum\limits_{j=1}^N m_j} \int [ds] \, \nu_{N-1}(s) \, Q_{N,N-1}(m ,s ) \\ \times  \cZ_{U(N-1)} \left(s, \{ t_1-t_N,\ldots,t_{N-1}-t_N \} \right)
\eea
where we use shorthand notation $m = \{m_1,\ldots,m_N\}$, $t=\{t_1,\ldots,t_N\}$ for mass parameters of the $U(N)$ symmetries and $s = \{s_1,\ldots,s_{N-1}\}$ for the mass parameters of the $U(N-1)$ symmetry. The contour is a small deformation away from the imaginary axis. 

Let us now act with the difference operator $T_r(x)$. This passes through the exponential prefactor leaving behind a phase $\tau_N^r$. It now acts on the bifundamental hypermultiplet contribution $Q_{N,N-1}(m,s)$ and can be exchanged for the sum of difference operators $T_r(-s)+T_{r-1}(-s)$ according to the intertwining relation~\eqref{intgen}. Finally, we integrate by parts inside the integral to find the sum of difference operators $T_r(s) +T_{r-1}(s)$ acting on the partition function $\cZ_{U(N-1)}(s,\{t_1-t_N,\ldots,t_{N-1}-t_N\})$ which is an eigenfunction by assumption. Combining these factors we find that $\cZ_{U(N)}(m,t)$ is an eigenfunction of $T_r(m)$ with eigenvalue
\be
\tau_N^r \left[ \, \chi_r\left(\frac{\tau_1}{\tau_N},\ldots,\frac{\tau_{N-1}}{\tau_N}\right) + \chi_{r+1}\left(\frac{\tau_1}{\tau_N},\ldots,\frac{\tau_{N-1}}{\tau_N} \right) \, \right]
= \chi_r(\tau_1,\ldots,\tau_N) \, .
\ee
This completes the recursive step.

For completeness, let us write down the solution of the recursion relation for the partition function of $T[U(N)]$ in closed form. For this purpose, it is convenient to introduce the triangular array of mass parameters $m^{(\al)}_j$ where $\al=1,\ldots,N$ and $j=1,\ldots,\al$. For convenience we denote $m^{(N)}_j = m_j$. They are mass parameters for the nodes of a triangular quiver. We also introduce FI parameters $t_j$ with $j=1,\ldots,N$. The solution is then given by the formula
\begin{multline}
\cZ_{U(N)}(m,t) = \\
e^{2\pi i t_N(m_1+ \cdots m_N )}\int \prod_{\al=1}^{N-1} d\nu_{\al} (m^{(\al)} ) \, Q_{\al+1,\al} \left( m^{(\al+1)},m^{(\al)} \right) \, e^{2\pi i ( t_\al - t_{\al+1} )( m_1^{(\al)} + \cdots + m^{(\al)}_\al ) } 
\label{eq:TUNPartFuncCoul}
\end{multline}
where again the choice of the contour is explained later in \secref{Sec:VortAVort}.

Finally, we note that this mathematical argument has a pleasing interpretation in terms of interfaces. We can think of $T[U(N)]$ as obtained by colliding a series of symmetry breaking interfaces between a sequence $\cN=2^*$ theories with gauge groups $U(j)$, $j=1,\ldots N-1$ together with boundary FI parameters. Our argument simply brings the 't Hooft loop through the sequence of interfaces one-by-one.

$T[U(N)]$ is self-dual under mirror symmetry. A more precise statement is that the $S^3$ partition function obeys the symmetry property
\be
\cZ(\mu,\tau,\eta) = \cZ(\tau,\mu,q^{1/2}\eta^{-1}) \, .
\ee
This symmetry is extremely non-trivial from the integral expression that we have presented and we will not attempt to prove it directly. However, we note that it is equivalent to saying that the partition function is simultaneously an eigenfunction of Hamiltonians acting on the masses and FI parameters,
\bea
T_r(\tau,p_\tau,\eta) \cdot \cZ(\mu,\tau,\eta) = \chi_r(\mu) \, \cZ(\mu,\tau,\eta) \,, \\
T_r\left(\mu,p_\mu,q^{1/2}\eta^{-1}\right) \cdot \cZ(\mu,\tau,\eta) = \chi_r(\tau) \, \cZ(\mu,\tau,\eta) \,.
\label{eq:tRSEigenMirr}
\eea
The classical limit of trigonometric RS relations in the electric \eqref{eq:tRSW} and magnetic \eqref{eq:tRST} frames is reproduced from the above relations in the $q\to 1$ limit.

\subsection{Holomorphic Blocks}\label{Sec:VortAVort}
The squashed $S^3$ partition function of $T[U(N)]$ is simultaneously an eigenfunction of two sets of difference operators $T_r(m)$ and $\widetilde T_r(m)$ related by the transformation $b \leftrightarrow b^{-1}$. On the other hand, asking for an eigenfunction of a single set of difference operators, say $T_r(m)$, leads to a basis of solutions called holomorphic blocks~\cite{Pasquetti:2011fj,Beem:2012mb}. 

Let us denote the holomorphic blocks by $B_{j}$ where the index $j = 1,\ldots, N!$ labels the supersymmetric vacua of the theory on $S^1 \times \mathbb{R}^2$. i.e. the number of solutions of the Bethe equations. The $S^3$ partition function can be reconstructed from the holomorphic blocks as follows
\be
\cZ_{S^3_b}(m,t) = \sum_{j=1}^{N!} B_j(\mu,\tau) \, B_j(\widetilde\mu,\widetilde\tau) \, .
\ee
The holomorphic blocks can be computed a priori, up to some non-perturbative ambiguities, by systematically building formal solutions of the relevant difference equations and finding a basis of contours $C_j$ for the integrals. 

Here we follow a less sophisticated approach and derive the holomorphic blocks from the Coulomb branch integral expression for the $S^3$ partition function by explicitly evaluating the integral by residues in a certain regime of small FI parameters. This leads to an alternative Higgs branch representation of the partition function involving the vortex partition function.
Here we will concentrate on $T[U(N)]$ leaving the notationally more complicated case of $T[U(N)]_{\rho}$ in \appref{Sec:AppFormulae}.

\subsubsection{$T[U(2)]$ Theory}
Let us start with the simplest example, the partition function of $T[U(2)]$. The partition function is
\begin{equation}
\mathcal{Z}_{U(2)}^{[1^2]} = S(\varepsilon) \int dx \prod_{j=1}^2 S \left(\frac{\varepsilon}{2}^* \pm i(x-m_j)\right) 
e^{2\pi x(t_1-t_2)+2\pi t_2(m_1+m_2)} \, .
\end{equation}
The integrand has two sets of simple poles at $x=m_j+i\varepsilon^*/2+ikb+i\widetilde{k}b^{-1}$ and $x=m_j-i\varepsilon^*/2-ikb-i\widetilde{k}b^{-1}$ with  integers $k,\widetilde{k}\ge0$.
The integral contour can be chosen to enclose either set of poles, but the result is independent of this choice.
We take residues from the first set of poles. The residue sum is
\begin{align}
\mathcal{Z}_{U(2)}^{[1^2]} &= S(\varepsilon) \sum_{i=1}^2 \sum_{k,\widetilde{k}=0}^\infty \left( {\rm Res}_{x=0} S \big(ix-kb-\widetilde{k}b^{-1}\big) \right) \times \prod_{j\neq i}^2S \left(im_{ij}-kb-\widetilde{k}b^{-1}\right)  \nonumber \\
	& \hspace{1cm} \times \prod_{j=1}^2 S \left(-im_{ij}+\varepsilon^*+kb+\widetilde{k}b^{-1}\right)
	e^{2\pi (m_i+i\frac{\varepsilon^*}{2}+ikb+i\widetilde{k}b^{-1})(t_1-t_2)+2\pi  t_2(m_1+m_2)} \,.
\end{align}
Using the identity \eqref{eq:double-sine-identity}, one can evaluate the residue of the double sine function 
\be\label{eq:residue-double-sine}
	{\rm Res}_{x=0} S \big(ix-kb-\widetilde{k}b^{-1}\big) = i^{-k-\widetilde{k}}(-1)^{k+\widetilde{k}+k\widetilde{k}}q^{\frac{k(k+1)}{4}}\widetilde{q}^{\frac{\widetilde{k}(\widetilde{k}+1)}{4}}(q;q)_k^{-1} (\widetilde{q};\widetilde{q})_{\widetilde{k}}^{-1}
\ee
with $q\equiv e^{2\pi i b^2}, \widetilde{q} \equiv e^{2\pi i/ b^{2}}$.
Plugging this residue and using the identities \eqref{eq:double-sine-identity} we find
\begin{align}\label{eq:ZclZ1lZvZav}
\mathcal{Z}_{U(2)}^{[1,1]} 
&=	 \sum_{i=1}^2 e^{2\pi t_2(m_1+m_2)+2\pi (m_i+i\frac{\varepsilon^*}{2})(t_1-t_2)} 
	\prod_{j\neq i}\frac{S(im_{ij})}{S(im_{ij}+\varepsilon)} \nonumber \\
&	\hspace{1cm}\times \left(\sum_{k=0}^\infty \left(q\eta^{-2}\frac{\tau_1}{\tau_2}\right)^k
	\prod_{j=1}^2\frac{ (\eta^2\frac{\mu_i}{\mu_j};q)_k }
	{ (q\frac{\mu_i}{\mu_j};q)_k }\right)\times\left( \ 
	\sum_{\widetilde{k}=0}^\infty \left(q\widetilde{\eta}^{-2}\frac{\widetilde\tau_1}{\widetilde\tau_2}\right)^{\widetilde{k}}
	\prod_{j=1}^2\frac{ (\eta^2\frac{\widetilde{\mu}_i}{\widetilde{\mu}_j};\widetilde{q})_{\widetilde{k}} }
	{ (\widetilde{q}\frac{\widetilde{\mu}_i}{\widetilde{\mu}_j};\widetilde{q})_{\widetilde{k}} }\right) \,.
\end{align}
The result takes the form of the Higgs branch representation which is given by sum over the contributions from two supersymmetric vacua labelled by $i=1,2$.
Also, the second line shows the factorization into the vortex and anti-vortex partition functions.

\subsubsection{$T[U(N)]$ Theory}
Now we turn to the partition function of the $T[U(N)]$ theory. We shall perform the integral recursively starting with the recursive expression
\begin{equation}
\mathcal{Z}_{U(N)}(m,t) = \int [ds] \nu_{N-1}(s) Q_{N,N-1}(m,s) e^{2\pi (t_{N-1}-t_{N})\sum_{i=1}^{N-1}s_i} \mathcal{Z}_{U(N-1)}(s,\{t_1,\cdots,t_{N-1}\})
\label{eq:TUN-integral-partition-ftn}
\end{equation}
with $\mathcal{Z}_{U(1)}=e^{-2\pi i t_1 m^{(1)} }$.
This partition function differs by a prefactor $e^{2\pi t_N\sum_{i=1}^Nm_i} $ from the true partition function of the $T[U(N)]$ theory defined in the previous section. We shall multiply this prefactor to the last expression after integration. 

 The contour integrand has infinite simple poles and zeros from various contributions and also from the function $\mathcal{Z}_{U(N-1)}$.
We should first ask how the integral contour goes around those poles. We choose the contour along the real axis assuming that the mass parameters $m$ as well as the squashing parameter $b$ are real parameters. We then take residues from the poles above the real axis. One can also pick up the poles in the other side but the result will be the same.

Although pole structure seems to be complicated as they can be developed by various terms, it turns out that the relevant poles can arise only from the contributions of the bifundamental hypermultiplets $Q_{N,N-1}$. Indeed, we find that poles from $\mathcal{Z}_{U(N-1)}$ are completely cancelled by zeros of the vector multiplet contribution and poles from the vector multiplet are also cancelled by zeros of $\mathcal{Z}_{U(N-1)}$ and $Q_{N,N-1}$. Namely $\nu_{N-1} \mathcal{Z}_{U(N-1)}$ has no relevant pole inside the contour.  We will see this after computing $\mathcal{Z}_{U(N-1)}$ explicitly. However, before we get $\mathcal{Z}_{U(N-1)}$, we first assume that this is the case and evaluate the integral by picking up poles only from the hypermultiplet.

The contribution $Q_{N,N-1}$ has simple poles at $s_i = m_j+i\frac{\varepsilon^*}{2}+ik_ib+i\widetilde{k}_ib^{-1}$ for $k_i,\widetilde{k}_i \ge 0$ in the contour.
Summing over residues from these poles we obtain
\begin{align}
	&\mathcal{Z}_{U(N)}= \frac{1}{(N-1)!}\sum_{\sigma\in\mathcal{W}_{N}}
	 \sum_{k,\widetilde{k}\ge0} \left(\frac{\tau_{N-1}}{\tau_{N}}\right)^{|k|} \left(\frac{\widetilde\tau_{N-1}}{\widetilde\tau_{N}}\right)^{|\widetilde{k}|} e^{2\pi i(t_{N-1}-t_{N})(\frac{(N-1)\varepsilon^*}{2}-i\sum_{j=1}^{N-1} m_{\sigma(j)})} \nonumber \\
	 & \quad \times \frac{ \prod_{i,j=1}^{N-1} S\big(\varepsilon-im_{\sigma(i)\sigma(j)}+k_{ij}b+\widetilde{k}_{ij}b^{-1}\big) } 
	 { \prod_{i\neq j}^{N-1} S\big(-im_{\sigma(i)\sigma(j)}+k_{ij}b+\widetilde{k}_{ij}b^{-1}\big) }
	 \prod_{i=1}^{N-1} \prod_{j=1}^{N} S \left(\varepsilon^*-im_{\sigma(i)\sigma(j)}+k_ib+\widetilde{k}_ib^{-1}\right) \nonumber \\
	 &\quad \times \prod_{i=1}^{N-1} {\rm Res}_{x_i=0}  \bigg[\prod_{j=1}^{N} S \left(ix_i+im_{\sigma(i)\sigma(j)}-k_ib-\widetilde{k}_ib^{-1}\right) \bigg]
	 \ \mathcal{Z}_{U(N-1)}(s_*,\{t_1,\cdots,t_{N-1}\}) \,.\nonumber
\end{align}
We denote $k=\{k_i^{(N-1)}\}$ and $|k| = \sum_{i=1}^{N-1}k_i^{(N-1)}$ and similarly for $\widetilde{k}$.
Here $s_*$ stands for the integral variables $s_i$ evaluated at each pole and they replace the mass parameters in the partition function $\mathcal{Z}_{U(N-1)}$. In this expression, the (semi-)positive integer numbers $k_i$ and $\widetilde{k}_i$ correspond to the vortex and anti-vortex numbers of the $U(N-1)$ gauge group, which will become clearer later when we write the full partition function as a factorized form.
There is the summation over $\sigma$ in the Weyl group $\mathcal{W}_N$ and an overall factor $1/(N-1)!$ corresponding the the Weyl factor of the $U(N-1)$ gauge group. 
Combining these two, one notices that the number of supersymmetric vacua added to each recursion step is $N$. Therefore the total number of supersymmetric vacua of the $T[U(N)]$ partition function is $N!$ which is in a perfect agreement with the quaternionic dimension of the Higgs branch.

We can use the above formula to compute the function $\mathcal{Z}_{U(N-1)}$ at $s_*$. Using the Weyl group $\mathcal{W}_{N-1}$, one can set $\sigma^{(N-1)}=1$ and obtain
\begin{align}
	&\mathcal{Z}_{U(N\!-\!1)}(s_*)= (N-1)
	 \sum_{k,\widetilde{k}\ge0} \left(\frac{\tau_{N-2}}{\tau_{N-1}}\right)^{|k|} \left(\frac{\widetilde\tau_{N-2}}{\widetilde\tau_{N-1}}\right)^{|\widetilde{k}|} e^{2\pi i(t_{N-2}-t_{N-1})(\frac{(N-2)\varepsilon^*}{2}-i\sum_{j=1}^{N-2} m_{j})} \nonumber \\
	 & \quad \times \frac{ \prod_{i,j=1}^{N-2} \!S\big(\varepsilon\!-\!im_{ij}\!+\!k_{ij}b\!+\!\widetilde{k}_{ij}b^{-1}\big) } 
	 { \prod_{i\neq j}^{N-2} \!S\big(-\!im_{ij}\!+\!k_{ij}b\!+\!\widetilde{k}_{ij}b^{-1}\big) }
	 \prod_{i=1}^{N-2} \prod_{j=1}^{N-1} S \left(\varepsilon^*\!-im_{ij}\!+\!(k_i\!-\!k^{(N\!-\!1)}_j)b\!+\!(\widetilde{k}_i\!-\!\widetilde{k}_j^{(N\!-\!1)})b^{-1}\right) \nonumber \\
	 &\quad \times \prod_{i=1}^{N-2}\! {\rm Res}_{x_i=0}  \bigg[\prod_{j=1}^{N-1} \!S \!\left(ix_i\!+\!im_{ij}\!-\!(k_i\!-\!k_j^{(N\!-\!1)})b\!-\!(\widetilde{k}_i\!-\!\widetilde{k}_j^{(N\!-\!1)})b^{-1}\!\right)\! \bigg]
	 \, \mathcal{Z}_{U(N\!-\!2)}(s_*,\{t_1,\cdots,t_{N-2}\})\,. \nonumber
\end{align}
Here we have redefined the vortex numbers of the $U(N-2)$ gauge group as $k=\{k_i^{(N-2)}+k_i^{(N-1)}\}$ and $\widetilde{k}=\{\widetilde{k}_i^{(N-2)}+\widetilde{k}_i^{(N-1)}\}$ with $i=1,\cdots,N-2$. 
The second line and the residues in the third line can be further simplified using the equations in (\ref{eq:residue-double-sine}) and (\ref{eq:double-sine-identity}) as follows:	
\begin{align}
	&
	  e^{\pi i \sum_{i=1}^{N-2}\sum_{j=1}^{N-1}\left((k_i-k_j^{(N-1)})b(b-\varepsilon^*)+(\widetilde{k}_{i}-\widetilde{k}_j^{(N-1)})b^{-1}(b^{-1}-\varepsilon^*)\right)} \prod_{i=1}^{N\!-\!2}\!\frac{iS(im_{i,N-1})}{S(\varepsilon+im_{i,N-1})} \nonumber \\
	 & \prod_{i\neq j}^{N\!-\!2}\!\frac{ (q\eta^{-2}\frac{\mu_i}{\mu_j};q)_{k_{ij}}(\widetilde{q}\widetilde{\eta}^{-2}\frac{\widetilde{\mu}_i}{\widetilde{\mu}_j};\widetilde{q})_{\widetilde{k}_{ij}} }
	 { (\frac{\mu_i}{\mu_j} ;q)_{k_{ij}}( \frac{\widetilde{\mu}_i}{\widetilde{\mu}_j} ;\widetilde{q})_{\widetilde{k}_{ij}} }
	 \prod_{i=1}^{N\!-\!2}\prod_{j=1}^{N\!-\!1}\!\frac{ ( \eta^{2} \frac{\mu_i}{\mu_j} ;q)_{k_i\!-\!k_j^{(N\!-\!1)}}( \widetilde{\eta}^{2}\frac{\widetilde{\mu}_i}{\widetilde{\mu}_j} ;\widetilde{q})_{\widetilde{k}_i\!-\!\widetilde{k}_j^{(N\!-\!1)}} }
	 { (q\frac{\mu_i}{\mu_j} ;q)_{k_i\!-\!k_j^{(N\!-\!1)}}(\widetilde{q}\frac{\widetilde{\mu}_i}{\widetilde{\mu}_j} ;\widetilde{q})_{\widetilde{k}_i\!-\!\widetilde{k}_j^{(N\!-\!1)}} } \nonumber
\end{align}
We can consecutively apply this procedure and evaluate the partition functions $\mathcal{Z}_{U(n)}(s_*)$ for lower $n<N-2$ in the similar manner. Upon the redefinition of the vortex numbers for $U(n-1)$ gauge group as $k^{(n)} = \{ k_i^{(n)} = \sum_{j=n}^{N-1}k_i^{(j)}\}$ and $\widetilde{k}^{(n)} = \{ \widetilde{k}_i^{(n)} = \sum_{j=n}^{N-1}\widetilde{k}_i^{(j)}\}$ one can see that the partition function takes the same form as $\mathcal{Z}_{U(N-1)}$ but just the rank is reduced to $n-1$.

Finally we combine all integrated subpartition functions and find
\begin{align}
	\mathcal{Z}_{U(N)} &=  \sum_{\sigma\in\mathcal{W}_N} \!\! e^{-\pi i\varepsilon^*(Nt_N-\sum_{i=1}^Nt_i)+2\pi \sum_{i=1}^{N}t_i m_{\sigma(i)}}  
	\prod_{i<j}^N\frac{S(im_{\sigma(i)\sigma(j)})}{S(\varepsilon+im_{\sigma(i)\sigma(j)})} Z_V(m_{\sigma(i)})Z_{AV}(m_{\sigma(i)}) \nonumber \\
	Z_V &= \sum_{\{\vec{k}^{(1)},\cdots,\vec{k}^{(N-1)}\}\ge 0 } \prod_{n=1}^{N-1}\left(\frac{q}{\eta^2}\frac{\tau_{n}}{\tau_{n+1}}\right)^{|k^{(n)}|} 
	\prod_{i\neq j}^{n}
	\frac{\big(q\eta^{-2}\frac{\mu_i}{\mu_j};q\big)_{\!k^{(n)}_{ij}} }
	{\big(\frac{\mu_i}{\mu_j};q\big)_{\!k^{(n)}_{ij}} }
	\prod_{i=1}^{n} \prod_{j=1}^{n+1}
	\frac{\big(\eta^{2}\frac{\mu_i}{\mu_j};q\big)_{\!k^{(n)}_i\!-\!k^{(n+1)}_j} }
	{\big(q\frac{\mu_i}{\mu_j};q\big)_{\!k^{(n)}_i\!-\!k^{(n+1)}_j} }  
	\nonumber \\
	Z_{AV} &= Z_V\left((\tau,\mu,\eta,q) \rightarrow (\widetilde\tau,\widetilde\mu,\widetilde\eta,\widetilde{q})\right) \,.
\label{eq:S3-function-TU(N)}
\end{align}
where $|k^{(N)}|=|\widetilde{k}^{(N)}|=0$. Here we have multiplied the prefactor $e^{2\pi t_N\sum_{i=1}^Nm_i}$ which we omitted in~\eqref{eq:TUN-integral-partition-ftn}.
This is the Higgs branch representation of the $S^3$ partition function of the $T[U(N)]$ theory. We note that the partition function is factorized by the vortex and anti-vortex partition functions, $\CZ_V$ and $\CZ_{AV}$ respectively, as well as the perturbative contributions.

We still need to check the our assumption that the vector multiplet and the function $\mathcal{Z}_{U(N-1)}$ do not develop nontrivial poles in the contour integral.
To see this, it is convenient to first fix $\sigma=1$ in $\mathcal{Z}_{U(N-1)}$ using the Weyl group of the $U(N-1)$ gauge symmetry and perform the contour integral starting with $s_1$ and so on.
We then noticed that the 1-loop contribution in $\mathcal{Z}_{U(N-1)}(s)$ of the form
\begin{equation}
	\prod_{i<j}^{N-1}\frac{S(is_{ij})}{S(\varepsilon+is_{ij})}
\end{equation}
is completely cancelled by the vector multiplet contribution $\nu_{N-1}(s)$ of the $U(N-1)$. Thus the function $\mathcal{Z}_{U(N-1)}(s)$ can have poles only from the vortex and anti-vortex series. Furthermore, we find that possible poles from the vortex series are $s_i=m_j +i\frac{\varepsilon^*}{2}+inb+imb^{-1}$ with $n,m>0$ and they are also cancelled by zeros of the vector multiplet contribution at $s_i=s_j+inb+imb^{-1}$ for $i>j$. The remaining poles are at $s_i=s_j+i\varepsilon+ipb+iqb^{-1}$ with $p,q\ge0$ and $i>j$ in $\nu_{N-1}(s)$, which are also cancelled by zeros of $Q_{N,N-1}(s)$ at $s_i=m_j-i\frac{\varepsilon^*}{2}+inb+imb^{-1}$.
Therefore the integral contour involves no pole from the vector multiplet contribution and the function $\mathcal{Z}_{N-1}$. This proves that the relevant poles can come only from the contributions of the bifundamental hypermultiplets, and therefore the Higgs branch representation of the partition function (\ref{eq:S3-function-TU(N)}) is correctly derived.

\subsubsection{Givental J-functions}\label{eq:GivJFun}
Since the Higgs branch of 3d $\CN=2^*$ quiver theories can be identified with the cotangent bundle of the corresponding Nakajima quiver varieties, the vortex partition function thereof, which is the generating function of the BPS states on Higgs branch, gives the Givental J-function of the corresponding variety \cite{2001math8105G,2011arXiv1110.3117T}. 

The corresponding quiver variety for $T[U(2)]$ theory is $T^*\mathbb{P}^{1}$. According to \eqref{eq:ZclZ1lZvZav} the corresponding vortex partition function of the $T[U(2)]$ theory in one of the vacua (when, say, $i=1$) reads
\begin{equation}
\CZ_{V}^{[1,1]} = \displaystyle\sum_{k=0}^\infty \frac{\left(\eta^2;q\right)_k\left(\eta^2\frac{\mu_1}{\mu_2};q\right)_k}{(q\frac{\mu_1}{\mu_2};q)_k}\cdot\frac{\left(q\eta^{-2}\frac{\tau_1}{\tau_2}\right)^k}{(q;q)_k}
={}_2F_1\left(\eta^2,\eta^2\frac{\mu_1}{\mu_2};q\frac{\mu_1}{\mu_2};q;q\eta^{-2} z\right)\,,
\label{eq:TstP1JFun}
\end{equation}
where we used the definition of the Q-hypergeometric function (see \appref{Sec:AppFormulae}). 

\subsection{Difference Equations in Electric Frame}
It is instructive to explicitly demonstrate that holomorphic blocks, which we have computed above, satisfy difference relations in the electric frame (i.e. second relation in \eqref{eq:tRSEigenMirr}). We shall demonstrate this fact for 3d vortex partition functions  \eqref{eq:S3-function-TU(N)} for $T[U(2)]$ and $T[U(3)]$ theories.

\subsubsection{$T[U(2)]$ Theory}
We can explicitly check that \eqref{eq:TstP1JFun} satisfies the following difference relation
\begin{equation}
\left(\mu_1 \, \eta \, \frac{\eta  \tau _1-\eta^{-1}\tau _2}{\tau _1-\tau _2} p_{\tau}^1 +
\frac{\mu _2}{\eta} \, \frac{ \eta  \tau _2-\eta^{-1}\tau _1}{\tau _2-\tau _1} p_{\tau}^2\right)\CZ_{V}^{[1,1]}  = (\mu_1 +\mu_2)\CZ_{V}^{[1,1]} \,,
\label{eq:AlmostEigentRS2Sec3}
\end{equation}
and
\begin{equation}
 p_{\tau}^1 \, p_{\tau}^2 \CZ^{[1,1]}_V = \CZ^{[1,1]}_V\,,
\end{equation}
where momenta operators act as $p_\tau^i \tau_j = q^{\delta_{ij}}\tau_j p_\tau^i$. Note that the operator in the left hand side of the above equation differs from $T_1$ for trigonometric RS model \eqref{eq:tRSrelationsTU2} by prefactors which depend on $\mu_1,\mu_2$ and $\eta$. We can remove those factors by redefining 
\begin{equation}
\cZ_V =  \frac{\theta(\eta^{-1} \, \tau_1,q)\theta(\eta \, \tau_2,q) }{\theta(\mu_1\tau_1,q)\theta(\mu_2\tau_2,q)} \CZ_V^{[1,1]}\,, 
\label{eq:ZeignRedefineSec3}
\end{equation}
where $\theta(a,q) = (a,q)_\infty(qa^{-1},q)_\infty$ is one solution of the difference equation $p_{a} \theta(a,q) = -a^{-1}\theta(a,q)$. Our theta function conventions are given in \appref{sec:ThetaFunctions}. Then $\cZ_V$ obeys the following set of eigenfunction difference equations
\bea
 \left(\frac{ \eta  \tau _1-\eta^{-1}\tau _2}{ \tau _1-\tau _2} p_{\tau}^1 +
  \frac{\eta  \tau _2-\eta^{-1}\tau _1}{\tau _2-\tau _1} p_{\tau}^2\right)\cZ_V & = \left(\mu_1 +\mu_2\right)\cZ_V\,, \\
p_{\tau}^1 p_{\tau}^2 \,\cZ_V & =  \mu_1 \, \mu_2 \cZ_V\,,
\label{eq:tRSeigenfunctionsSec2}
\eea
which are nothing but conservation conditions for quantum two-body trigonometric RS Hamiltonians. The eigenvalues on right hand sides are the expectation values of supersymmetric Wilson loops in the fundamental and skew symmetric tensor representation of $U(2)$. Thus \eqref{eq:tRSeigenfunctionsSec2} provide the desired quantization of the twisted chiral ring of $T[U(2)]$ theory.

Had we chosen another vacuum of the $T[U(2)]$ theory in \eqref{eq:ZclZ1lZvZav} ($i=2$) we would have obtained the second vortex partition function which can be obtained from \eqref{eq:TstP1JFun} by replacing $\mu_2$ and $\mu_1$. After the redefinition similar to \eqref{eq:ZeignRedefineSec3} (again, with $\mu_2$ and $\mu_1$ interchanged) the partition function in the second vacuum also satisfies \eqref{eq:tRSeigenfunctionsSec2}.

Note that the prefactor in \eqref{eq:ZeignRedefineSec3} is not unique. For example, one may replace the factor $\theta(\eta^{-1} \, \tau_1) / \theta(\mu_1\tau_1) $ by another one $\theta(\tau_1) / \theta( \mu_1 \tau_1 \eta )$ that obeys the same difference equation. They correspond to factorizations of mixed Chern-Simons terms (or FI terms) for the 3d theory living on the surface defect, i.e. $e^{\pi i\varepsilon^*t_1 + 2\pi t_1m_1}$ in this case. 
In the classical limit $b\rightarrow0$ (or $b \rightarrow \infty$), this prefactor reproduces the classical FI terms up to proper rescaling of the parameters.

At this step let us make the following observation. Vortex partition functions which were computed in \eqref{eq:S3-function-TU(N)} are infinite series in FI parameters. Note, however, that for some values of the coefficients these series truncate and become polynomials. 
Remarkably, if we combine them with the corresponding prefactors containing theta-functions, like \eqref{eq:ZeignRedefineSec3}, we can reproduce the so-called Macdonald polynomials, which recently appeared in many different contexts in mathematical physics. For example, if in \eqref{eq:ZeignRedefineSec3} one puts $q^{-k}=\eta^2\mu_1/\mu_2$ for $k=0,1,2$ the series become degree-$k$ symmetric Macdonald polynomials in $\tau_1$ and $\tau_2$ with parameters $q$ and $\eta$. In \cite{2012arXiv1206.3131B} it was shown that Macdonald polynomials with proper normalization are eigenfunctions of difference operators of tRS type.

\subsubsection{$T[U(3)]$ Theory}
The vortex partition functions of $T[U(3)]$ theory are slightly cumbersome, so we shall only specify difference operators which they satisfy. There are $3!=6$ vacua in the $T[U(3)]$ theory, meaning that there are as many different vortex partition functions which are 
related to each other by interchanging mass parameters $\mu_1,\mu_2$ and $\mu_3$ between each other. Thus in one of the vacua the partition function satisfies the following difference equation
\begin{align}
\Bigg(&\mu_1 \, \eta^2 \, \frac{\eta  \tau _1-\eta^{-1}\tau _2}{\tau _1-\tau _2} \frac{\eta  \tau _1-\eta^{-1}\tau _3}{\tau _1-\tau _3} p_{\tau}^1 + \mu _2 \, \frac{ \eta  \tau _2-\eta^{-1}\tau _1}{\tau _2-\tau _1} \frac{ \eta  \tau _2-\eta^{-1}\tau _3}{\tau _2-\tau _3} p_{\tau}^2 \notag\\
&+ \frac{\mu_3}{ \eta^2 } \, \frac{ \eta  \tau _3-\eta^{-1}\tau _1}{\tau _3-\tau _1} \frac{ \eta  \tau _3-\eta^{-1}\tau _2}{\tau _3-\tau _1} p_{\tau}^3\Bigg)\CZ_{V}^{[1,1,1]}  = (\mu_1 +\mu_2+\mu_3)\CZ_{V}^{[1,1,1]} \,.
\label{eq:AlmostEigentRS3Sec3}
\end{align}
Again, in order to convert the above equation into the momentum conservation of the three-particle trigonometric RS model we redefine the vortex partition function as
\begin{equation}
\CZ_{V}=\frac{ \theta(\eta^{-2} \, \tau_1,q)\theta(\tau_2,q)\theta(\eta^{2} \, \tau_3,q) }{\theta(\mu_1\tau_1,q)\theta(\mu_2\tau_2,q)\theta(\mu_3\tau_3,q)}\CZ_{V}^{[1,1,1]}\,.
\end{equation}
The prefactors again correspond to factorization of the FI terms.

We have checked in the series expansion in the FI parameters that \eqref{eq:AlmostEigentRS3Sec3} with the above redefinition is the true eigenfunction of the tRS model. We have also verified that it is an eigenfunctions of the quadratic tRS operator $T_2$. Interestingly, one can reproduce the entire vortex partition function using the perturbation theory in FI parameters given the difference relations of the form \eqref{eq:AlmostEigentRS3Sec3}.

\pagebreak
\section{5d/3d Partition Functions}\label{Sec:5dRamification}
The three-dimensional $T[U(N)]$ theory can be coupled to maximally supersymmetric $U(N)$ Yang-Mills theory in five dimensions to form a surface defect. Following references~\cite{Gaiotto:2009fs,Gaiotto:2013sma}, if the surface defect is supported on the subspace $S^1 \times \mathbb{R}^2 \subset S^1 \times \mathbb{R}^4$, we expect that the twisted chiral ring of $T[U(N)]$ is deformed by an additional parameter that we denote by $Q$. This parameter is related to the dimensionless combination of the five-dimensional gauge coupling $g^2$ and the radius $R$ of $S^1$ by the formula 
\be
Q=e^{-\frac{8\pi^2 R}{g^2}} \, .
\ee
This deformed twisted chiral ring of the defect theory is expected to provide a representation of the Seiberg-Witten curve of the five-dimensional theory on $S^1\times\mathbb{R}^4$.
The 5d gauge coupling can be considered as the scalar field in the background vector multiplet coupled to the topological instanton charge. When the 5d gauge theory is compactified on a circle, we can turn on a background holonomy in the vector multiplet and it complexifies the 5d gauge coupling.

The Higgs branch $U(N)$ symmetry of $T[U(N)]$ is gauged in coupling to five-dimensional maximal SYM. In particular, the twisted masses $\mu_j$ of the three-dimensional theory are identified with the vacuum expectation value of the five-dimensional real vectormultiplet scalar (complexified by the holonomy around $S^1$). Furthermore, the three-dimensional $\cN=2^*$ deformation $\eta$ can be identified with the five-dimensional $\cN=1^*$ deformation. The FI parameters $\tau_j$ are additional parameters associated to the remaining $U(N)$ symmetry of the defect. We claim that the twisted chiral ring of $T[U(N)]$ is deformed as follows
\be
\sum_{\substack{\mathcal{I}\subset\{1,\dots, N\} \\ |\mathcal{I}|=r}}\prod_{\substack{i\in\mathcal{I} \\ j\notin\mathcal{I}}} \frac{\theta_1(\tau_i/\eta^2\tau_j;Q)}{\theta_1(\tau_i/\tau_j;Q)}\prod\limits_{i\in\mathcal{I}}p_\tau^i = \chi_r(\mu_1,\ldots,\mu_N,Q) \,,
\label{eq:TwistChirRingDef5d}
\ee
where $\theta_1(x,q)$ is the first Jacobi theta function (see \appref{sec:ThetaFunctions}). The functions $\chi_r(\mu)$ are the expectation values of supersymmetric Wilson loops wrapping $S^1$ in the anti-symmetric tensor representations of $U(N)$ of rank $r=1,\ldots,N$, including instanton corrections. It is straightforward to see that
\bea
\frac{\theta_1(\tau_i/\eta^2\tau_j;Q)}{\theta_1(\tau_i/\tau_j;Q)}  & = \frac{\eta^{-1}\tau_i-\eta \tau_j}{\tau_i-\tau_j} + \cO(Q)\,, \\
\chi_r(\mu_1,\ldots,\mu_N;Q) & = \sum_{i_1<\cdots <i_r} \mu_{i_1} \cdots \mu_{i_r} + \cO(Q)\,,
\eea
and hence that the twisted chiral ring of $T[U(N)]$ is reproduced in the limit $Q \to 0$ where we turn of the coupling to the five-dimensional degrees of freedom. 

The combinations appearing on the left are the Hamiltonians of a classical integrable system: the complexified elliptic RS system. This reduces to the complex trigonometric RS system in the limit $Q\to0$.  The same integrable system describes the Seiberg-Witten geometry of five-dimensional maximal SYM. In particular, the expectation values of supersymmetric Wilson loops $\chi_r(\mu_1,\ldots,\mu_N,Q)$ can be taken to parametrize the Coulomb branch of the five-dimensional theory on $S^1 \times \mathbb{R}^4$, over which equation~\eqref{eq:TwistChirRingDef5d} defines a family of elliptic curves. The Seiberg-Witten differential is identified with the symplectic potential $\lb = \sum_j \log p_{\tau_j} \wedge d \log \tau_j $ for the phase space of the elliptic RS system.

The classical integrable system can be quantized by studying this coupled 3d - 5d system in a curved background preserving some supersymmetry. A natural extension of our discussion in section~\ref{Sec:3dpf} would be to study the partition function on $S^5$ with the surface defect supported on an $S^3$. This provides a formidable technical challenge. However, in the absence of defects, there is much evidence to suggest that the partition function on $S^5$ can be reconstructed from knowledge of the Nekrasov partition function in the Omega background $S^1 \times \mathbb{R}^4_{\ep_1,\ep_2}$. Recently, there is some evidence that the same conclusion can be reached in the presence of a surface defect by enriching the set of building blocks to include the Nekrasov partition function in the presence of a surface defect supported on $S^1 \times \mathbb{R}^2_{\epsilon_1}$. We focus solely on this case in what follows.

In principle, one should be able to perform an exact localization computation for the coupled 5d/3d system. This computation is beyond the scope of the current paper~\footnote{Some work is this direction is now being done in \cite{GL2014}.}. Instead, we will use an alternative description of the surface defect as a monodromy defect labelled by the partition $[1^N]$, whose partition function can be computed using ramified instanton counting~\cite{Alday:2010vg,Kanno:2011fw}. We emphasize that this equivalence has a conjectural status.  Therefore, we first check that the partition function in the presence of a defect holomorphic blocks of $T[U(N)]$ in the limit $Q \to 0$,
\be
\cZ^{(i)}(\vec m, \vec t , \ep_1,\ep_2 , Q) =  \cB^{(i)}(\vec m , \vec t,\ep_1)  + \cO(Q) \, .
\ee
The monodromy defect of type $[1^N]$ has Gukov-Witten parameters $\tau_1,\ldots,\tau_N$ which are identified with FI parameters of $T[U(N)]$ in the decoupling limit. The leading term is independent of the equivariant parameter $\ep_2$ in the plane orthogonal of the defect.

The complete partition function depends on two quantization parameters, $\ep_1$ and $\ep_2$. At finite $Q$ we will consider the Nekrasov-Shatashvili limit $\epsilon_2\to 0$ of the normalized expectation value of the defect
\be
\mathcal{D}^{(i)}_{\rho}(\vec{a},Q,\ep_1,\epsilon_2,\vec t) = \lim_{\ep_2 \to 0} \frac{\CZ^{(i)}_{\rho}(\vec{a},Q,\ep_1,\ep_2,\vec t)}{\CZ(a,Q,\ep_1,\ep_2)}\,,
\label{eq:eRSEgenFunGen}
\ee
We will be able to check to a finite order in the instanton expansion in $Q$ that~\eqref{eq:eRSEgenFunGen} form a basis of solutions to the quantum elliptic RS Hamiltonians. We should emphasize that these are so far formal solutions: we have neither specified a measure nor checked normalizability with respect to it. This point is discussed further below.

We shall momentarily discuss the construction of expectation values \eqref{eq:eRSEgenFunGen}, but first we need to build up the necessary ingredients for instanton calculus in the presence of monodromy defects.

\subsection{Instanton Counting}\label{Sec:Instantons}
Below we review only essential ingredients for our computations further in this section. We refer the reader to \cite{Losev:1997tp,2014arXiv1401.6782N,Nekrasov:2013xda}. 

Let us consider the hyper K{\" a}hler quotient construction for the instanton moduli space $\cM_{N,k}$ with gauge group $U(N)$ and instanton number $k$. We introduce vector spaces $V = \bC^k$ and $W=\bC^N$ and matrices $A,B \in \mathrm{Hom}(V,V)$, $P\in \mathrm{Hom}(W,V)$ and $Q \in \mathrm{Hom}(V,W)$. An element $g \in U(k)$ acts on these matrices by
\be
(A,B,P,Q) \to (gAg^{-1},gBg^{-1},gP,Qg^{-1})\,.
\ee
The hyperkahler moment maps are
\bea
\mu_{\bC} & = [A,B]+PQ\,, \\
\mu_{\bR} & = [A,A^\dagger]+[B,B^\dagger]+PP^\dagger+QQ^\dagger\,,
\eea
which are valued in $\mathrm{Hom}(V,V)$. The instanton moduli space $\cM_{N,k}$ is then given by the hyper K{\" a}hler quotient construction. As a complex manifold, the instanton moduli space can be constructed as a K{\" a}hler quotient by discarding the real moment map in favor of a stability condition and dividing by the complex group $GL(k,\mathbb{C})$.

Consider the following action of $(\mathbb{C}^*)^2 \times GL(N,\mathbb{C})$ on the matrices
\be
(A,B,P,Q) \to ( t_1 A , t_2 B , P f^{-1}, t_1 t_2 f Q)\,,
\ee
where $t_1,t_2\in \bC^*$ and $f\in GL(N,\bC)$. Note that $(\bC^*)^2$ corresponds to rotations in the two coordinate planes. This commutes with the action of $GL(k,\bC)$ and preserves the stability condition, therefore it descends to an action on the instanton moduli space $\cM_{N,k}$. The fixed points of $\cM_{N,k}$ under this action are labelled by an $N$-tuple of Young tableaux $\vec \lb = (\lb_1,\ldots,\lb_N)$ where the total number of boxes is $k$. 

Let us introduce the equivariant parameters $\ep_1$, $\ep_2$ and $\vec a =(a_1,\ldots,a_N)$ that are valued in the Cartan subalgebra of the symmetry group action, i.e. such that $t_1 = e^{\ep_1}$ and $t_2 = e^{\ep_2}$ and $f =\mathrm{diag} (e^{a_1},\ldots,e^{a_N})$. Then the equivariant character for the action of the global symmetry $(\mathbb{C}^*)^2 \times GL(N,\mathbb{C})$ on the vector spaces $W$ and $V$ at a fixed point $\vec\lb$ is
\bea
\chi_{\vec\lb}^{(W)} & = \sum_{I=1}^N e^{a_I} \\
 \chi^{(V)}_{\vec\lb} & = \sum_{I=1}^N e^{a_I} \sum_{(i,j)\in \lb_I} e^{(1-i)\ep_1 + (1-j)\ep_2} \, .
\eea
The summation $(i,j) \in \lb_I$ refers to a sum over the boxes of Young tableaux $\lb_I$. For example, the Young tableaux $\yng(3)$ has three boxes with the labels $(1,1)$, $(1,2)$ and $(1,3)$.

Next we introduce equivariant Chern character 
\be
\chi_{\vec\lb}^{(\cE)} = \chi_{\vec\lb}^{(W)} - (1-e^{-\epsilon_1})(1-e^{-\epsilon_2}) \chi_{\vec\lb}^{(V)} \,.
\label{eq:CharFramedSheaf}
\ee
Using $\chi_{\vec\lb}^{(\cE)}$ we can now consider the equivariant character of the tangent space to the universal bundle over the instanton moduli space at a fixed point $\lb$,
\bea
\chi_{\vec\lb}^{\cN=1} = & -  \frac{\chi_{\vec\lb}^{(\cE)} \chi_{\vec\lb}^{(\cE^*)}}{(1-e^{-\epsilon_1})(1-e^{-\epsilon_2})} \\
 = & - \frac{\chi_{\vec\lb}^{(W)} \chi_{\vec\lb}^{(W^*)}}{(1-e^{-\epsilon_1})(1-e^{-\epsilon_2})}  \\
& - (1-e^{\ep_1})(1-e^{\ep_2}) \chi^{(V)}_{\vec\lb}\chi^{(V^*)}_{\vec\lb} +e^{\ep_1+\ep_2}
 \chi^{(W)}_{\vec\lb} \chi^{(V^*)}_{\vec\lb} +  \chi^{(V)}_{\vec\lb} \chi^{(W^*)}_{\vec\lb} \, .
\label{eq:N1Charachter}
\eea
and the operation ${}^*$ corresponds to reversing the sign of all of the equivariant parameters. The second line is the perturbative contributions to the Nekrasov partition function. The third line is the equivariant character of the tangent space to the instanton moduli space, that is ignoring the universal bundle. They provide the instanton contributions to the Nekrasov partition function. This is the relevant character for pure $\cN=1$ $U(N)$ gauge theory in five dimensions.

Let us now consider the maximally supersymmetric $\cN=2$ $U(N)$ gauge theory in five dimensions. This corresponds to an $\cN=1$ $U(N)$ vectormultiplet together with an adjoint hypermultiplet. Due to the presence of the adjoint matter we must now consider a vector bundle on the instanton moduli space. The relevant character is therefore
\be
\chi_{\vec\lb}^{\cN=2}  = (1-e^{m - \epsilon_1-\epsilon_2}) \, \chi_{\vec\lb}^{\cN=1}\,,
\ee
where $m$ is a mass parameter for the adjoint hypermultiplet. The equivariant character can always be written in the following form
\be
\chi^{\cN=2}_{\vec\lb} = \sum_{\al} n_{\al} e^{w_\al}\,,
\label{eq:N2CharacterGen}
\ee
where we sum the exponents of the weights at all the fixed points. From this we compute the five-dimensional instanton partition function
\be
\CZ_{\text{inst}} = \sum_{\lb} Q^{|\vec\lb|} \prod_{\al}   \left( 2 \sinh \left( \frac{w_\al}{2}  \right) \right)^{-n_\al}  \, .
\label{eq:PartFuncInstSum}
\ee
In what follows we use exponentiated parameters $\mu_j = e^{a_j}$, $e^{\ep_1} = q$ and $e^m = \eta^{-2} q$. Since we will mainly be interested in the limit $\epsilon_2 \to 0$ we will not need to introduce a parameter for $\ep_2$. The coefficients in the expansion are then rational functions of these exponentiated parameters.

\subsection{Ramified Instantons}\label{Sec:RamInstantons}
Let us now consider ramified instanton counting, namely we want to compute instanton partition function in the presence of a monodromy defect~\cite{Gukov:2006jk}. It was shown in~\cite{Alday:2010vg,Kanno:2011fw} that the desired answer can be obtained by applying a simple orbifolding procedure to the above computation. 

The possible monodromy defects are labelled by a partition $\rho = [n_1,n_2,\ldots,n_s]$ where we choose to order $n_1 \geq n_2 \geq \ldots \geq n_s$ and $\sum_{i=1}^s n_i = N$. This determines the subgroup $\mathbb{L} =  U(n_1) \times U(n_2) \times \cdots \times U(n_s)$ of $U(N)$ gauge group which is left unbroken by the defect. The gauge field has a singularity\footnote{This is a so-called \textit{tame} ramification, namely when $z_2=0$ is a regular singular point, as opposed to a \textit{wild} ramification with higher degree singularities.} in the complex plane orthogonal to the defect which can be described as follows
\be
\oint_{|z_2|=\epsilon} A^a = 2 \pi m^a\,, \quad a = 1,\dots, N\,,
\ee
where
\be
	m^a = (\underbrace{m_1,\cdots,m_1}_{n_1},\underbrace{m_2,\cdots,m_2}_{n_2},\cdots,\underbrace{m_s\cdots,m_s}_{n_s}) \, .
\ee
There is an additional label $\sigma$ which determines how $\mathbb{L}$ is embedded into $U(N)$. Each $\sigma$ corresponds to permutation of the monodromy parameters $m$ that are not simply permutations within each block, that is $\sigma \in \cW / \cW_{\mathbb{L}}$ where $\cW_{\mathbb{L}}$ is the Weyl group of $\mathbb{L}$. The number of such permutations is clearly $N_\rho = N! / ( n_1! \ldots n_s! )$.

To compute the ramified instanton partition function, we quotient the standard construction of the instanton moduli space we have reviewed earlier in \secref{Sec:Instantons} by a $\bZ_s$ - action where $s$ is the length of the partition $\rho$. The $\bZ_s$ - action is embedded inside the $(\bC^*)^2 \times GL(N,\bC)$ symmetry of the instanton moduli space. The component in $(\bC^*)^2$ acts on the complex coordinates by $(z_1,z_2) \to (z_1,\omega z_2)$ where $\omega^s=1$. The component in $GL(N,\bC)$ acts on the vector space $W$ such that it decomposes
\bea
W  = \oplus_{j=1}^s W_j\,,  \qquad n_j  = \dim_{\bC}W_j\,
\eea
into eigenspaces of the $\bZ_s$ - action. Our convention is that the generator of $\bZ_s$ acts on the vector space $W_j$ by $W_j \to \omega^j W_j$. In the sector with instanton number $k$, we must make an additional choice of the decomposition of the other vector space
\be
V = \oplus_{j=1}^s V_j\,, \qquad \dim_\bC V_j = k_j\,, \qquad \sum_{j=1}^sk_j = k \, .
\label{eq:OrbADHMData}
\ee
Each of these choices corresponds to a distinct topological sector and hence to a distinct ramified instanton moduli space $\cM_{\rho,k_1,\ldots,k_s}$. In summary, the ramified instanton moduli space $\cM_{\rho,k_1,\ldots,k_s}$ can be obtained as a $\bZ_s$ quotient of the standard instanton moduli space $\cM_{N,k}$ with $N=\sum_{j=1}^sn_j$ and $k=\sum_{j=1}^s k_j$.

Let us now explain how to compute the answer. The first statement is that each fixed point $\vec\lb$ of the standard instanton moduli space $\cM_{N,k}$ is also a fixed point of one and only one ramified instanton moduli space $\cM_{\rho,k_1,\ldots,k_s}$. The hardest part of the computation is to identify which sector $\{k_1,\ldots,k_s\}$ a given fixed point $\vec\lb$ contributes to. It is clear that the total number of boxes in $\vec \lb$ must add up to $k = \sum_{j=1}^sk_j$. Introduce the following labels for the Young tableaux
\bea
\vec \lb = \left\{ \lb_{j,\al} \right\}\,, \quad j=1,\ldots,s\,, \quad \al = 1,\ldots,n_s \,.
\eea
Then the boxes in the $i$-th column of $\lb_{j,\al}$ contribute to the instanton number $k_{i+j-1}$. If $i+j-1>N$ then we count modulo $N$.  We can denote the sector associated to a fixed point $\vec\lb$ by $k_j(\vec\lb)$.

For example, let us consider $U(2)$ theory in the two-instanton sector $k_1+k_2=k=2$, so we need to describe three spaces $\mathcal{M}_{2,1,1},\mathcal{M}_{2,2,0}, \mathcal{M}_{2,0,2}$. They are correspondingly generated by the following tuples of Young tableaux 
\begin{align}
\mathcal{M}_{2,1,1}:&\quad \left\{\yng(2),\emptyset\right\}\,,\quad \left\{\yng(1),\yng(1)\right\}\,,\quad \left\{\emptyset,\yng(2)\right\}\,,\notag \\
\mathcal{M}_{2,2,0}:&\quad \left\{\yng(1,1),\emptyset\right\}\,, \qquad \mathcal{M}_{2,0,2}:\quad \left\{\emptyset,\yng(1,1)\right\}\,.
\end{align}

Let us now compute the equivariant character of the tangent space to the ramified instanton moduli space at a fixed point $T_{\vec\lb}\cM_{\rho,k_1,\ldots,k_s}$. Let us denote this character by $\chi_{\rho,\vec\lb}(\vec a,\ep_1,\ep_2)$, whilst the character of unramified case is as above $\chi_{\vec\lb}(\vec a,\ep_1,\ep_2)$. According to the decomposition of space $W$ we have introduced the following notation for the equivariant parameters
\be
\vec a = \{ a_{j,\al} \}\,, \qquad j=1,\ldots,s\,, \quad \al = 1,\ldots,n_s \, .
\ee
Then we compute the character of the ramified moduli space of instantons via averaging over the orbifold action, therefore the answer is automatically invariant, which we can always write using a summation over weights
\bea
\chi_{\rho,\vec\lb}(\vec a,\ep_1,\ep_2) & = \frac{1}{s} \sum_{r=1}^s \chi_{\lb} \left(a_{j,\al} - \frac{\ep_2+2\pi i r}{s}j,\ep_1,\frac{\ep_2+2\pi i r}{s}\right)  \\
& = \sum_\al n_\al e^{w_\al}\,. 
\eea
Analogously to the unramified case we define
\begin{equation}
w_{\rho,\vec\lb} = \prod_\al \left( 2 \sinh(w_\al / 2) \right)^{-n_\al} \,.
\end{equation}
In addition, for $\cN=1^*$ theories there is the additional equivariant mass parameter $m$. We claim that this mass parameter is invariant under the orbifold action. Finally, the ramified instanton partition function is given by
\be
\CZ_{\rho} = \sum_{\vec \lb} Q_1^{k_1(\vec \lb)} \ldots Q_s^{k_s(\vec\lb)} w_{\rho,\vec{\lb}} \,.
\label{eq:ZramrhoGen}
\ee
Note that the product of instanton parameters for each sector is equal to the unramified instanton parameter 
\begin{equation}
Q_1\cdot\dots\cdot Q_s=Q\,, 
\label{eq:InstChargeConstr}
\end{equation}
which is also implied by the last equality in \eqref{eq:OrbADHMData}.

\subsection{3d Decoupling Limit}\label{Sec:tRS}
\subsubsection{$U(2)$ Theory}
We shall consider $\CN=1^*$ theory with gauge group $U(2)$ in great detail. In this case, there is only one non-trivial monodromy defect labelled by the partition $\rho=[1,1]$, which is expected to correspond to the surface defect obtained by coupling to $T[U(2)]$. The additional label $\sigma$ corresponds to permutations $\pm$ of the monodromy parameters, corresponding to the two massive supersymmetric vacua of $T[U(2)]$.

 It is convenient to make the following replacements
\begin{alignat}{3}
(+) & \quad Q_1 = z\,,\qquad && Q_2 = \frac{Q}{z}\,, \\
(-) & \quad Q_1 =  \frac{Q}{z}\,, \qquad &&  Q_2 = z\,.
\end{alignat}
in order to satisfy \eqref{eq:InstChargeConstr}. We shall immediately see physical meaning of both choices. As we have already mentioned, variable $Q$ can be identified with the holomorphic scale the five-dimensional bulk theory while $z$ can be understood as a ratio of the FI parameters of the 3d defect theory coupled to the bulk 5d theory $z=\frac{\tau_2}{\tau_1}$. 

Let us compute the instanton contributions to the Nekrasov partition function \eqref{eq:ZramrhoGen} and send $Q\to0$ in order to decouple the five-dimensional degrees of freedom. The first few terms are
\bea
\CZ^{(+)}_{[1,1]} = 1 & + \frac{q \left(\eta ^2-1\right) \left(\eta ^2 \mu _2 -\mu _1 \right)}{\eta ^2 (q-1) \left(q \, \mu _2 -\mu _1\right)}  z \\
& + \frac{q^2\left(\eta ^2-1\right)  \left(\eta ^2 q-1\right)  \left(\eta ^2 \mu _2-\mu _1\right)  \left(\eta ^2 \mu _2 q-\mu _1\right)}{\eta ^4 (q-1) (q^2-1)  \left(q \, \mu _2-\mu _1\right) \left(q^2 \, \mu _2 -\mu _1\right)} z^2  + \cO(z^3)\,,
\eea
\bea
\CZ^{(-)}_{[1,1]} = 1 & + \frac{q \left(\eta ^2-1\right) \left(\eta ^2 \mu _1 -\mu _2 \right)}{\eta ^2 (q-1) \left(q \, \mu _1 -\mu _2\right)}  z \\
& + \frac{q^2\left(\eta ^2-1\right)  \left(\eta ^2 q-1\right)  \left(\eta ^2 \mu _1-\mu _2\right)  \left(\eta ^2 \mu _1 q-\mu _2\right)}{\eta ^4 (q-1) (q^2-1)  \left(q \, \mu _1-\mu _2\right) \left(q^2 \, \mu _1 -\mu _2\right)} z^2 + \cO(z)^3 \,,
\eea
where we identify $q=e^{\epsilon_1}$. Note that the dependence on the equivariant parameter $\epsilon_2$ in the plane orthogonal to the defect has dropped out in this limit. The two expressions are related by $\mu_1 \leftrightarrow \mu_2$. It is straightforward to recognize them as q-hypergeometric series
\bea
\CZ^{(+)}_{[1,1]} = {}_2F_{1} \left(\eta ^2,\eta^2 \frac{ \mu _2}{\mu _1},q \frac{\mu _2 }{\mu _1},q, q\eta^{-2}z \right)\,, \\
\CZ^{(-)}_{[1,1]} = {}_2F_{1} \left(\eta ^2,\eta^2 \frac{ \mu _1}{\mu _2},q \frac{\mu _1}{\mu _2},q,q \eta^{-2} z \right)\,.
\label{eq:vortex-q-hypergeometric-series}
\eea
These are indeed the equivariant vortex partition function of $U(1)$ theory with two flavors, evaluated in each of the supersymmetric vacua. Alternatively they are the non-perturbative contributions to the two independent holomorphic blocks.  We can now see that $\CZ^{(+)}_{[1,1]}$ exactly matches with the vortex partition function of the $T[U(2)]$ theory in one of the vacua \eqref{eq:TstP1JFun}, whereas $\CZ^{(-)}_{[1,1]}$ coincides with the the vortex partition function in the other vacuum. Recall that both expressions are related merely by interchanging $\mu_1$ and $\mu_2$.

After redefinition \eqref{eq:ZeignRedefineSec3} the defect partition function satisfies trigonometric RS difference relations \eqref{eq:tRSeigenfunctionsSec2}, where momenta act as $p_{\tau_j} = e^{\ep_1 \tau_j\del_{\tau_j}}$.  

\subsubsection{$U(N)$ Theory}
When we consider the decoupling limit of the 5d $U(N)$ $\CN=1^*$ theory with maximal monodromy defect $[1^N]$ 
we can eliminate each of $Q_i$ in \eqref{eq:ZramrhoGen} from in favor of $Q$ and other $Q_i$. Thus there are $N$ different partition functions $\CZ_{i}^{[1^N]}$ which can be reproduced in the decoupling limit. Similarly to the $T[U(2)]$ example, they can be mapped onto vortex partition functions in \eqref{eq:S3-function-TU(N)}.

\subsection{Wilson Loops}
An essential ingredient of our construction is the computation of the vacuum expectation values of supersymmetric Wilson loops wrapping $S^1$. Using the language of equivariant characters we can easily modify formula \eqref{eq:PartFuncInstSum} in order to insert the fundamental Wilson loop operator inside
\be
\la W_{(1)} \ra =  \frac{\sum_{\vec\lb} q^{|\vec\lb|} \chi^{(\cE)}_{\vec\lb} \prod_{\al}   \left( 2 \sinh \left( \frac{w_\al}{2}  \right) \right)^{-n_\al} }{\sum_{\vec\lb} q^{|\vec\lb|} \prod_{\al}   \left( 2 \sinh \left( \frac{w_\al}{2}  \right) \right)^{-n_\al}} \, .
\label{eq:WilsonCharact}
\ee
The additional character $\chi^{(\cE)}_{\vec\lb}$ \eqref{eq:CharFramedSheaf} in the numerator represents the contributions from a heavy charged BPS particle propagating around $S^1$. It turns out that the numerator has the same universal divergence as the denominator in the limit $\epsilon_2 \to 0$. Thus the expectation value is finite and we can denote it by
\be
E_{(1)} = \lim_{\ep_2 \to \, 0} \la W_{(1)} \ra \, .
\ee
Remember that this is the Wilson loop expectation value of $U(N)$ gauge group, which differs from that of the $SU(N)$ gauge group by the overall $U(1)$ contribution. The overall $U(1)$ contribution can be interpreted as a contribution from a heavy free BPS particle and one can compute it using the Wilson loop expectation value of unit charge in the abelian gauge theory~\cite{BK2014}. 
The $U(1)$ contribution is given by
\be
	\la W^{U(1)}_{(1)}\ra = \frac{(Q/\eta^2,Q)_\infty(\eta^2Q/q,Q)_\infty}{(Q,Q)_\infty (Q/q,Q)_\infty} \ .
\ee
The fundamental Wilson loop expectation value of $SU(N)$ gauge group is then given by
\be
	\la W^{SU(N)}_{(1)}\ra = \frac{ \la W^{U(N)}_{(1)} \ra } { \la W^{U(1)}_{(1)}\ra }
\ee
with a condition $\prod_i \mu_i = 1$.
For later convenience we define 
\be\label{eq:WilsonLineVEVU1}
	\mathcal{N}_{(1)} \equiv \lim_{\epsilon_2\rightarrow 0}\la W^{U(1)}_{(1)}\ra^{-1} \ ,
\ee
which will be used in the difference equations below.

Let us consider some examples. For $U(2)$ $\CN=1^*$ theory with a fundamental Wilson loop around $S^1$ we find the following expectation value
\begin{equation} 
E_{(1)}^{U(2)}  =  (\mu_1+\mu_2) \left[ \, 1 - (1-\eta^2)(q-\eta^2)\frac{\mu _1 \mu _2 \left(\eta ^2+q \left(\eta ^4+\eta ^2+q\right)\right) - \left(\mu _1+\mu _2\right)^2 \eta^2 q}{\eta ^4 q \left(\mu _1 q-\mu _2\right) \left(\mu _2 q-\mu _1\right)}Q + \cO(Q^2) \, \right] \ .
\label{eq:WilsonLineVEVU2}
\end{equation}
Similarly, the Wilson loop expectation value in the fundamental representation of $U(3)$ is given by
\begin{align}
E_{(1)}^{U(3)} &= \mu_1\left[ \, 1-\frac{(1-\eta^2)(q-\eta^2)(\mu_1-\eta^2\mu_2)(\mu_1-\eta^2\mu_3)(\eta^2\mu_1-q\mu_2)(\eta^2\mu_1-q\mu_3)}{\eta^6q(\mu_1-\mu_2)(\mu_1-\mu_3)(\mu_1-q\mu_2)(\mu_1-q\mu_3)}Q  +\mathcal{O}(Q^2)  \right] \nonumber \\
&\quad +(\mu_1,\mu_2,\mu_3 \ {\rm cyclic \ permutations}) \ .
\end{align}
In order to insert a Wilson loop in the $r$-th skew symmetry power of the fundamental representation of $U(N)$ we must modify the computation. It will be discussed elsewhere.

\subsection{5d/3d Coupled System}\label{eq:eRS}
\subsubsection{$U(2)$ Theory}
Turning on the parameter $Q$, the partition function in the presence of the defect is no longer finite in the limit $\epsilon_2 \to 0$. The divergence exponentiates and the anomalous dimension is universal, that is, it is the same divergence without the defect. The saddle point equations arising from the anomalous dimension will fix the mass parameters $\mu_1$ and $\mu_2$ to be some discrete solutions. For now we will ignore this issue and concentrate on the normalized expectation value of the surface defect with unconstrained $\mu_1$ and $\mu_2$.

As the divergence is universal, the normalized expectation value of the monodromy defect is well defined in the limit $\epsilon_2\to0$. We introduce the notation
\be
\cD^{(\pm)}_{[1,1]} = \lim_{\ep_2\to0} \frac{\CZ_{[1,1]}^{(\pm)} }{\CZ} \,.
\label{eq:EigenFunctionsD}
\ee 
The first few terms of the expansion are
\bea
\cD^{(+)}_{[1,1]} & = 1+ \frac{\left(\eta ^2-1\right) q  \left(\eta ^2 \mu _2 -\mu _1\right)}{\eta ^2 (q-1)  \left(\mu _2 q-\mu _1\right)} z
+ \frac{ \left(\eta ^2-1\right)q \left(\eta ^2 \mu _1-\mu _2\right)}{\eta ^2 (q-1) \left(\mu _1 q-\mu _2\right)} \frac{Q}{z}  + \cdots\,, \\
\cD^{(-)}_{[1,1]} & = 1+ \frac{ \left( \eta ^2-1\right) q \left(\eta ^2 \mu _1-\mu _2\right)}{\eta ^2 (q-1)  \left(\mu _1 q-\mu _2\right)} z
+ \frac{\left(\eta ^2-1\right) q  \left(\eta ^2 \mu _2 -\mu _1 \right)}{\eta ^2 (q-1)  \left(\mu _2 q-\mu _1\right)}\frac{Q}{z} + \cdots\,.
\label{eq:eRSeigenfunctionsExp}
\eea
One may check that the above expressions reduce to the vortex partition functions in the limit $Q \to 0$. It is very important that we have a regular expansion in $Q_1$ and $Q_2$ and hence at higher orders in the $Q$ expansion there are negative powers of $z=\tau_2 / \tau_1$. It appears that $\cD^{(+)}$ and $\cD^{(-)}$ are related by interchanging the mass and FI parameters $(\mu_1,\tau_1)\leftrightarrow(\mu_2,\tau_2)$.

As a regular expansion in $Q_1$ and $Q_2$ we have checked up to order $\cO(Q_1^{n_1}Q_2^{n_2})$ with $n_1+n_2=5$ that the above normalized expectation values obey the following difference equations
\bea
\left(\frac{\mu _1}{\eta} \frac{  \theta_1 \left( \tau _1 / \eta ^2 \tau _2, Q \right)}{\theta_1 \left(\tau_1 / \tau_2,Q \right)} p_{\tau}^1 +
\eta  \mu _2\frac{  \theta_1 \left( \tau _2 / \eta ^2 \tau _1 , Q \right)}{\theta_1 \left(\tau _2 /\tau _1, Q \right)} p_{\tau}^2\right)\cD^{(+)}
& = \cN_{(1)} E_{(1)} \cD^{(+)}_{[1,1]}\,, \\
\left(\frac{\mu_2}{\eta} \frac{  \theta_1 \left( \tau _1 / \eta ^2 \tau _2, Q \right)}{\theta_1 \left(\tau _1 / \tau _2,Q \right)} p_{\tau}^1 +
\eta \mu _1 \frac{  \theta_1 \left( \tau _2 / \eta ^2 \tau _1 , Q \right)}{\theta_1 \left(\tau _2 /\tau _1, Q \right)} p_{\tau}^2\right)\cD^{(-)}
& = \cN_{(1)} E_{(1)} \cD^{(-)}_{[1,1]}\,, \eea
where $E_{(1)}$ is the normalized expectation value of a Wilson loop in the fundamental representation of $U(2)$ in the limit $\epsilon_2 \to0$ which we have computed earlier in \eqref{eq:WilsonLineVEVU2} and $\cN_{(1)}$ is the inverse of the overall $U(1)$ contribution defined in \eqref{eq:WilsonLineVEVU1}.
Therefore if we impose the traceless condition $\prod_i\mu_i=1$, the eigenvalues of the difference equations are the fundamental Wilson loop expectation value of $SU(2)$ gauge group.

As before, we can define new partition functions 
\bea
\cD^{(+)} =  \frac{ \theta(\eta \, \tau_1,q)\theta(\eta^{-1} \, \tau_2,q) }{\theta(\mu_1\tau_1,q)\theta(\mu_2\tau_2,q)} \cD^{(+)}_{[1,1]}\,, \\
\cD^{(-)} =  \frac{ \theta(\eta^{-1} \, \tau_1,q) \theta(\eta \, \tau_2,q) }{\theta(\mu_2\tau_1,q)\theta(\mu_1\tau_2,q)} \cD^{(-)}_{[1,1]}\,,
\eea
which obey the same difference equations
\bea
\left(\frac{  \theta_1 \left( \tau _1 / \eta ^2 \tau _2, Q \right)}{\theta_1 \left(\tau _1 / \tau _2,Q \right)} p_{\tau}^1 +
\frac{  \theta_1 \left( \tau _2 / \eta ^2 \tau _1 , Q \right)}{\theta_1 \left(\tau _2 /\tau _1, Q \right)} p_{\tau}^2\right) \cD^{(\pm)}  & = \cN_{(1)} E_{(1)} \, \cD^{(\pm)}\,, \\
p_{\tau}^1 \, p_{\tau}^2\,\cD^{(\pm)}  = \mu_1 \mu_2\, \cD^{(\pm)}\,. 
\label{eq:eRS-Hamiltonian}
\eea
The same non-uniqueness caveats as before holds also here. One should also check that the perturbative contributions obtained from orbifolding factorize nicely into contributions from the defect and the bulk. At this point we claim to have found formal eigenfunctions and eigenvalues of the two-body elliptic (complexified) Ruijsenaars-Schneider integrable system, at least as a series expansion. 

\subsubsection{$U(3)$ Theory}
The generalization to many-body system is straightforward -- one needs to compute ramified instanton partition function of $U(N)$ $\CN=1^*$ theory in the presence of full $\rho=[1,\dots, 1]:=[1^N]$ monodromy defect. In this section we will present the results for $U(3)$ theory.
There are $3!$ different embeddings of the Levi subgroup $\mathbb{L}=U(1)^3$ into $U(3)$ labelled by a permutation $\sigma$.
We shall focus on a particular embedding $\sigma=1$ in what follows.

We are interested in the normalized partition function in the presence of a monodromy defect in the limit $\epsilon_2\rightarrow1$.
As discussed in the previous sections, this limit is well defined and we end up with a finite expression.
The partition function is expanded as
\begin{align}
	\lim_{\epsilon_2\rightarrow0}\frac{\cZ_{[1^3]}}{\cZ} &= 1 + \frac{q(1-\eta^2)}{\eta^2(1-q)} \bigg[\frac{(\mu_1-\eta^2\mu_2)}{(\mu_1-q\mu_2)} z_1 + \frac{(1-\eta^2)(\eta^2\mu_1-\mu_3)(\eta^2\mu_2-q\mu_3)}{\eta^2(1-q)(q\mu_1-\mu_3)(\mu_2-\mu_3)} z_2z_3 \nonumber \\
	& \qquad + \frac{(\eta^2\mu_1-\mu_2)(\eta^2\mu_2-q\mu_3)(\mu_2-\eta^2\mu_3)}{\eta^2(q\mu_1-\mu_2)(\mu_2-\mu_3)(\mu_2-q\mu_3)} z_2z_3 + \frac{ (1-q\eta^2)(\mu_1-\eta^2\mu_2)(\mu_1-q\eta^2\mu_2) }{ \eta^2(1-q^2)(\mu_1-q\mu_2)(\mu_1-q^2\mu_2) } z_1^2 \nonumber \\
	& \qquad + (\mu_1,\mu_2,\mu_3 \ {\rm and} \ z_1,z_2,z_3 \ {\rm cyclic \ permutations}) \bigg] + \cdots \ ,
\end{align}
where $z_1=\frac{\tau_2}{\tau_1}, z_2 = \frac{\tau_3}{\tau_2}, z_3=\frac{\tau_1}{\tau_3}Q$. We have also assumed that $|z_1|,|z_2|,|z_3| < 1$ in the expansion. In the decoupling limit $Q\rightarrow 0$, this partition function precisely reproduces the vortex partition function of the $T[U(3)]$ theory in~\eqref{eq:S3-function-TU(N)} in one of the vacua, after a relabelling $(\mu_i,\tau_i) \rightarrow (\mu_i^{-1},\tau_i^{-1})$.

Let us define a new function as
\begin{equation}
	\cD_{[1^3]} = \frac{ \theta(\eta^2 \, \tau_1,q)\theta(\tau_2,q)\theta(\eta^{-2} \, \tau_3,q) }{\theta(\mu_1\tau_1,q)\theta(\mu_2\tau_2,q)\theta(\mu_3\tau_3,q)}\lim_{\epsilon_2\rightarrow0}\frac{\cZ_{[1^3]}}{\cZ} \ .
\end{equation}
It follows that this function satisfies three difference equations:
\begin{align}\label{eq:eRSSp}
&\left( \frac{\theta_1 (\tau_2/\eta^2\tau_1,Q)}{\theta_1 (\tau_2/\tau_1,Q)}\frac{\theta_1 (\tau_3/\eta^2\tau_1,Q)}{\theta_1 (\tau_3/\tau_1,Q)}p_{\tau}^1 \right.
+\frac{\theta_1 (\tau_1/\eta^2\tau_2,Q)}{\theta_1 (\tau_1/\tau_2,Q)}\frac{\theta_1 (\tau_3/\eta^2\tau_2,Q)}{\theta_1 (\tau_3/\tau_2,Q)}p_{\tau}^2   \\
& \hspace{4cm} +\left.\frac{\theta_1 (\tau_1/\eta^2\tau_3,Q)}{\theta_1 (\tau_1/\tau_3,Q)}\frac{\theta_1 (\tau_2/\eta^2\tau_3,Q)}{\theta_1 (\tau_2/\tau_3,Q)}p_{\tau}^3 \right) \cD_{[1^3]} = \cN_{(1)}E_{(1)} \cD_{[1^3]}  \ ,\nonumber \\
&\left( \frac{\theta_1 (\tau_1/\eta^2\tau_2,Q)}{\theta_1 (\tau_1/\tau_2,Q)}\frac{\theta_1 (\tau_1/\eta^2\tau_3,Q)}{\theta_1 (\tau_1/\tau_3,Q)}p_{\tau}^2p_{\tau}^3 \right.
+\frac{\theta_1 (\tau_2/\eta^2\tau_3,Q)}{\theta_1 (\tau_2/\tau_3,Q)}\frac{\theta_1 (\tau_2/\eta^2\tau_1,Q)}{\theta_1 (\tau_2/\tau_1,Q)}p_{\tau}^3p_{\tau}^1  \nonumber \\
& \hspace{3.6cm} +\left.\frac{\theta_1 (\tau_3/\eta^2\tau_1,Q)}{\theta_1 (\tau_3/\tau_1,Q)}\frac{\theta_1 (\tau_3/\eta^2\tau_2,Q)}{\theta_1 (\tau_3/\tau_2,Q)}p_{\tau}^1 p_{\tau}^2 \right) \cD_{[1^3]} = \cN_{(1)}E_{(1,1)} \cD_{[1^3]} \nonumber \,
\end{align}
and $p_\tau^1p_\tau^2p_\tau^3 \cD_{[1^3]} = \mu_1\mu_2\mu_3\cD_{[1^3]}$. Here $\cN_{(1)}$ is the inverse of the overall $U(1)$ Wilson loop factor and $E_{(1)}$ is the expectation value of a Wilson loop in the fundamental representation of $U(3)$.
$E_{(1,1)}$ is the Wilson loop expectation value in the rank-two antisymmetric tensor representation. We can compute $E_{(1,1)}$ using the fundamental Wilson loop expectation value by a simple replacement of the gauge fugacities such as 
\begin{equation}
	E_{(1,1)}(\mu_1,\mu_2,\mu_3) = E_{(1)}\big|_{(\mu_1,\mu_2,\mu_3) \rightarrow (\mu_2\mu_3,\mu_3\mu_1,\mu_1\mu_2)} \ ,
\end{equation}
which holds only for $U(3)$. The difference equations have been checked in a series expansion in $z_i$'s up to $\mathcal{O}(z_1^{n_1}z_2^{n_2}z_3^{n_3})$ with $n_1+n_2+n_3=5$.

The difference equations are the eigenvalue equations of the three-particle eRS integrable system. Therefore we found that the monodromy defect partition functions are eigenfunctions of the integrable Hamiltonians and the Wilson loops in rank $1$ and $2$ antisymmetric representations are their eigenvalues. In the decoupling limit $Q\rightarrow0$ one can also notice that these equations reduce to the equations for three-body tRS Hamiltonians in (\ref{eq:AlmostEigentRS3Sec3}).

\subsubsection{Comments on Normalizability of Wavefunctions}
The complexification plays a crucial role here. Indeed, normally one may expect that elliptic system has a discrete spectrum, whereas our eigenfunctions are parametrized by parameters variables $\mu_1$ and $\mu_2$. The point is that we have found only formal eigenfunctions in that they may not be normalizable wavefunctions. In solving the additional Bethe equations arising from the saddle point analysis of the effective twisted superpotential $\cW$ of the defect theory coupled to the 5d theory in the divergence $\log Z \to \frac{1}{\ep_2}\cW + \cO(1)$ we will fix the parameters $\mu_1$ and $\mu_2$ to discrete set of values. It is possibly the case that for these values we obtain normalizable wavefunctions and hence obtain a discrete spectrum.  

The quantum elliptic Ruijsenaars-Schneider model in connection with supersymmetric gauge theories has been discussed in the
the literature before. For instance, in \cite{Gaiotto:2012xa}, some approximate solutions to the elliptic RS eigenvalue problem are labeled by a discrete parameter. We therefore expect that extremization of eigenfunctions \eqref{eq:EigenFunctionsD} with respect to the mass parameters $\mu_i$ (putting the solution on shell) will resolve the normalizability issues. We plan to discuss it elsewhere.

\subsection{5d Theory Coupled to 3d Hypermultiplets}\label{Sec:FreeHypers5d}
We now consider a different type of co-dimension two defect in five dimensions. We shall couple 3d hypermultiplets to the 5d maximal supersymmetric $SU(N)$ gauge theory. The 3d hypermultiplets have $U(1)\times SU(N)$ flavor symmetry and we couple this $SU(N)$ symmetry to the bulk gauge symmetry. Our motivation is that the $U(2)$ gauge theory with this type of defect turns out to be related to the $U(2)$ gauge theory with a monodromy defect of type $\rho=[1^2]$ by {\it bispectral} duality. 

In this section, however, we carry out the computation using $U(N)$ and we set $\prod_{i=1}^N\mu_i=1$ to get the results for $SU(N)$.
Naively $U(N)$ and $SU(N)$ are the same under the constraint $\prod_{i=1}^N\mu_i=1$ as the diagonal $U(1)$ part decouples in the field theory limit. However, this is no longer true in the instanton computation. The subtle issues related to the $U(1)$ part in the 5d partition functions are discussed in \cite{Bao:2013pwa,Hayashi:2013qwa,Hwang:2014uwa}.
Here we assume that the $U(1)$ part does not affect the gauge theory dynamics and its contribution to the partition function can be subtracted by hand.

In order to compute the partition function in the presence of this 3d defect we shall follow the prescription given in \cite{GK2014}. The defect introduces an additional three dimensional vector bundle on the instanton moduli space of the 5d gauge theory. The 3d fields of the defect theory are in the fundamental and anti-fundamental representations of the $SU(2)$ bulk gauge group and thus the corresponding bundles are the universal bundle and its conjugation, respectively. The equivariant characters of the 3d hypermultiplets can be computed using the equivariant character $\chi^{(\mathcal{E})}$ of the universal bundle at $k$ instantons. We get
\begin{align}\label{eq:3d-characters}
\chi^{3d}_{\rm chiral} &= \frac{e^{\frac{m-\epsilon_1}{2}-x}\chi_k^{(\mathcal{E})}}{(1-e^{-\epsilon_1})} = e^{\frac{m-\epsilon_1}{2}-x}\left[\frac{\sum_{I=1}^2 e^{a_I}}{(1-e^{-\epsilon_1})} - (1-e^{-\epsilon_2}) \sum_{i=1}^k e^{\phi_i}\right]\,, \cr
\chi^{3d}_{\rm anti} &= \frac{e^{\frac{m-\epsilon_1}{2}+x}\chi_k^{(\mathcal{E}^*)}}{(1-e^{-\epsilon_1})} = e^{\frac{m-\epsilon_1}{2}+x}\left[\frac{\sum_{I=1}^2 e^{-a_I}}{(1-e^{-\epsilon_1})} - (1-e^{-\epsilon_2}) e^{\epsilon_1+\epsilon_2}\sum_{i=1}^k e^{-\phi_i}\right]\,,
\end{align}
where $\chi_{\rm chiral}$ and $\chi_{\rm anti}$ are the equivariant characters of the 3d chiral and anti-chiral multiplets respectively, and $x$ is the equivariant parameter for the $U(1)$ flavor symmetry. These characters are expressed in term of the equivariant parameter $e^{\phi_i}$ for the auxiliary gauge group $U(k)$ of the $k$ instanton moduli space. Due to the 3d contribution the saddle point value of $\phi_i$ is not fully classified by the $N$-tuple of Young tableaux.
We note that the path integral on the instanton moduli space has extra saddle points from the 3d factors other than the previous saddle points labeled by Young tableaux.

We find from the above character formulae the extra contribution of the 3d fields to the $k$ instanton partition function in the integral expression. It is given by
\begin{equation}\label{eq:3d-inst}
\CZ^{3d}_k = \prod_{i=1}^k\frac{(1-\eta^{-1}\tau^{-1}\rho_i)(1-pq\eta^{-1}\tau\rho_i^{-1})}{p(1-p^{-1}\eta^{-1}\tau^{-1}\rho_i)(1-q\eta^{-1}\tau\rho_i^{-1})}\,,
\end{equation}
where $\eta^2 = e^{-m+\epsilon_1}, \rho_i = e^{\phi_i}, \tau = e^{x}, p=e^{\epsilon_2}$.

There are subtleties in the perturbative contribution related to the boundary condition on $\partial (S^1\times \mathbb{R}_{\epsilon_1}^2) \cong T^2$ due to the presence of the Omega background. The superpotential and the Chern-Simons term in the 3d theory are in general not invariant under the supersymmetry in the presence of a boundary \cite{Gadde:2013sca,Yoshida:2011au}.
We will discuss the latter in the next subsection when we gauge the 3d flavor symmetry.

Let us focus on the standard $\mathcal{N}=4$ superpotential.
For being supersymmetric with boundary we should impose the relevant boundary conditions for the fields. 
For the adjoint chiral multiplet in the $\mathcal{N}=4$ vector multiplet, we impose the Dirichlet boundary condition. The boundary conditions for the hypermultiplets, we impose the Neumann boundary condition on the chiral multiplets and the Dirichlet boundary condition on the anti-chiral multiplets, or vice versa.
This choice guarantees the supersymmetry invariance of the superpotential in the presence of the boundary. 

The partition function of 3d theories on solid torus $S^1\times D^2$ with specified boundary conditions was recently computed in \cite{Yoshida:2014ssa} using localization. We will compute the 1-loop determinant for the 3d hypermultiplets using character formula (\ref{eq:3d-characters}) and the appropriate boundary conditions. The terms independent of the instanton number $k$ give the perturbative contributions. See \appref{app:perturbative} for detailed 1-loop computations.
The 1-loop determinant of the 3d hypermultiplets is given by
\begin{equation}
Z_{\rm 1-loop}^{3d} =e^{-\frac{m}{\epsilon_1}(\pi i-x)}
\prod_{I=1}^2 \frac{ (q\eta^{-1}\tau\mu_I^{-1};q)_\infty } { (\eta\tau\mu_I^{-1};q)_\infty } \ .
\label{eq:partition-function-3dhyper}
\end{equation}
The prefactor comes from regularization.
Note that this 1-loop contribution includes the prefactor $e^{\frac{mx}{\epsilon_1}}$ which amounts to the mixed Chern-Simons term between the background $U(1)$ flavor gauge field and R-symmetry current.

The full partition function of the $\mathcal{N}=1^*$ theory in the presence of the 3d hypermultiplets is given by
\begin{equation}
\CZ^{3d/5d} =  Z_{\rm 1-loop}^{3d}Z_{\rm 1-loop}^{5d} \sum_{k=0}^\infty Q^k \ Z_k \,, \quad Z_k= \oint \prod_{i=1}^k \frac{d\rho_i}{2\pi i\rho_i} Z_k^{5d}\, Z_k^{3d}\ .
\label{eq:partition-function-3d5d}
\end{equation}
$Z^{5d}_{\rm 1-loop}$ and $Z^{5d}_k$ are the 1-loop and $k$ instanton partition function of the original 5d theory without the 3d fields,
\begin{align}
Z^{5d}_{\rm 1-loop} \sim & \frac{(p;p,q)_\infty (q;p,q)_\infty}{ (p\eta^{-2};p,q)_\infty (q\eta^{-2};p,q)_\infty } \prod_{I\neq J}^2\left[\frac{(\mu_I/\mu_J;p,q)_\infty (pq\mu_I/\mu_J;p,q)_\infty}{ (p\eta^{-2}\mu_I/\mu_J;p,q)_\infty (q\eta^{-2}\mu_I/\mu_J;p,q)_\infty } \right]^{1/2} \cr
Z^{5d}_{k}=&\prod_{I=1}^2\prod_{i=1}^k\frac{(1-p^{-1}\eta^{-2}\rho_i/\mu_I)(1-q\eta^{-2}\mu_I/\rho_i)}{(1-\rho_i/\mu_I)(1-pq\mu_I/\rho_i)} \prod^k_{i\neq j}(1-\rho_i/\rho_j)\,, \cr
& \times \prod_{i=j}^k\frac{(1-pq\rho_i/\rho_j)}{(1-p\rho_i/\rho_j)(1-q\rho_i/\rho_j)}\frac{(1-\eta^{-2}\rho_i/\rho_j)(1-qp^{-1}\eta^{-2}\rho_i/\rho_j)}{(1-p^{-1}\eta^{-2}\rho_i/\rho_j)(1-q\eta^{-2}\rho_i/\rho_j)}\,.
\end{align}

One can obtain the instanton part using the quantum mechanics on the instanton moduli space. For the $\mathcal{N}=1^*$ theory with $U(N)$ gauge group, the Witten index of the $\mathcal{N}=(4,4)$ ADHM gauged quantum mechanics gives the instanton partition function, which was computed in \cite{Kim:2011mv} using localization.
When we couple the 3d fields, the supersymmetry reduces to $\mathcal{N}=(2,2)$ and one can deduce from (\ref{eq:3d-inst}) that there would be extra degrees of freedom, a chiral and an anti-chiral multiplets, added to the quantum mechanics.

The contour integral \eqref{eq:partition-function-3d5d} over $\rho_i$ can be evaluated using the Jeffrey-Kirwan (JK) residue prescription introduced in \cite{1993alg.geom..7001J,Benini:2013xpa}. The JK prescription has also been applied to the localization of the index in quantum mechanics, which we will briefly review now. See \cite{Hwang:2014uwa,Cordova:2014oxa,Hori:2014tda} for more detailed explanations.

We can first define a hyperplane in the $\phi$ plane for each charge vector $Q_i \in \mathbb{R}^k$ of the multiplets where the integrand in $Z_k$ \eqref{eq:partition-function-3d5d} becomes singular
\begin{equation}
	H_i = \{ \phi \in \mathbb{C}^k \, \big| \, Q_i(\phi) + z = 0\}\,.
\end{equation}
Here $z$ denotes other chemical potentials (or log of fugacities).
Let us then consider when $n\ge k$ hyperplanes meet at a point $\phi=\phi^*$ and denote by $\mathrm{Q}(\phi^*) \equiv \{Q_i \big| \phi\in H_i\}$ a set of $n$ charge vectors at the point. The residue at the singular point can be computed using the JK prescription.

The integrand of $Z_k$ is Laurent expanded around the singular point $\phi^*$ in negative powers of $Q_i(\phi-\phi^*)$. Among others, the JK residue yields nonzero result only at simple poles of the form
\begin{equation}
\frac{1}{Q_{i_1}(\phi-\phi^*)\cdots Q_{i_k}(\phi-\phi^*)}\,,
\end{equation}
where $Q_{i_1},\cdots,Q_{i_k}$ are in $\mathrm{Q}(\phi^*)$. 

We will choose a reference vector $\eta$ arbitrarily in $\mathbb{R}^k$. This vector $\eta$ should not be confused with $\mathcal{N}=1^*$ mass parameter. The JK residue defined in~\cite{1993alg.geom..7001J} is:
\begin{equation}
\text{JK-Res}(\mathrm{Q}_*,\eta)\frac{d Q_{i_1}(\phi)\wedge \cdots \wedge d Q_{i_k}(\phi)}{Q_{i_1}(\phi)\cdots Q_{i_k}(\phi)}   = 
\left\{ \begin{array}{cl} \left|{\rm det}(Q_{i_1}\cdots Q_{i_k})\right|^{-1} & \ \text{if} \ \eta\in \text{Cone}(Q_{i_1},\cdots,Q_{i_k}) \\ 0 & \ \text{otherwise} \end{array}\right.
\end{equation}
where 'Cone' denotes the cone formed by the $k$ independent charge vectors. The result turns out to be independent of the choice of $\eta$. The JK prescription can be applied when the projective condition is satisfied \cite{1993alg.geom..7001J}. This turns out to be the case with our $Z_k$. 
The partition function now takes the form of
\begin{equation}
Z_k  = \frac{1}{|W|}\sum_{\phi_*} \text{JK-Res} (Q_*,\eta) \ Z_k^{5d}(\phi,z)Z_k^{3d}(\phi,z)
\end{equation}
where $W$ is the Weyl group of the $U(k)$ gauge group.

For the 5d $U(N)$ gauge theory without coupling the 3d fields, the JK prescription reproduces the Young tableaux sum rule of the instanton partition function. To check this, we need to choose the reference vector like $\eta = (1,1,\cdots,1)$. However, after coupling the 3d fields, the residue prescription implies that there would be nontrivial contributions at the extra poles developed by the 3d factors as well as the Young tableaux summation.

For instance, at the single instanton sector, we should pick up all the poles from the fields of positive charge which give the factors of the form $\frac{1}{\sinh\frac{Q_i\phi_1}{2}}$ with $Q_i>0$. From the formula for $Z_{k=1}$ given above one can see that the poles at $\{\rho_1 = \mu_1,\rho_1 = \mu_2, \rho_1=p\eta\tau\}$ are inside the contour. Note that the first two poles are from the 5d factors which were labeled by the $N$-tuple of Young tableaux, whereas the last pole is the new pole arising from the 3d factors. The JK residue at the last pole is nontrivial and should be involved in the $Z_{k=1}$ computation.

At two instanton sector, we have nontrivial poles due to the 3d factors at
\begin{equation}
(\rho_1=p\eta\tau,\rho_2=\mu_1),\ (\rho_1=p\eta\tau,\rho_2=\mu_2), \ (\rho_1=p\eta\tau,\rho_2=q^{-1}\rho_1), \ (\rho_1=p\eta\tau,\rho_2=p\eta^2\rho_1)
\end{equation}
apart from the poles in the Young tableaux classification.
Summing over all JK residues at the these poles, one can compute the two instanton partition functions $Z_{k=2}$ in the presence of the 3d matter fields. We expect that the Jeffrey-Kirwan method also works for higher instanton sectors $Z_{k>2}$.

Let us now define the normalized partition function of the 3d-5d coupled system and take the limit $\epsilon_2\rightarrow 0$
\begin{equation}
\cD^{3d/5d} = e^{-\frac{mx}{\epsilon_1}}\lim_{\epsilon_2\rightarrow0} \frac{\CZ^{3d/5d}}{\CZ}\,.
\label{eq:DeRSEigenFunction}
\end{equation}
Here we turned on the classical mixed Chern-Simons term $e^{-\frac{mx}{\epsilon_1}}$ to cancel the induced Chern-Simons term.
This normalized partition function obeys the difference equation
\begin{equation}
\left(\eta\tau \theta(q^2 \eta^{-4} p_\tau,Q) + \eta^{-1}\tau^{-1} \theta(p_\tau,Q)\right)\cD^{3d/5d}  = \theta(q \eta^{-2} p_\tau,Q)\mathcal{N}_{(1)}E_{(1)}\, \cD^{3d/5d} \,,
\end{equation}
where, as before, we denoted by $p_\tau$ the difference operator satisfying $p_\tau \tau = q \tau p_\tau$ so that $p_\tau$ becomes the conjugate momentum of $\tau$. 
The factor $\cN_{(1)}$ is given in \eqref{eq:WilsonLineVEVU1} and $E_{(1)}$ is the Wilson loop expectation value in \eqref{eq:WilsonLineVEVU2}.
We have checked this relation by expanding both sides up to two instanton order.

One can also recast the above difference equation as an eigenfunction equation
\begin{equation}
\left(\tau \frac{\theta_1(q^2 \eta^{-4} p_\tau,Q)}{\theta_1(q^2 \eta^{-2} p_\tau,Q)} + \frac{1}{\tau} \frac{\theta_1(p_\tau^{-1},Q)}{\theta_1(\eta^2 p_\tau^{-1},Q)}\right)\cD^{3d/5d} = \mathcal{N}_{(1)}E_{(1)}\,\cD^{3d/5d}\,,
\label{eq:DualeRSEigenfunction}
\end{equation}
To verify this equation we should expand both sides first in $Q$ and later in $\frac{1}{\eta}$. We have checked for some lowest orders in these expansions. Therefore we have computed the spectrum of the two-body \textit{dual elliptic Ruijsenaars-Schneider} Hamiltonian (l.h.s. of \eqref{eq:DualeRSEigenfunction}), whose eigenvalue is the expectation value of the Wilson loop \eqref{eq:WilsonLineVEVU2} and the eigenfunction, up to a normalization, is the partition function of the 5d $\cN=1^*$ theory coupled to two free 3d hypermultiplets \eqref{eq:DeRSEigenFunction}. 

\subsection{S-Transformation}\label{Sec:STransform}
We will now show that {\it S-transformation} which is an extension of the S-duality in three dimensions can relate two different types of defects in five dimensions: a monodromy defect and a defect of 3d hypermultiplets. More precisely, we will act the S operation on the partition function $Z^{\rm 3d/5d}$ of (\ref{eq:partition-function-3d5d}) and show that the result agrees with the partition function of the $U(2)$ gauge theory in the presence of a monodromy defect.

Let us first review the S transformation in three dimension.
There is a natural $SL(2,\mathbb{Z})$ action on 3d CFTs with $U(1)$ symmetry \cite{Witten:2003ya}. This action maps a theory to other theories which could be inequivalent to the original theory. It is sufficient to understand two generators $T$ and $S$ in the $SL(2,\mathbb{Z})$ action.
The $T$ transformation simply shifts by one unit the Chern-Simons level of the background gauge field $A$ for the $U(1)$ symmetry.
The $S$ transformation is more complicated. Firstly we gauge the $U(1)$ global symmetry and make the background gauge field $A$ as a dynamical gauge field. The theory then has a new global symmetry whose conserved current is the magnetic flux of $A$. Secondly we introduce a mixed Chern-Simons term $AdB$ at unit level with background gauge field $B$ for the new global symmetry. 

When the theory is on the manifold with boundary, the $S$ transformation should be carefully taken. If there exists the effective Chern-Simons term of the background gauge field $A$, the naive gauging procedure fails to work due to the gauge anomaly. The Chern-Simons term makes the gauge invariance anomalous on the boundary. To keep the gauge invariance one need to couple relevant boundary degrees of freedom such that the flavor anomaly of the 2d theory compensates the gauge anomaly on the boundary.
In addition, the mixed Chern-Simons term introduced by the $S$ transformation also breaks the gauge invariance. The 2d theory should be chosen to cancel this anomaly as well.

In the partition function, the gauge invariance is associated to the periodicity of the holonomy parameter, $a \sim a + 2\pi i$. The Chern-Simons coupling induces the terms violating this periodicity. To cancel these gauge non-invariant terms, we should multiply the elliptic genus of the proper 2d theory. We will now see this with our example.

Let us first consider the $S$ transformation on the 3d partition function of the defect after decoupling the 5d theory.
The 3d theory consists of free hypermultiplets of the $SU(2)$ doublet.
The $S$ transformation acts on the $U(1)$ flavor symmetry with parameter $x$.
We turn off all the classical background Chern-Simons terms.
However, the 1-loop effect generates the nontrivial effective mixed Chern-Simons term between the $U(1)$ flavor symmetry and the $U(1)$ R-symmetry. In the 3d partition function (\ref{eq:partition-function-3dhyper}), the prefactor $e^{\frac{m x}{\epsilon_1}}$ induced from the 1-loop determinant amounts to the dynamically generated Chern-Simons coupling. This prefactor violates the periodicity of the parameter $x$. Moreover, the mixed Chern-Simons term $AdB$ in the $S$ transformation adds a term like $e^{-\frac{yx}{\epsilon_1}}$ where $y$ is the holonomy parameter for the new $U(1)$ flavor symmetry, which also violates the periodicity of $x$. These gauge non-invariant terms must be canceled. In the previous section, we cancel the former anomalous factor by turning on the classical Chern-Simons term corresponding to $e^{-\frac{mx}{\epsilon_1}}$. In this section we couple instead boundary $\mathcal{N}=(0,2)$ multiplets and cancel the anomalous boundary terms.

The $\mathcal{N}=(0,2)$ theory on the boundary $T^2$ consists of two fermi and one chiral multiplets. We refer the reader to \cite{Gadde:2013dda,Gadde:2013sca} for details of elliptic genera of the $\mathcal{N}=(0,2)$ theories. The regularized elliptic genus is summarized in \appref{app:perturbative}. We introduce the boundary theory whose the elliptic genus is given by
\begin{align}
Z^{2d}&=\exp\left(-2\pi i \zeta_2(0,-\frac{m+x-y}{2\pi i}|1,\frac{\epsilon_1}{2\pi i})+2\pi i \zeta_2(0,-\frac{x}{2\pi i}|1,\frac{\epsilon_1}{2\pi i})+2\pi i \zeta_2(0,-\frac{m-y}{2\pi i}|1,\frac{\epsilon_1}{2\pi i})\right)\notag \\
&\cdot\frac{ \theta(\tau^{-1};q) \theta(q^{-1}\eta^2u;q)}{\theta(q^{-1}\eta^2u\tau^{-1};q)}\,
\end{align}
with $u=e^y$. One can check that the prefactor precisely cancels the gauge non-invariant terms of the 3d theory on the boundary.

We now gauge the $U(1)$ symmetry parametrized by $\tau$. Then the S transformation on the 3d partition function can be written as \cite{Beem:2012mb}
\begin{align}
Z^{3d}_S(u) &= \int\frac{d\tau}{\tau} e^{\frac{yx}{\epsilon_1}} Z^{2d}(u,\tau) \, Z^{3d}_{\rm 1-loop}(\tau) \nonumber \\
=&\int\frac{d\tau}{\tau} e^{-\mathcal{F}}\, \frac{ \theta(\tau^{-1};q) \theta(q^{-1}\eta^2u;q) }
{ \theta(q^{-1}\eta^2u\tau^{-1} ;q) } 
\prod_{I=1}^2 \frac{ (q\eta^{-1}\tau\mu_I^{-1};q)_\infty } { (\eta\tau\mu_I^{-1};q)_\infty }\,,
\end{align}
where the prefactor is $\mathcal{F} = \big(\pi im - \pi i\epsilon_1/2 +\pi^2/3 -\epsilon^2/12\big)/\epsilon_1$ and $\tau=e^x$.

The integral can be evaluated using the residue theorem with an appropriate pole prescription. We propose that the integral contour encloses the poles at $\tau = q^{-k}\eta^{-1}\mu_I,\, k\ge 0$ which come from the 1-loop determinant of the 3d fundamental chiral multiplet.
It may be possible to justify this pole prescription by the JK-like residue prescription, but we will not attempt to do it.
Let us focus on the residues at poles of $\tau = q^{-k}\eta^{-1}\mu_1$. 
Applying the residue theorem, we obtain
\begin{equation}
	Z^{3d,(+)}_S(u) = Z_0	\times \frac{(q\eta^{-2};q)_\infty(q\eta^{-2}\mu_1/\mu_2;q)_\infty}{(q;q)_\infty(\mu_1/\mu_2;q)_\infty} \times
	{}_2F_{1} \left(\eta ^2,\eta^2 \frac{ \mu _2}{\mu _1},q \frac{\mu _2 }{\mu _1};q, q\eta^{-2}u \right)\,,
\end{equation}
where
\begin{equation}
	Z_0 = -e^{-\mathcal{F}}\frac{\theta(\eta\mu_1^{-1};q)\theta(q^{-1}\eta^2u;q)}{\theta(q^{-1}\eta^3u\mu_1^{-1};q)}\,.
\end{equation}
Apart from $Z_0$, this $S$ transformed partition function reproduces the holomorphic block of the 3d $U(1)$ gauge theory with 2 flavors in one of the supersymmetric vacua. The q-hypergeometric series ${}_2F_{1}$ is precisely the vortex partition function $Z^{(+)}_{[1^2]}$ of the monodromy defect $\rho=[1^2]$ in (\ref{eq:vortex-q-hypergeometric-series}) upon the identification $u=\tau_2/\tau_1$.
This result shows that the 3d theories on the defects of two different types are related by the S transformation. The residues at the poles $\tau = q^k\eta^{-1}\mu_2$ yields the same result with $\mu_1$ and $\mu_2$ exchanged which corresponds to the holomorphic block in the second vacuum.

The S transformation can be promoted to five dimensions. It acts on the $U(1)$ flavor symmetry of the 3d theory on the defect as we did above in the decoupling limit.  The $S$ transformation on the surface defect partition function is defined as in \cite{GK2014}
\begin{equation}
Z^{3d/5d}_S(u,Q) = \int\frac{d\tau}{\tau} e^{\frac{yx}{\epsilon_1}} Z^{2d}(u,\tau) \, Z^{3d/5d}(\tau,Q) \,.
\label{eq:STransform}
\end{equation}
The contour is chosen to enclose the poles at $\tau = q^{-k}\eta^{-1}\mu_I,\, k\in \mathbb{Z}$. We note that the instanton contribution also has poles at $\tau = q^k\eta^{-1}\mu_I$ for negative integer $k<0$ and the contributions from the residues at these poles are crucial to compare with the partition function of the dual theory.

We sum over all residues of the poles at $\tau = q^{-k}\eta^{-1}\mu_1$ and find
\begin{equation}
Z^{3d/5d,(+)}_S(u,Q) = Z_0	\times \frac{(q\eta^{-2};q)_\infty(q\eta^{-2}\mu_1/\mu_2;q)_\infty}{(q;q)_\infty(\mu_1/\mu_2;q)_\infty} \times Z^{\rm inst,(+)}_S(u,Q)\,.
\end{equation}
The instanton partition function after $S$ transformation is expanded in $u$ and $Q/u$ and the first few terms are given by
\begin{equation}
Z^{\rm inst, (+)}_S(u,Q) = 1+\frac{(\eta^2-1)q(\eta^2\mu_2-\mu_1)}{\eta^2(q-1)(\mu_2q-\mu_1)}u + \frac{(\eta^2-1)q(\eta^2\mu_1p-\mu_2)}{\eta^2(q-1)(\mu_1pq-\mu_2)} Q/u +\cdots
\end{equation}
Remarkably, having identified the parameter as $u=\tau_2/\tau_1$, this instanton partition function reproduces the ramified instanton partition function $Z^{(+)}_{[1^2]}$ in \Secref{Sec:5dRamification}
\begin{equation}
	Z_S^{\rm inst,(+)}(u=\frac{\tau_2}{\tau_1},Q) = \widetilde{\mathcal{N}} \, Z^{(+)}_{[1^2]}
\end{equation}
up to the factor
\begin{equation}
	\widetilde{\mathcal{N}} = {\rm PE}\left[\frac{Q}{1-Q}\frac{(\eta^2-1)(\eta^4p-q)}{(q-1)(\eta^2p-1)}\right] \,,
\end{equation}
which is independent of the mass and FI parameters\footnote{{\rm PE} denotes the Plethystic exponential defined as {\rm PE}$\left[f(x)\right] = {\rm exp}\left(\sum_{n=1}^\infty f(x^n)\right)$}.
We have confirmed this relation up to $u^2(Q/u)^2$ order.
This result provides a strong evidence that the two types of defects, the monodromy defect and the defect of 3d hypermultiplets, are dual to each other under the $S$ transformation: i.e. {\it bispectral dual}.
One can also obtain the second ramified partition function $Z^{(-)}_{[1^2]}$ from the sum over residues at the second set of poles $\tau = q^{-k}\eta^{-1}\mu_2$.

The $S$ transformation exchanges the mass parameter and the corresponding momentum as follows:
\begin{equation}
	\tau \rightarrow p_u^{-1} \,, \quad p_\tau \rightarrow q^{-1}\eta^2 u \,.
\end{equation}
Therefore the normalized partition function defined as
\begin{equation}
	\cD_S^{(+)} =\lim_{\epsilon_2\rightarrow0}\frac{Z^{3d/5d,(+)}_S}{Z}
\end{equation}
obeys the difference equation
\begin{equation}
\left(\frac{\theta_1(\eta^{-2}u,Q)}{\theta_1(u,Q)}p_u^{-1} + \frac{\theta_1(\eta^{-2}u^{-1},Q)}{\theta_1(u^{-1},Q)}p_u \right)\cD_S^{(+)} = \mathcal{N}_{(1)}E_{(1)}\cD_S^{(+)} \,,
\label{eq:ellRS2bodydual}
\end{equation}
which is in perfect agreement with the eigenfunction equation of the 2-body elliptic Ruijsenaars system in \eqref{eq:eRS-Hamiltonian}.

\pagebreak
\section{Gauge Theories with Chiral Matter and 4d Reduction}\label{Sec:Toda}
In this section we discuss chiral limits of our 5d/3d construction. Gauge theories in three dimensions which we have considered so far included both chiral and antichiral matter fields whose masses were split via the twisted mass of the axial $U(1)$ subgroup of the $\CN=4$ R-symmetry. Since the supersymmetry is already broken down to $\CN=2$ by this axial mass, nothing prevents us from examining the possibility of more extreme mass splitting when some of the mass parameters become large. This is the goal of the current section where we shall study chiral limits of three-dimensional theories and their 5d/3d completions along the lines of the previous sections.

\subsection{Complete Flags and Open Toda Chains}
\begin{figure}[!h]
\begin{center}
\includegraphics[scale=0.5]{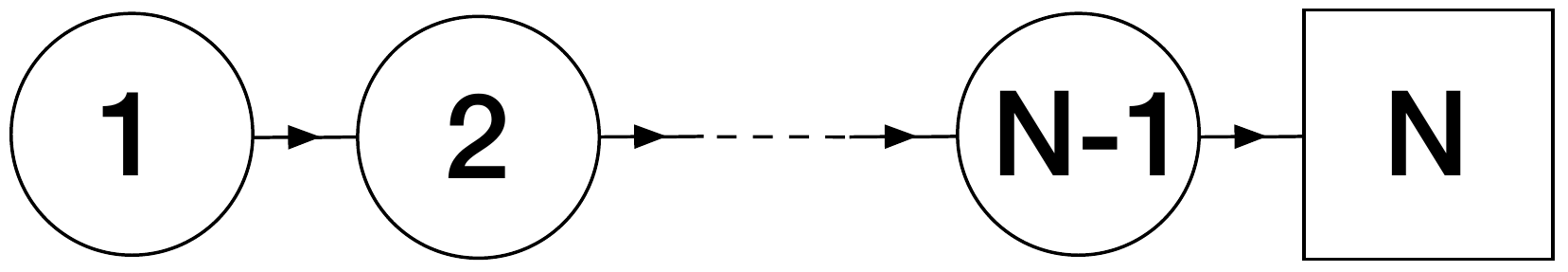}
\caption{A Lagrangian description of the $U(1)\times\dots \times U(N-1)$ theory with fundamental matter and $N$ chiral multiplets at the final node.}
\label{Fig:completeflag}
\end{center}
\end{figure}

The chiral version of the quiver theory from \figref{Fig:NakajimaComplete} is formulated using complete N-dimensional complex flag variety $\mathbb{F}_N$: $0\subset\mathbb{C}\subset\mathbb{C}^2\subset\dots\subset\mathbb{C}^N$. We start with the connection of equivariant quantum K-theory of complete flags and twisted chiral rings of the corresponding 3d gauge theories along the lines of \secref{Sec:EqQuntKThNak}. In physics literature complete N-flags arise as target spaces of supersymmetric sigma models with chiral matter (see \figref{Fig:completeflag}). 

In order to see how the vacua equations for chiral quivers arise form vacua equations of $T[U(N)]$ theories let us rewrite equation \eqref{eq:XXZGen} as follows 
\begin{equation}
\frac{\tau_{j}}{\tau_{j+1}}\prod_{n'=1}^{N_{j-1}}\frac{\eta \varepsilon\sigma^{(j)}_n- \sigma^{(j-1)}_{n'}}{\eta \sigma^{(j-1)}_{n'}-\varepsilon\sigma^{(j)}_n} \cdot \prod_{n'\neq n}^{N_{j}}\frac{\eta^{-1} \sigma^{(j)}_n - \eta \sigma^{(j)}_{n'}}{\eta^{-1}\sigma^{(j)}_{n'}-\eta \sigma^{(j)}_n} \cdot \prod_{n'=1}^{N_{j+1}}\frac{\eta \sigma^{(j)}_n - \varepsilon\sigma^{(j+1)}_{n'}}{\eta\varepsilon\sigma^{(j+1)}_{n'}-\sigma^{(j)}_n}\cdot\prod_{a=1}^{M_j}\frac{\eta \varepsilon\sigma^{(j)}_n-\mu^{(j)}_a}{\eta \mu^{(j)}_a-\varepsilon\sigma^{(j)}_n}=(-1)^{\delta_j}\,, 
\label{eq:XXZGenScaled}
\end{equation}
where we scaled 
\begin{equation}
\sigma^{(j)}_i\to \varepsilon^j\sigma^{(j)}_i\,, \quad \mu^{(j)}_a\to \varepsilon^{j-1}\mu^{(j)}_a\,,
\label{eq:sigmascaling}
\end{equation}
and then take the limit
\begin{equation}
\varepsilon\to 0\,, \quad \eta\to\infty\,,
\label{eq:Etascaling}
\end{equation}
such that $\varepsilon\eta=1$ to get the following
\begin{equation}
\frac{\tau_{j}}{\tau_{j+1}}\prod_{n'=1}^{N_{j-1}}\frac{\sigma^{(j)}_n- \sigma^{(j-1)}_{n'}}{\eta \sigma^{(j-1)}_{n'}} \cdot \prod_{n'\neq n}^{N_{j}}\frac{\sigma^{(j)}_{n'}}{\sigma^{(j)}_n} \cdot \prod_{n'=1}^{N_{j+1}}\frac{\eta \sigma^{(j)}_n}{\sigma^{(j+1)}_{n'}-\sigma^{(j)}_n}\cdot\prod_{a=1}^{M_j}\left(\sigma^{(j)}_n-\mu^{(j)}_a\right)=(-1)^{\delta_j}\,, 
\label{eq:XXZGenLimit}
\end{equation}
For flag manifold we have $N_j=j$, so, after additional scaling 
\begin{equation}
\tau_j\to \epsilon^{2j}\tau_j\,,
\label{eq:Tauscaling}
\end{equation}
we get
\begin{equation}
\frac{\tau_{j}}{\tau_{j+1}}\prod_{n'=1}^{j+1}\frac{\sigma^{(j)}_n- \sigma^{(j-1)}_{n'}}{\sigma^{(j-1)}_{n'}} \cdot \prod_{n'\neq n}^{j} \sigma^{(j)}_{n'}\cdot \prod_{n'=1}^{j-1}\frac{1}{\sigma^{(j+1)}_{n'}-\sigma^{(j)}_n}\cdot\prod_{a=1}^{M_j}\left(\sigma^{(j)}_n-\mu^{(j)}_a\right)=(-1)^{\delta_j}\,.
\label{eq:XXZGenLimit}
\end{equation}
We have just described how to obtain a chiral theory which lives on a complete flag from the theory supported on $T^*\mathbb{F}_N$. In order to derive other chiral theories corresponding to non complete flags one needs start from theories of type $T[U(N)]_\rho$ (see \appref{Sec:AppFormulae}, where we complete the corresponding vortex partition functions) and take a limit similar to \eqref{eq:Etascaling}. 

\subsubsection{Chiral Limit of $T[U(2)]$}
The simplest example of a complete flag variety is $\mathbb{P}^1$. Let us see how to describe its quantum cohomology starting from $T^\ast\mathbb{P}^1$. From \cite{Gaiotto:2013bwa} we get
\begin{equation}
\frac{\tau_1}{\tau_2}\frac{\eta\sigma-\mu_1}{\eta\mu_1-\sigma}\frac{\eta\sigma-\mu_2}{\eta\mu_2-\sigma}=1\,, \quad p_\tau^1=\sigma\,, \quad p_\tau^2 = \frac{\tau_1\tau_2}{\sigma}\,,
\label{eq:TU2Bethe}
\end{equation}
The set of the above three equations can be rewritten as 
\begin{equation}
\frac{\eta\tau_1-\eta^{-1}\tau_2}{\tau_1-\tau_2}p_\tau^1 + \frac{\eta\tau_2-\eta^{-1}\tau_1}{\tau_2-\tau_1}p_\tau^2=\mu_1+\mu_2\,,\quad p_\tau^1p_\tau^2= \tau_1\tau_2\,.
\end{equation}
Performing scaling limits in \eqref{eq:sigmascaling} and \eqref{eq:Tauscaling} we get for \eqref{eq:TU2Bethe}
\begin{equation}
\frac{\tau_1}{\tau_2}\varepsilon^{-2}\frac{\eta\varepsilon\sigma-\mu_1}{\eta\mu_1-\varepsilon\sigma}\frac{\eta\varepsilon\sigma-\mu_2}{\eta\mu_2-\varepsilon\sigma}=1\,.
\end{equation}
or
\begin{equation}
(\sigma-\mu_1)(\sigma-\mu_2) = \mu_1\mu_2\frac{\tau_2}{\tau_1}=e^{2\pi R\Lambda}=\ell\,,
\end{equation}
where $\Lambda$ is the dynamically generated scale. By expanding the quadratic expression in $\sigma$ in the l.h.s. of the above expression and using the definition of the momenta we arrive at 
\begin{equation}
\frac{\mu_1\mu_2-\ell}{\tau_1\tau_2}p_\tau^1+p_\tau^2 = \mu_1+\mu_2\,,
\end{equation}
which can be identified with the trace of two-body open Toda Lax matrix. From \eqref{eq:TU2Bethe} we can derive
\begin{equation}
p^2 (\mu_1\mu_2-\ell)-p(\mu_1+\mu_2)\tau_1\tau_2+\tau_1^2\tau_2^2=0 \,,
\end{equation}
where we put $p_\tau^1=p$. One can recognize in the above equation equivariant quantum K-ring relation for $\mathbb{P}^1$ (cf. \eqref{eq:HypergeomEqClass} for $N=2$).

\subsubsection{Effective Twisted Superpotential}
Equivalently we can derive the above equations from the first principles by considering $\cN=2$ supersymmetric gauge theories in three dimensions built around a triangular quiver. Let us fix and integer $N \geq 2$. Similarly to \eqref{eq:T[U(N)]w} we introduce real mass parameters $s^{(i)}_j$ with $i,j=1,\ldots,N$ such that $s^{(i)}_j=0$ if $i<j$ and $s^{(N)}_j = m_j$. The parameters $\{ s^{(j)}_1,\ldots, s^{(j)}_j \}$ are gauge parameters for the $j$-th node and $\{m_1,\ldots,m_N\}$ are mass parameters for $U(N)$ flavor symmetry. We also introduce FI parameters $\{t_1,\ldots,t_N\}$. 

The effective twisted superpotential for theory in \figref{Fig:completeflag} reads
\bea
\cW & = \sum_{I=2}^N \sum_{i=1}^{I-1} \sum_{j=1}^I \ell\left(s^{(I-1)}_i-s^{(I)}_j \right) \\
& + \sum_{I=1}^{N-1} \sum_{i \neq j}^I \ell\left(s^{(I)}_i-s^{(I)}_j \right) \\
& + \sum_{j=1}^{N} \left(t_{j}-t_{j+1}+\frac{i\delta_j}{2R} \right)\sum_{i=1}^j s^{(j)}_i\\
& + \frac{1}{2} \sum_{I=1}^{N-1} \left[ \left( \sum_{i=1}^I s_{i}^{(I)} \right)^2 - \sum_{i,j=1}^N s^{(I+1)}_is^{(I)}_j  \right]\,.
\label{eq:TwEffSPChiral}
\eea
Each line of the above formula corresponds to chiral multiplets, vectormultiplets, FI terms and Chern-Simons terms respectively, which become manifest in this form of the twisted superpotential\footnote{Recall that each function $\ell(s)$ contains a Chen-Simons term $s^2/4$ which is generated in one loop, and it is the same for chiral and anti-chiral fields. After we take the chiral limit from \eqref{eq:T[U(N)]w} and anti-chiral fields become infinitely massive, quadratic contributions remain in the last line of \eqref{eq:TwEffSPChiral}.}.

The vacuum equations can be readily derived from \eqref{eq:TwEffSPChiral} and read as
\be
\frac{\tau_j}{\tau_{j+1}} Q_{j+1}(\sigma^{(j)}_i)  \prod_{\substack{k=1 \\ k \neq i} }^j  \sigma^{(j)}_k= (-1)^{\delta_j} Q_{j-1}(\sigma^{(j)}_i)  \prod_{k=1}^{j+1} \sigma^{(j+1)}_k\,, \qquad i = 1 \ldots, j\,,
\ee
where $Q_j(u) = \prod\limits_{i=1}^j(u - \sigma^{(j)}_i) $. In what follows we denote $Q_{N}(u) = M(u) = \prod\limits_{i=1}^N(u- \mu_i)$. These equations describe \eqref{eq:XXZGenLimit} for the complete $N$-flag. As in the $T[U(N)]$ case they may appear to be slightly complicated, but it is favorable to express them the in functional form similarly to \eqref{eq:QQTUN}. Using the recursive nature of conjugate momenta $p_{\tau}^1 = \sigma^{(1)}_1$, $p_{\tau}^2 = \sigma^{(2)}_1\sigma^{(2)}_2 / \sigma^{(1)}_1 $, and so on it is easy to see that they are recovered from the following functional equations
\be
Q_{j+1}(u) -(-1)^{\delta_j} \frac{\tau_{j+1}}{\tau_j} Q_{j-1}(u) \, \sigma \, p_\tau^{j+1}  = Q_j(u) \, \widetilde{Q}_j (u)\,,
\ee
where $\widetilde{Q}_j(u)$ is monic monomial of degree one. As before, these equations serve as a powerful toll for deriving the Lax matrix of Toda systems from the vacuum equations. 

Again, let us consider $N=2$ for simplicity. There is a single equation
\be
M(u) -(-1)^{\delta_1} \frac{\tau_2}{\tau_1} \, \sigma \, p_{\tau}^2 = Q_1(u) \, \widetilde{Q}(u) \, .
\ee
By evaluating the equation at $u=0$ it is straightforward to see that $Q_1(u) = u - p_{\tau}^1$ and $\widetilde{Q}_1(u) = u- p_{\tau}^2$ and therefore
\bea
M(u) & = (u-p_{\tau}^1)(u-p_{\tau}^2) + (-1)^{\delta_1} u \frac{\tau_2}{\tau_1} p_{\tau}^2 \\
& = u^2 - u \left( p_{\tau}^1 + p_{\tau}^2 - (-1)^{\delta_1} \frac{\tau_2}{\tau_1} p_{\tau}^2  \right)   + p_{\tau}^1p_{\tau}^2 \,.
\eea
Thus we have recovered the Q-Toda Hamiltonians.

Now we can graduate on to $N=3$ case. We have two equations
\bea
Q_2(u) -(-1)^{\delta_1} \frac{\tau_2}{\tau_1} \, u \, p_{\tau}^2 & = Q_1(\sigma) \, \widetilde{Q}_1(u) \\
M(u) - (-1)^{\delta_2} Q_1(u) \frac{\tau_3}{\tau_2} \, u \, p_\tau^3 & = Q_2(u) \, \widetilde Q_2(u)
 \, .
\eea
We can recycle the information from the $N=2$ case so that $Q_1(u) = u - p_{\tau}^1$ and $\widetilde{Q}_1(u) = u- p_{\tau}^2$. Furthermore, by evaluating the second equation at $u=0$ we find that $\widetilde Q_2(u) = u - p_\tau^3$. Recall that $Q_2(u)$ is the matter polynomial of the $N=2$ example. Thus we have
\bea
M(u) & = (-1)^{\delta_2} u (u-p_{\tau}^1) \frac{\tau_3}{\tau_2} \, \, p_\tau^3  + (u-p_\tau^3) \left( (u-p_{\tau}^1)(u-p_{\tau}^2) + (-1)^{\delta_1} u \frac{\tau_2}{\tau_1} p_{\tau}^2
\right) \\
& = u^3 - u^2 \left[p_{\tau}^1+p_{\tau}^2+p_\tau^3 - (-1)^{\delta_1} \frac{\tau_2}{\tau_1} p_{\tau}^2 - (-1)^{\delta_2} \frac{\tau_3}{\tau_2} p_\tau^3 \right] \\
& + u \left[ p_{\tau}^1p_{\tau}^2 +p_{\tau}^2p_\tau^3 +p_\tau^3p_{\tau}^1 - (-1)^{\delta_1} \frac{\tau_2}{\tau_1} p_{\tau}^2p_\tau^3 - (-1)^{\delta_2} \frac{\tau_3}{\tau_2} p_\tau^3 p_{\tau}^1 \right] - p_{\tau}^1p_{\tau}^2p_{\tau}^3\,,
\eea
which are Hamiltonians of the three-body relativistic Toda system. 

\subsubsection{Open Toda Lax Matrix}
Now we systematize the above computation in the general case $N>2$ to find the Lax matrix of the open relativistic Toda chain. We should find the matrix $L(u)$ with non-zero components
\be
L_{i+1,i} = 1\,, \qquad L_{i,i} = u - p_{i}\,, \qquad L_{i,i+1} = -(-1)^{\delta_{i} } u  \frac{\tau_{i+1}}{\tau_i} p_{\tau_{i+1}}\,,
\ee
and then $M(u) = \det(L(u))$ analogously to \eqref{eq:tRSW}. Thus we have constructed the limit when trigonometric Ruijsenaars-Schneider model is reduced to relativistic Toda lattice using gauge theories.

\subsection{Pure Super Yang Mills Theories with Defects and Closed Toda Chains}
Thus far we have identified parameter space of supersymmetric vacua of quiver theories with chiral matter with global symmetry of rank $N$ and the phase space of the (complexified) open Toda chain with $N$ particles. In other words, we have described the `chiral' version of the XXZ/tRS duality from \secref{Sec:TwistChiralRings}. Now, along the lines of \secref{Sec:5dRamification}, we shall investigate how the above construction is modified when the same theory is coupled as a surface defect in pure $U(N)$ $\cN=1$ SYM in five dimensions. We expect a deformation of the twisted chiral ring depending on the dynamical scale of the five dimensional theory. 
In what follows, we use a dimensionless quantity $\Lambda$ obtained by multiplying by the dynamical scale by the radius of the circle.

We therefore claim that the twisted chiral ring of quiver theory in \figref{Fig:completeflag} is described by the functional equations
\be
Q_{j+1}(u) -(-1)^{\delta_j} \frac{\tau_{j+1}}{\tau_j} Q_{j-1}(u) \, u \, p_{\tau_{j+1}}  = Q_j(u) \, \widetilde{Q}_j (u)
\ee
for $j=1,\ldots,N-2$ and with $Q_0(u) =1$, whereas at the final node there is a modification of the equation to
\bea
M(u) -(-1)^{\delta_{N-1}} \frac{\tau_{N}}{\tau_{N-1}} Q_{N-2}(u) \, u \, p_{\tau_N}  - (-1)^{\delta_{N}} \frac{\tau_1}{\tau_N} Q_{N-1}(u) u \, p_{\tau}^1 \\
= Q_{N-1}(u) \, \widetilde{Q}_{N-1} (u) \, .
\eea
For example, for $N=3$ we would find
\bea
M(u) & = u^3 - u^2 \left[p_{\tau}^1+p_{\tau}^2+p_\tau^3 - (-1)^{\delta_1} \frac{\tau_2}{\tau_1} p_{\tau}^2 - (-1)^{\delta_2} \frac{\tau_3}{\tau_2} p_\tau^3 - (-1)^{\delta_3} \frac{\tau_1}{\tau_3} p_{\tau}^1 \right] \\
& + u \left[ p_{\tau}^1p_{\tau}^2 +p_{\tau}^2p_\tau^3 +p_\tau^3p_{\tau}^1 - (-1)^{\delta_1} \frac{\tau_2}{\tau_1} p_{\tau}^2p_\tau^3 - (-1)^{\delta_2} \frac{\tau_3}{\tau_2} p_\tau^3 p_{\tau}^1 - (-1)^{\delta_3} \frac{\tau_1}{\tau_3} p_{\tau}^1 p_{\tau}^2 \right] \\
& - p_{\tau}^1p_{\tau}^2p_\tau^3\,,
\eea
which are nothing but closed Toda spectral relations. Thus we claim that coupling the theory to five dimensions transforms open relativistic Toda chain to \textit{closed} relativistic Toda chain -- a natural generalization of the similar statement found in \cite{Gaiotto:2013sma} for nonrelativistic Toda systems.

We now discuss the computation of ramified instanton partition functions of $\CN=1$ SYM theories in five dimensions. For simplicity we start with gauge group $U(2)$. In order to compute the corresponding instanton partition function we use character for $\CN=1$ theory \eqref{eq:N1Charachter}. Then we introduce the monodromy defect according to the prescription of \secref{Sec:RamInstantons}. For $U(2)$ theory the instanton partition function will have the form \eqref{eq:ZramrhoGen} labelled by the partition $\rho=[1,1]$. We make the replacement
\begin{alignat}{3}
(+) & \quad Q_1 = \frac{1}{\sqrt{\mu_1\mu_2}} \, z\,, \qquad && Q_2 = \frac{1}{\sqrt{\mu_1\mu_2}} \, \frac{Q}{z}\,, \\
(-) & \quad Q_1 = \frac{1}{\sqrt{\mu_1\mu_2}}\, \frac{Q}{z}\,, \qquad &&  Q_2 = \frac{1}{\sqrt{\mu_1\mu_2}} \, z\,.
\label{eq:Q12zchiral}
\end{alignat}
Again, later we shall replace $z=\frac{\tau_2}{\tau_1}$. The normalization factors of $\sqrt{\mu_1\mu_2}$ are included to remove fractional powers of $\mu_1$ and $\mu_2$ in the formulae for the Nekrasov partition function in the presence of the defect, which we shall address later in this section. The same normalization will be used for the difference equations. Note that they would cancel if we move to the center of mass frame i.e. consider gauge group $SU(2)$ instead of $U(2)$.

\subsubsection{Open Toda from Decoupling Limit}
Analogously to \secref{Sec:tRS} we discuss the decoupling limit first, so we send $Q\to0$ and decouple the five-dimensional degrees of freedom\footnote{One can also study `chiral' version of \secref{Sec:3dpf}, namely study 3d partition functions of chiral quivers and prove that they satisfy Q-Toda difference equations. This approach (albeit without reference to gauge theories in three dimensions) was pursued in \cite{Kharchev:2001rs}.}. In this limit we expect to find contributions that can be accounted for by degrees of freedom supported on the defect. The answer should be consistent with chiral limits \eqref{eq:Etascaling, eq:Tauscaling} of holomorphic blocks and vortex partition functions which we have computed earlier in the paper. Indeed, this turns out to be the case.

For example, in order to find the vortex partition function for $\mathbb{P}^1$ sigma model we can start with Givental J-function \eqref{eq:TstP1JFun} for $T^*\mathbb{P}^1$ and perform the chiral limit in order to see that
\begin{equation}
\CZ_V \to {}_2F_1\left(0,0;q\frac{\mu_1}{\mu_2};q;z\right)=\sum_{k=0}^\infty\frac{\left(qz\right)^k}{\prod_{l=1}^k\left(\mu_2-\mu_1 q^l\right)(1-q^l)}\,,
\label{eq:GiventalJfuncLimit}
\end{equation}
which leads us to the formula for equivariant K-theoretic J-function for $\mathbb{P}^1$ from \cite{2001math8105G}. At the last step we rescaled $z\to z/\mu_2$. The above result can be understood as hypergeometric series of type ${}_0 F_1$.

Equivalently, from the computation of the ramified instanton partition functions in the $Q\to 0$ limit using notations \eqref{eq:Q12zchiral} we get
\bea
\CZ^{(+)}_{[1,1]} & = \sum_{n=1}^{\infty} \frac{q^{n(n+1)/2}}{\prod_{j=1}^n (1-q^j)(\mu_1-q^j\mu_2)}z^n\,, \\
\CZ^{(-)}_{[1,1]} & = \sum_{n=1}^{\infty} \frac{q^{n(n+1)/2}}{\prod_{j=1}^n (1-q^j)(\mu_2-q^j\mu_1)} z^n\,,
\label{eq:Z11chiralOpen}
\eea
which are related by interchanging $\mu_1\leftrightarrow \mu_2$. The above solutions represent the non-perturbative contributions to the two independent holomorphic blocks of $U(1)$ theory with two chiral multiplets.

It is straightforward to prove that expansions \eqref{eq:Z11chiralOpen} obey the following difference equations
\bea
\left(\mu_1 \, p_{\tau}^1 + \mu_2 \, p_{\tau}^2 - \frac{\tau_2}{\tau_1}\right) \CZ^{(+)}_{[1,1]} & = \left(\mu_1+\mu_2\right) \CZ^{(+)}_{[1,1]}\,,  \qquad p_{\tau}^1 \, p_{\tau}^2 \CZ^{(+)}_{[1,1]}= \CZ^{(+)}_{[1,1]}\,, \\
\left(\mu_2 \, p_{\tau}^1 + \mu_1 \, p_{\tau}^2 - \frac{\tau_2}{\tau_1}\right) \CZ^{(-)}_{[1,1]} & = \left(\mu_1+\mu_2\right) \CZ^{(+)}_{[1,1]}\,, \qquad p_{\tau}^1 \, p_{\tau}^2\CZ^{(-)}_{[1,1]} = \CZ^{(-)}_{[1,1]} \, .
\eea
The second equations in each line impose that the answer depends only on the ratio $\tau_2/\tau_1$. In order to remove the dependence on $\mu_1$ and $\mu_2$ on the left we must include contributions to the classical action from FI parameters. These are exactly the same ambiguities we have addressed in \eqref{eq:ZeignRedefineSec3}. We could choose, for example
\bea
\cZ^{(+)} & = \frac{\theta(\tau_1,q)\theta(\mu_1,q) }{\theta(\tau_1\mu_1,q)}  \frac{\theta(\tau_2,q)\theta(\mu_2,q) }{\theta(\tau_2\mu_2,q)}\CZ^{(+)}_{[1,1]} \\
\cZ^{(-)} & = \frac{\theta(\tau_1,q)\theta(\mu_2,q) }{\theta(\tau_1\mu_2,q)}  \frac{\theta(\tau_2,q)\theta(\mu_1,q) }{\theta(\tau_2\mu_1,q)} \CZ^{(-)}_{[1,1]}\,,
\label{eq:Z11ChiralScaling}
\eea
where, we recall, $\theta(a,q)$ obeys $p_a\theta(a,q) = -a^{-1}\theta(a,q)$. Now both of the functions $\cZ^{(\pm)}$ obey the same difference equations
\be
\left(p_{\tau}^1 + p_{\tau}^2 - \frac{\tau_2}{\tau_1}\right)\cZ^{(\pm)}  = (\mu_1+\mu_1)\cZ^{(\pm)}\,, \qquad p_{\tau}^1\,p_{\tau}^2\cZ^{(\pm)}=\mu_1\,\mu_2\,\cZ^{(\pm)}\,,
\label{eq:OpenTodaEigenEq}
\ee
which are the Hamiltonians of the two-body open Q-Toda integrable system. The holomorphic blocks in this example were discussed in detail in \cite{Beem:2012mb}. 

In \cite{Braverman:2011hf} the K-theoretic J-function of the flag variety was shown to be the (universal) eigenfunction of the relativistic Toda system for any simply laced group and in \cite{2014arXiv1410.2365B} the analysis was extended to non-simply laced groups.

\subsubsection{Closed Toda Chain}
Turning on the parameter $Q$, the partition function the presence of the defect diverges in the limit $\epsilon_2 \to0$ since we have degrees of freedom propagating in this plane. However, the divergence exponentiates and is universal i.e. it is independent of the presence of the defect. Thus the expectation value of the defect is finite in the limit
\be
\CR^{(\pm)} = \lim_{\ep_2\to0} \frac{\CZ_{\rho=[1^2]}^{(\pm)}}{\CZ} \, .
\ee
The first few terms in the expansion are
\bea
\CR^{(+)} & = 1 + \frac{q }{(1-q) \left(\mu _1 -\mu _2 q \right)}z+ \frac{q }{(1-q) \left(\mu _2-\mu _1q\right)}\frac{Q}{z} + \cdots \\
\CR^{(-)} & = 1+ \frac{q}{(q-1)\left(\mu _2-\mu _1 q\right)}z + \frac{q}{(1-q)\left(\mu _1 -\mu _2q\right)}\frac{Q}{z} + \cdots
\label{eq:ClosedQTodaEigenExpan}
\eea
We can see that, unlike the elliptic RS eigenfunctions, these two functions are no longer related by $\mu_1 \leftrightarrow \mu_2$ at higher order in the $Q$ expansion. This fact is not surprising since one needs to scale $Q$ as we obtain the above expressions via taking the limit from the elliptic RS eigenfunctions. Yet, the expressions are symmetric if one interchanges $z=\tau_1/\tau_2$ with $Q\tau_2/\tau_1$. Starting from \eqref{eq:eRSeigenfunctionsExp} and taking the limit described after \eqref{eq:XXZGenScaled} we arrive to the above expressions provided that $Q$ is also rescaled as follows
\begin{equation}
\frac{\tau_1}{\tau_2} \to \epsilon^2\frac{\tau_1}{\tau_2}\,,\quad Q\to Q\epsilon^4\,,\quad \eta\epsilon =1\,,
\end{equation}
as $\epsilon\to 0$. In addition to the above scaling one can redefine the FI parameters as 
\begin{equation}
\frac{\tau_1}{\tau_2} \to \frac{1}{\mu_2}\frac{\tau_1}{\tau_2}\,,\quad Q\to \frac{Q}{\mu_1\mu_2}\,,
\end{equation}
then the elliptic RS eigenfunctions take the form of \eqref{eq:ClosedQTodaEigenExpan}.

We have checked to order $\cO(Q_1^{n_1} Q_2^{n_2})$ with $n_1+n_2=5$ that \eqref{eq:ClosedQTodaEigenExpan} obey the difference equations
\bea
\left(\mu_1 \, p_{\tau}^1 + \mu_2 \, p_{\tau}^2 - \frac{\tau_2}{\tau_1} - \frac{\tau_1}{\tau_2}Q\right)\CR^{(+)} & = (\mu_1+\mu_2)\CR^{(+)}\,,  \qquad p_{\tau}^1 \, p_{\tau}^2\, \CR^{(+)} = \CR^{(+)}\,, \\
\left(\mu_2 \, p_{\tau}^1 + \mu_1 \, p_{\tau}^2 - \frac{\tau_2}{\tau_1} - \frac{\tau_1}{\tau_2}Q \right)\CR^{(-)}& = (\mu_1+\mu_2)\CR^{(-)}\,, \qquad p_{\tau}^1 \, p_{\tau}^2\, \CR^{(-)} = \CR^{(-)} \, .
\eea
We now multiply $\CR^{(\pm)}$ by the same factors as in \eqref{eq:Z11ChiralScaling}. From a three-dimensional perspective they were contributions to classical action from FI parameters. The final result is that both functions obey
\be
 \left(p_{\tau}^1 +  p_{\tau}^2 - \frac{\tau_2}{\tau_1} - \frac{\tau_1}{\tau_2}Q \right) \CR^{(\pm)}= (\mu_1+\mu_2)\CR^{(\pm)}\,,  \qquad p_{\tau}^1 \, p_{\tau}^2\, \CR^{(\pm)} = \mu_1 \, \mu_2\,\CR^{(\pm)} \, .
\label{eq:ClosedTodaEigenEq}
\ee
Thus we can see how turning on the parameter $Q$ in the difference equation controls the five-dimensional instanton corrections. We have found formal eigenfunctions of the two-body closed Q-Toda Hamiltonians. Note that the spectrum should be discrete, whereas we have continuous parameters $\mu_1$ and $\mu_2$. Again, as in the non chiral case, they should be fixed by Bethe equations coming from the divergent prefactor. It could be that this is related to normalizability of the wavefunctions. 

\subsection{Connections to Kapustin-Willett Results} 
Kapustin and Willett \cite{Kapustin:2013hpk} computed quantum K-ring of a family of line bundles of Grassmanninas $T(M,N,k)$, which naturally appear in the study of twisted chiral rings of 3d $\CN=2$ $U(N)$ theories with $M$ fundamental twisted chirals and Chern-Simons level $k$. In other words, the authors computed the quantum K-ring for $M$ copies of the tautological bundle  $\mathcal{O}(-1)^{M}\to \text{Gr}(M-k,N)$ over the Grassmannian\footnote{We changed $k$ to $-k$ for convenience}. 

Let us write Bethe equations for 3d $N=2^*$ $U(N)$ theory with $M$ flavors
\begin{equation}
\frac{\tau_2}{\tau_1}\prod_{a=1}^M\frac{\eta\sigma_i-\mu_a}{\eta\mu_a-\sigma_i}\prod_{j\neq i}^N \frac{\eta^{-1}\sigma_i-\eta \sigma_j}{\eta^{-1}\sigma_j-\eta \sigma_i}=1\,.
\end{equation}
As we know these are quantum ring relations for $T^*(\text{Gr}(M,N))$. For simplicity let us consider $N=1$ and $M=3$ first. The variables in the above equation then can be rescaled as follows
\begin{equation}
\eta\to \epsilon^{-1}\eta\,,\quad \sigma\to\epsilon\sigma\,,\quad\, \mu_3\to \epsilon^2\mu_1\,\quad \mu_2\to\epsilon\mu_2\,,
\end{equation}
to get
\begin{equation}
q\sigma^2\frac{\eta-\mu_1\sigma^{-1}}{\eta\mu_1\sigma^{-1}-1}=1\,,\quad q = \frac{\tau_2}{\tau_1\mu_1\mu_2}\,,
\end{equation}
which is the chiral ring relation for $T(3,2,1)$.

\subsection{4d/2d Construction}\label{Sec:2d4dToda}
For completeness we exhibit 2d analogues of some properties of 3d theories we have used in the main text.
Quantum cohomology of complete flag varieties and their cotangent bundles in connection with monodromy defects were discussed in \cite{Nawata:2014nca}. Here we discuss chiral limit of the twisted chiral ring of $(2,2)^*$ theory in two dimensions.

Supersymmetric vacua equations can be obtained from \eqref{eq:XXZGen} and read as follows
\begin{equation}
\frac{\tau_{j}}{\tau_{j+1}}\prod_{n'=1}^{N_{j-1}}\frac{s^{(j)}_n- s^{(j-1)}_{n'}+\frac{\epsilon}{2}}{s^{(j-1)}_{n'}- s^{(j)}_{n}+\frac{\epsilon}{2}} \cdot \prod_{n'\neq n}^{N_{j}}\frac{s^{(j)}_n- s^{(j)}_{n'}-\epsilon}{s^{(j)}_{n'}- s^{(j)}_{n}-\epsilon} \cdot \prod_{n'=1}^{N_{j+1}}\frac{s^{(j)}_n - s^{(j+1)}_{n'}+\frac{\epsilon}{2}}{s^{(j+1)}_{n'} - s^{(j)}_{n}+\frac{\epsilon}{2}}\cdot\prod_{a=1}^{M_j}\frac{s^{(j)}_n-m^{(j)}_a+\frac{\epsilon}{2}}{m^{(j)}_a-s^{(j)}_n+\frac{\epsilon}{2}}=(-1)^{\delta_j}\,, 
\label{eq:XXXGen}
\end{equation}
Now we make some shifts $s^{(j)}_n\to s^{(j)}_n-j\frac{\epsilon}{2}$
\begin{equation}
\frac{\tau_{j}}{\tau_{j+1}}\prod_{n'=1}^{N_{j-1}}\frac{s^{(j)}_n- s^{(j-1)}_{n'}}{s^{(j-1)}_{n'}- s^{(j)}_{n}+\epsilon} \cdot \prod_{n'\neq n}^{N_{j}}\frac{s^{(j)}_n- s^{(j)}_{n'}-\epsilon}{s^{(j)}_{n'}- s^{(j)}_{n}-\epsilon} \cdot \prod_{n'=1}^{N_{j+1}}\frac{s^{(j)}_n - s^{(j+1)}_{n'}+\epsilon}{s^{(j+1)}_{n'} - s^{(j)}_{n}}\cdot\prod_{a=1}^{M_j}\frac{s^{(j)}_n-m^{(j)}_a+(1-j)\frac{\epsilon}{2}}{m^{(j)}_a-s^{(j)}_n+(1+j)\frac{\epsilon}{2}}=(-1)^{\delta_j}\,, 
\label{eq:XXXGenShifted}
\end{equation}
We now take the limit $\epsilon\to \infty$ combined with $m^{(j)}_a\to \infty$ such that $-m^{(j)}_a+(1-j)\frac{\epsilon}{2}=-M^{(j)}_a$ is kept fixed. In order to keep the whole expression finite the FI couplings have to run thereby generating a scale $\Lambda_j$ for each gauge group, so we have the following chiral ring relations
\begin{equation}
\prod_{n'=1}^{N_{j-1}}(s^{(j)}_n- s^{(j-1)}_{n'})\cdot\prod_{a=1}^{M_j}(s^{(j)}_n-M^{(j)}_a)=\Lambda_j^{M_j+N_{j-1}-N_{j+1}}\prod_{n'=1}^{N_{j+1}}(s^{(j)}_{n}-s^{(j+1)}_{n'})\,, 
\label{eq:XXXGenLimit}
\end{equation}
which is the quiver generalization of a $\mathbb{CP}^{N}$ sigma model. We can also write the above equation in Baxter form
\begin{equation}
Q_{i-1}M_i=\Lambda_i^{D_i}Q_{i+1}\,,
\label{eq:vacuaeqchiralgen}
\end{equation}
where
\begin{equation}
D_i=M_j+N_{j-1}-N_{j+1}\,.
\end{equation}
For a complete flag case only $M_1=Q_0$ is nonzero, the rest $M_i$ vanish and vacua equations \eqref{eq:vacuaeqchiralgen} can be written as
\begin{equation}
Q_{i-1}-\Lambda_i^2 Q_{i+1}= Q_i \widetilde{Q}_i\,,
\end{equation}
using auxiliary polynomial $\widetilde{Q}_i$. It is assumed that $Q_{N}=1$. Following \cite{Gaiotto:2013sma} we rewrite the above equation using Toda Lax matrix $L$
\begin{equation}
L_{i,j} Q_j = \delta_{i,1}M\,,
\end{equation}
where 
\begin{equation}
L_{i,j}=\widetilde{Q}_i \delta_{ij} + \Lambda_i^2 \delta_{i,j+1} - \delta_{i,j-1}\,,
\label{eq:LaxTodaLambda}
\end{equation}
and it also follows that 
\begin{equation}
M = \text{det} L\,.
\end{equation}
Interestingly the momenta in Toda are the auxiliary polynomials $\widetilde{Q}_i$ and the coordinates are dynamical scales $\Lambda_i$ at each gauge group.

\subsubsection{Equivariant Quantum Cohomology}
One of the motivating results for this note is the famous theorem by Givental and Kim, which we list below for completeness. 

\begin{theorem}[Givental, Kim 1993]
The equivariant quantum cohomology ring of the complete manifold of flags inside $\mathbb{C}^N$ is isomorphic  to
\begin{equation}
QH^\bullet_T(\mathbb{F}_N)\simeq \mathbb{C}[p_1,\dots,p_N,\tau_1,\dots \tau_{N-1},\mu_1,\dots\,\mu_N]/\mathcal{I}_{\text{Toda}}\,
\end{equation}
where ideal $\mathcal{I}_{\text{Toda}}$ is generated by the coefficients of the following polynomial 
\begin{equation}
\det\left(u-L_{\text{Toda}}(p,\tau)\right)=\prod_{j=1}^N(u-\mu_i)\,,
\label{eq:GenRelToda}
\end{equation}
where matrix $L_{\text{Toda}}$ (Lax matrix of open Toda chain)
\begin{equation}
L_{\text{Toda}}=
\begin{pmatrix}
p_1 & \tau_1 & 0          & \dots & 0 \\
-1   &  p_2     & \tau_2  & \dots & 0 \\
0    & -1&  p_3  & \dots & 0 \\
\cdots\\
 0& 0 & 0 & -1 & p_N
\end{pmatrix}\,.
\end{equation}
\label{Th:GiventalKim}
\end{theorem}

In the Lax matrix above we have identified momenta $p_i$ and (exponential of) relative coordinates between Toda particles $\tau_i=e^{t_i-t_{i+1}}$ with dynamical scales $\Lambda_i$ and polynomials $\widetilde Q_i$ respectively in \eqref{eq:LaxTodaLambda}.
This theorem can now be viewed as an important limiting case of our construction. In order to obtain the above result from Proposition \ref{Th:NakajimaComplete} we need to first, shrink the compactification radius $R\to 0$, and, second, take the chiral limit in mass parameters \eqref{eq:Etascaling}. 

A detailed analysis of twisted chiral rings of $(2,2)$ and $(2,2)^*$ quiver theories in two dimensions in the context of ramified instanton counting and equivariant quantum cohomology was done in \cite{Nawata:2014nca}. In particular, two-dimensional vortex partition functions for the theories on two-dimensional defects inside four-dimensional $\CN=2$ and $\CN=2^*$ gauge theories were computed, and the identification with Givental J-functions was presented. One can also obtain these expressions by taking the 2d limit of 3d holomorphic blocks \eqref{eq:S3-function-TU(N)}. 

Note that currently in mathematical literature all necessary tools for computing quantum cohomology of vector bundles over complex projective spaces and their generalizations is already available. Thus using the results of \cite{2001math.....10142C} one can derive quantum J-function for the cotangent bundle to the flag variety in the framework of \cite{2007InMat.171..301C} (see \cite{Nawata:2014nca} for details).

Further connections to integrability in $\CN=(2,2)^*$ theories were studied in \cite{Benini:2014mia}, in particular, $Q\widetilde Q$-type relations, similar to \eqref{eq:QQrelations} were examined using cluster algebra methods.

\subsubsection{Coupled 4d/2d Systems}
Using the results of the previous section we can compute twisted chiral rings of two dimensional quiver theories coupled to four dimensional $\CN=2^*$ gauge theory by taking the radius of the compact circle in the 5d computation to zero. Classical analysis has been conducted in \cite{Gaiotto:2013sma}, in particular, the 4d analogue of the deformed twisted chiral ring relation \eqref{eq:TwistChirRingDef5d} was constructed in the cited paper for the $SU(N)$ $\CN=2^*$ theory.

\pagebreak
\section{Summary and Outlook}\label{Sec:IntApplications}

\subsection{Integrable Systems}
In this section we outline how the results of this paper fit into the wider context of integrable many-body systems. 

There exists a correspondence between the spectral curves of a family of many-body integrable systems (Calogero-Moser-Sutherland \cite{Donagi:1995cf,Itoyama:1995nv,Itoyama:1995uj}, Ruijsenaars-Schneider \cite{Braden:1999zpa}, and Double-elliptic \cite{Braden:1999aj,Gorsky:2000px, Braden:2001yc, Braden:2003gv}) and Seiberg-Witten curves of supersymmetric gauge theories with adjoint matter in four, five and six dimensions (see \cite{Gorsky:2000px} and references therein). These are the examples that we have focussed on in this paper.

Recently, in \cite{Nekrasov:2012xe} this correspondence was extended and systematized to a large class four dimensional quiver gauge theories of finite and affine ADE type. Later in \cite{Nekrasov:2013xda} the five and six-dimensional versions of the Seiberg-Witten geometry together with its quantum deformation are also discussed.


\subsubsection{Calogero--Ruijsenaars--Dell Family}
\begin{figure}[h]
\begin{center}
\includegraphics[scale=0.55]{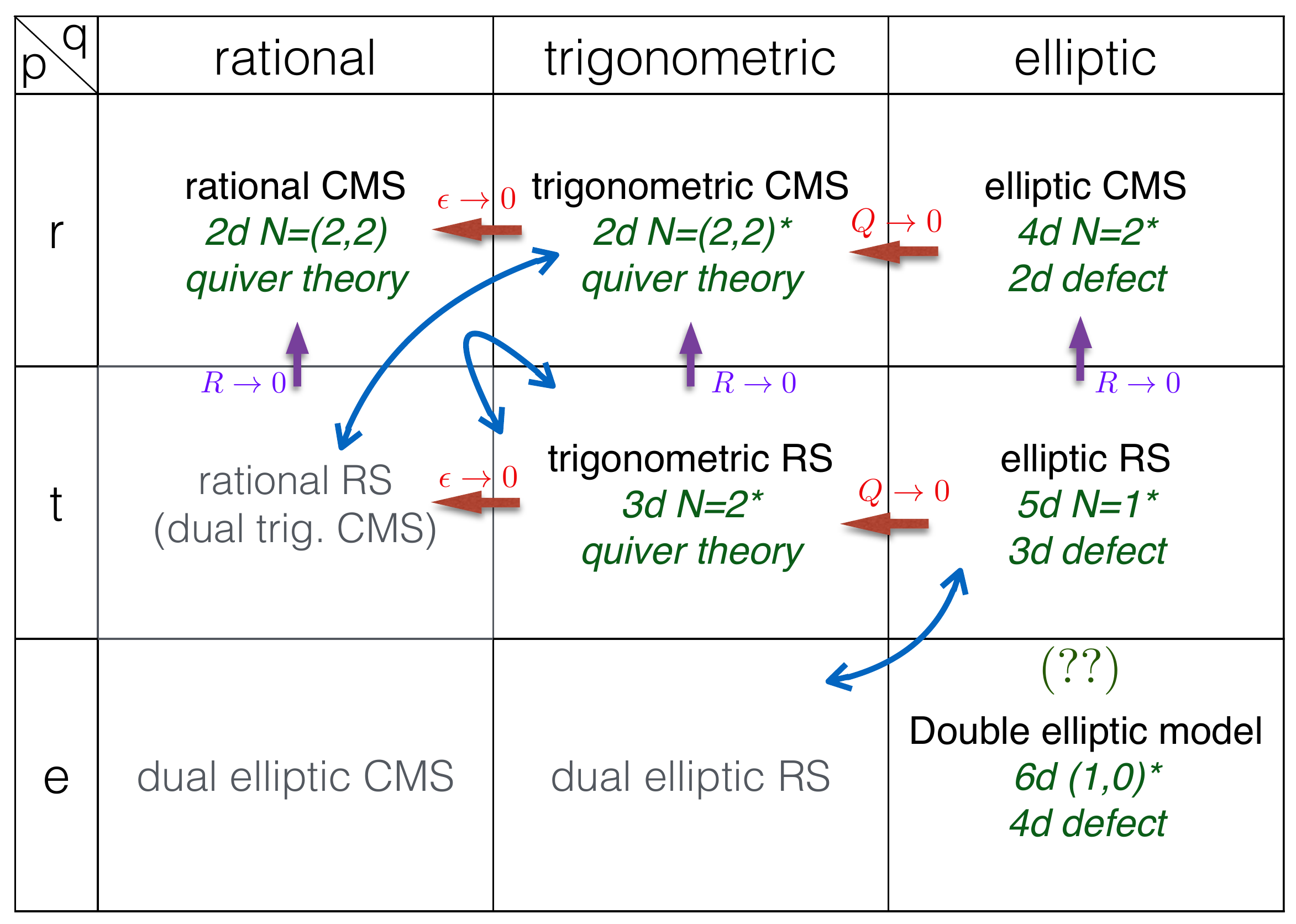}
\caption{Classification of integrable many-body systems according to their periodicity properties in coordinates $q$ (columns) and momenta $p$ (rows).}
\label{Fig:CalRSDell}
\end{center}
\end{figure}

\figref{Fig:CalRSDell} exhibits a family on many-body classical integrable systems and our proposal for how their eigenfunctions are realized by the partition functions of supersymmetric gauge theories in various dimensions. The integrable models are characterized by how the periodicity properties of the coordinates and momenta: each may be rational, trigonometric and elliptic. The most general system studied in this paper is the elliptic RS system (corresponding to 5d $\cN=2$ gauge theory with a 3d surface defect), which has elliptic dependence on positions and trigonometric dependence on momenta. We have indicated how some of the remaining members of the family can be obtained by various limits.

In \figref{Fig:CalRSDell} the first row contains Calogero-Moser-Sutherland (CMS) family, the second row describes Ruijsenaars-Schneider (RS) family, whereas the last row contains models that are bispectrally dual to eCMS, eRS, and, finally, the double elliptic (Dell) system. Blue double arrows describe bispectral dualities between the models in the table. Because the properties of the Dell system are largely unexplored, the anticipated duality between Dell and itself is designated by the dotted blue double arrow. Various limits are shown by arrows. Thus the right column describes elliptic models, whose ellipticity parameter plays a role of the bulk (4,5, or 6-dimensional) instanton fugacity $Q$. When we send $Q\to 0$ bulk degrees of freedom decouple and we are left with the corresponding defect theories which are located in the middle column. These models are trigonometric in coordinates. Next, one may take a limit when the adjoint mass $\epsilon$ vanishes. Provided that an appropriate scaling is chosen (see \cite{Gaiotto:2013bwa}), one obtains a family of rational models from the $\epsilon\to 0$ limit. Equivalently, one can take the limits row-wise. First, starting from the Dell model and its descendants in the last row take a limit $\widetilde R \to 0$ thereby shrinking one of the extra compact direction of the 6d theory. This limit moves us to the RS family. Further on, shrinkage of the other circle (again, with a properly chosen scaling of parameters) $R\to 0$ descends us further to the CMS family. One can use the arrows we described above to navigate through the diagram starting from Dell.

Recall that the classical tRS model was derived from the twisted chiral ring of $T[U(N)]$ theory \eqref{eq:tRSRelationsEl}. After quantization these relations become operator equations which we solved in \eqref{eq:tRSEigenMirr}, where the eigenfunctions are written in \eqref{eq:TUNPartFuncCoul} as a partition function on the Coulomb branch of $T[U(N)]$ theory, then in terms of vortex partition functions in \eqref{eq:S3-function-TU(N)}, and finally as a decoupling limit of the 5d/3d partition function in \secref{Sec:tRS}. The eigenvalues are characters of antisymmetric tensor representation of $U(N)$.

Next, we have found the eigenfunctions of elliptic RS system for two and three particles as partition functions of 5d $\CN=1^*$   $U(2)$ \eqref{eq:eRS-Hamiltonian} and $U(3)$ \eqref{eq:eRSSp} gauge theories in the presence of monodromy defects of maximal type \secref{eq:eRS}. The eigenvalues in this case can be computed in \eqref{eq:WilsonCharact} and correspond to VEVs of Wilson loops in the skew powers of the fundamental representation of $U(N)$.

Finally, we were able to find the spectrum of the two-body dual eRS model using the $U(2)$ 5d gauge theory coupled to 3d free hypermultiplets \eqref{eq:DualeRSEigenfunction}. By applying the S-transformation \eqref{eq:STransform} we have shown that the eigenfunctions of the elliptic RS model \eqref{eq:ellRS2bodydual} can be reproduced from the eigenfunctions of the dual model.

\subsubsection{Toda Family}
\begin{figure}[!h]
\begin{center}
\includegraphics[scale=0.46]{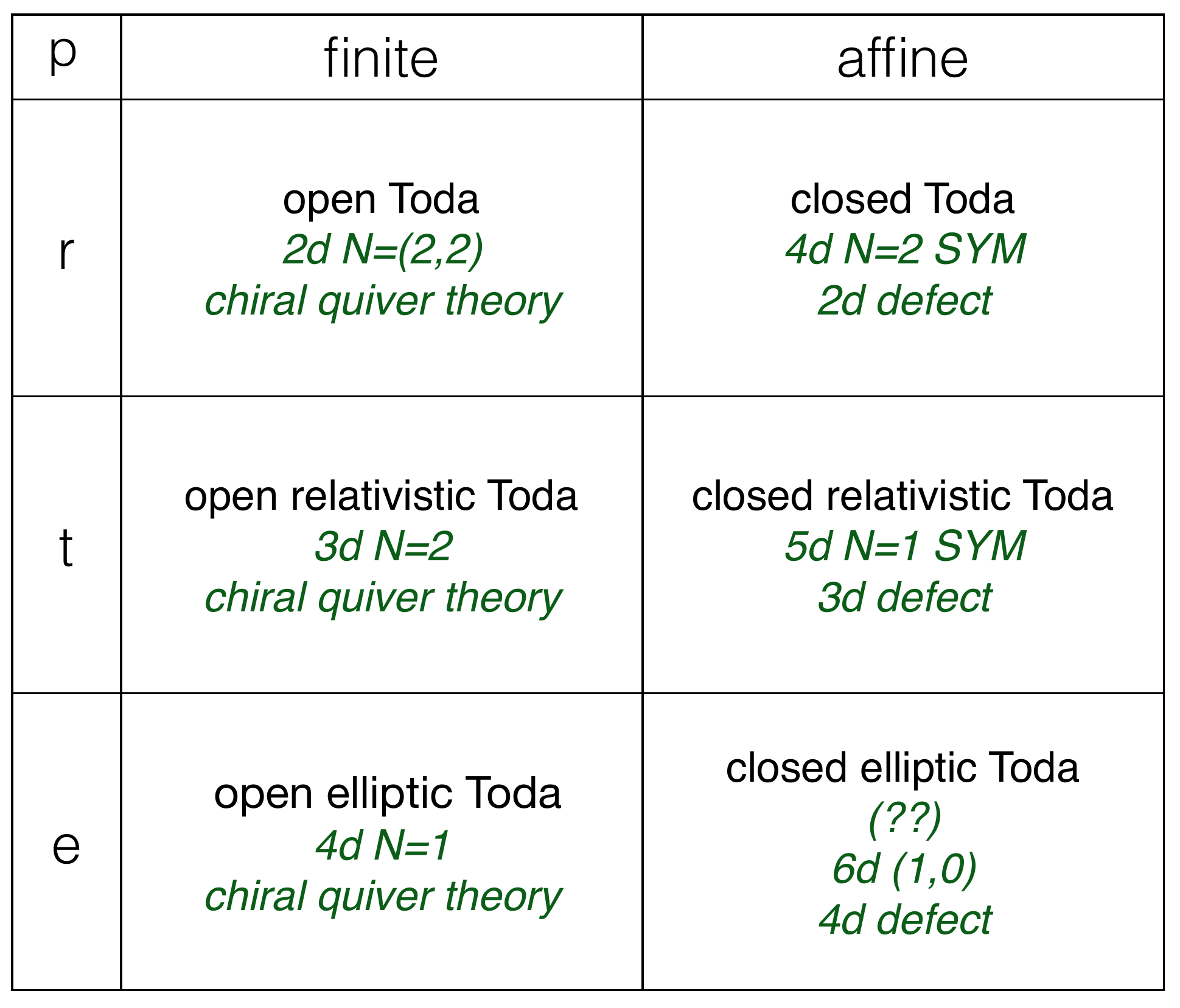}
\caption{Classification of rational, trigonometric (relativistic) and elliptic Toda integrable systems. Closed Toda chains exhibit affine Lie algebra symmetry, whereas open chain are symmetric under the action of the Lie algebra itself.}
\label{Fig:todafamily}
\end{center}
\end{figure}

In \secref{Sec:Toda} we have discussed 3d quiver gauge theories with chiral matter and pure 5d SYM theories with defects which support chiral quiver theories. We have found quantum spectra of relativistic open \eqref{eq:OpenTodaEigenEq} and closed \eqref{eq:ClosedTodaEigenEq} Toda chains. In the limit when the compactification radius of the 3d theory becomes small we recover nonrelativistic Toda chains (see \secref{Sec:2d4dToda}). The classification is presented in \figref{Fig:todafamily}.
In the last row of the table one can find elliptic Toda systems which were studied in \cite{Krichever:1999fj}.

\subsubsection{Gaudin--Spin Chain Family}
\begin{figure}[!h]
\begin{center}
\includegraphics[scale=0.5]{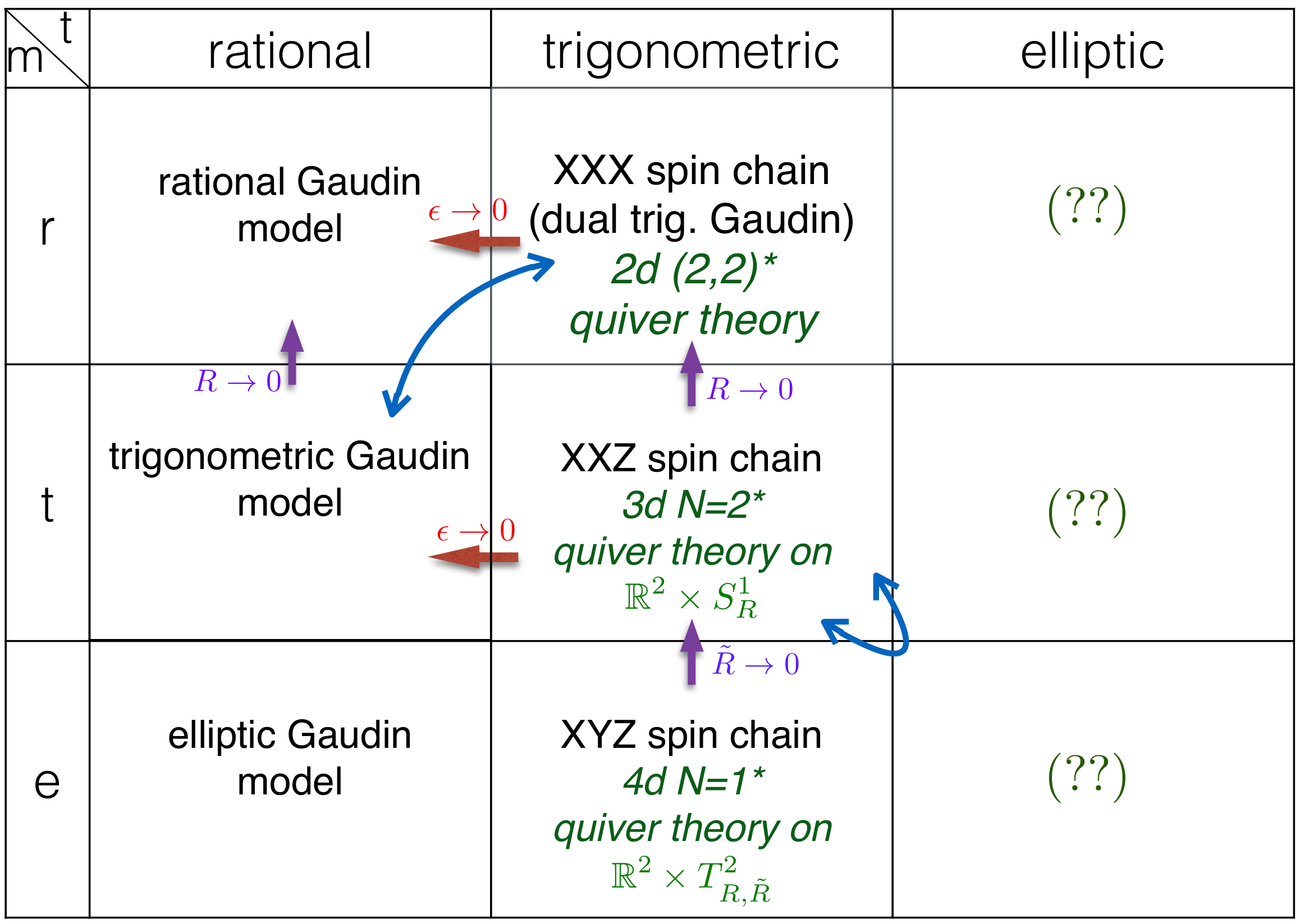}
\caption{Classification of integrable models of Gaudin and spin chain type according to the periodicity properties in twists $t$ (columns) and anisotropies $m$ (rows). Blue double arrows describe bispectral dualities between the models in the table.}
\label{Fig:GaudXXX}
\end{center}
\end{figure}

For completeness we also discuss the quantum integrable models whose Bethe equations reproduce the twisted chiral rings of the supersymmetric gauge theories appearing (after decoupling $Q \to 0$) on codimension two surface defects in the four, five and six-dimensional theories we have discussed above. This is illustrated in \figref{Fig:GaudXXX} and can be compared to \figref{Fig:CalRSDell}. 

The spin chain family (XXX, XXZ, and XYZ) in the middle column of \figref{Fig:GaudXXX} can be described using the Nekrasov-Shatashvili correspondence using quiver gauge theories in dimensions two, three, and four respectively. The Gaudin family in the first column can be obtained from the spin chain family by turning off the `Planck constant', which in our construction is represented by the adjoint mass $\epsilon$ (see \cite{Gaiotto:2013bwa} for the details about XXX and XXZ models). We refer the reader to \cite{Sklyanin:1996pr} and references therein for the discussion about XYZ spin chain and XYZ (elliptic) Gaudin model. The limit $\epsilon\to 0$ should be supplemented by scaling of anisotropy parameters such that $\epsilon m_i$ remain constant.

To the best of our knowledge there are no known quantum models with elliptic dependence on twists which could be consistently placed in the right column of \figref{Fig:GaudXXX}. Moreover, our analysis involving defects in higher dimensional gauge theories with adjoint matter fields does not immediately suggest any candidates for such models. Their existence, is certainly an intriguing question on its own and should be pursued independently\footnote{Some work understanding these integrable models is being done in \cite{Litvinov:2013zda,Alfimov:2014qua,NOtoapp}.}.

\subsection{Open Problems}
Let us now articulate what we believe to be some important questions unanswered by the present work. We list them below in the random order.

\begin{itemize}
\item In this paper we have conjectured the structure of formal eigenfunctions of the elliptic Ruijsenaars-Schneider model and (in the 2-body case) its bispectral dual. We have checked our conjectures by expanding the eigenfunctions (instanton partition functions of 5d theories with codimension two defects) to first several orders. It should be possible to \textit{prove} these conjectures. We understand that work in this direction is being pursued in \cite{NP2014}.

\item The Hamiltonians of trigonometric Ruijsenaars-Schneider model form a commutative subalgebra inside double affine Hecke algebras (DAHA) and Cherednik algebras \cite{Cherednik:1991}. Further understanding of the role of 
Cherednik algebras in gauge theories is due.

\item We have analyzed in this paper five-dimensional theories with defects. An immediate generalization would be to study six dimensional $(1,1)$ theory with a four dimensional defect and use it to find formal eigenfunctions of the double-elliptic (Dell) model. It would be interesting to understand the action of bispectral duality in this case.


\item In the presence of monodromy defects, it is known that the Alday-Gaiotto-Tachikawa \cite{Alday:2009aq} correspondence is modified to conformal blocks of affine algebras. Understanding of five-dimensional gauge theories with surface defects might lead to new ideas in quantum affine algebras (see, e.g. \cite{Nieri:2013yra,Nieri:2013vba,Aganagic:2014oia, Aganagic:2013tta} for a related study of q-Toda degenerate conformal blocks). 

\item There is a notion of regular and irregular monodromy defects in gauge theories. In this paper we have only addressed the regular case and computed instanton partition functions in the presence of such monodromy defects. One may wonder what happens in the case of irregular singularities. 

\item The Calogero-Moser-Sutherland family of integrable models is connected to the Korteweg -- de Vries/Benjamin-Ono/Intermediate Long Wave hierarchy \cite{LebRadu2,Dorey:2007zx,Abanov:2008ft,Litvinov:2013zda,Bonelli:2014iza,Alfimov:2014qua} in the limit of large number of particles. One may study generalizations of the correspondence to the Ruijsenaars-Schneider family. Some work in this direction has already been done in \cite{2009arXiv0911.5005T,2009JPhA...42N4018S,2011ntqi.conf..357S}.

\item In this paper we discussed the relationship between the many-body trigonometric Ruijsenaars-Schneider system and Bethe ansatz of quantum XXZ spin chains. It is a legitimate question to ask how the correspondence is generalized to the related elliptic many-body system. 

\end{itemize}

\section*{Acknowledgements}
We would like to thank Davide Gaiotto, Jaume Gomis, Nikita Nekrasov, Vasily Pestun, Mina Aganagic, Shamil Shakirov, Nathan Haouzi, Satoshi Nawata, Amihay Hanany, Andrey Zayakin, Mikhail Bershtein, Aleksey Morozov, Antonio Sciarappa, Alessandro Tanzini, Giulio Bonelli, Heeyeon Kim for fruitful discussions.

MB gratefully acknowledges support from the Perimeter Institute for Theoretical Physics and IAS Princeton through the Martin A. and Helen Choolijan Membership. PK thanks W. Fine Institute for Theoretical Physics at University of Minnesota, Kavli Institute for Theoretical Physics at University of California Santa Barbara, University of California at Berkeley, and Simons Center for Geometry and Physics where part of his work was done, for kind hospitality. HC gratefully acknowledges support from the Perimeter Institute for Theoretical Physics, the organizers of ``Exact Results in SUSY Gauge Theories in Various Dimensions'' at CERN and also CERN-Korea Theory Collaboration funded by National Research Foundation (Korea) for the hospitality and support. This work was presented at a number of conferences and seminars including ``IV SISSA Workshop on Dualities and Geometric Correspondences'' and KITP program ``New Methods in Nonperturbative Quantum Field Theory''. We are grateful to the organizes for invitations and kind hospitality.

Our research was partly supported by the Perimeter Institute for Theoretical Physics. Research at Perimeter Institute is supported by the Government of Canada through Industry Canada and by the Province of Ontario through the Ministry of Economic Development and Innovation.

\pagebreak
\appendix
\section{Conventions of Special Functions}
We summarize our conventions on various functions we have used in the main text.

\subsection{Q-Hypergeometric Functions}
We use the following definition for Q-hypergeometric functions
\begin{equation}
\, _2F _1(a,b;c;q,z)=\sum_{k=1}^\infty \frac{(a;q)_k (b;q)_k}{ (c;q)_k}\frac{z^k}{(q;q)_k}\,.
\end{equation}

\subsection{Double Sine Functions}\label{appendix:doublesine}
Here we use the double sine function $S_2(z | \vec\omega)$ where $\vec\omega=(\omega_1,\omega_2)$ with $\omega_1,\omega_2>0$ defined for example in the appendix of. Let us summarize the properties written there. This function has an integral representation
\be
\log S_2(z| \vec\omega) = \frac{i\pi}{2}B_{2,2}(z| \vec\omega) + \int\limits_{\mathbb{R}+i0} \frac{e^{zt}}{(e^{\omega_1t}-1)(e^{\omega_1 t}-1)} \frac{dt}{t}
\ee
where $B_{2,2}(z|\vec\omega)$ is the multiple Bernoulli polynomial. The most important property for this paper is the functional relations
\be
\frac{S_2(z+\omega_1|\vec\omega)}{S_2(z|\vec\omega)}\,.
\ee

\subsection{Theta Functions}\label{sec:ThetaFunctions}
We have used basic theta-function in the computations in the main text
\begin{equation}
\theta(a,q) = (a,q)_\infty(qa^{-1},q)_\infty=\prod\limits_{m\geq 0}(1-aq^m)\left(1-\frac{q^{m+1}}{a}\right)\,,
\end{equation}
where, we recall, the q-Pochhammer symbol is defined as follows
\be
	(x;q)_n \equiv \left\{ \begin{array}{cl}\prod_{i=0}^{n-1} (1-xq^i)& {\rm for} \ n>0\,, \\ 1 & {\rm for} \ n=0 \,, \\ \prod_{i=1}^{-n}(1-xq^{-i})^{-1} & {\rm for } \ n <0\,. \end{array}\right.
\ee
It is related to Kac-Weyl character for $\hat{A}_1$ as
\begin{equation}
\theta_{KW}(a,q)=\theta(a,q)\cdot\prod\limits_{m\geq 0}(1-q^{m+1})\,,
\end{equation}
which in turn is related to conventional elliptic theta functions associated to the elliptic curve constructed on complex vector $q$.  For instance
\begin{equation}
\theta_1(a,q)=-iq^{-\frac{1}{8}}a^{\frac{1}{2}}\theta_{KW}(a,q)\,.
\end{equation}

\section{Factorization of $T_\rho$ Partition Functions}\label{Sec:AppFormulae}
\begin{figure}[!h]
\begin{center}
\includegraphics[scale=0.5]{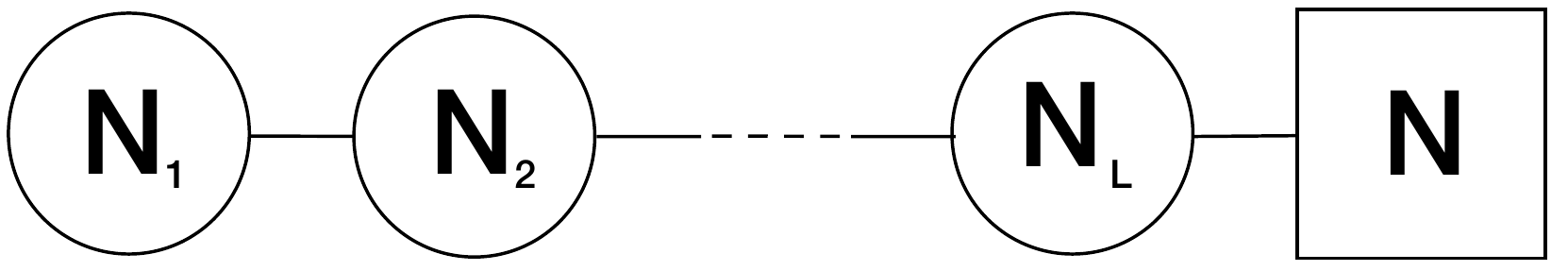}
\caption{Quiver diagram for $T[U(N)]_\rho$ theory where $\rho^T_i = N_{i+1}-N_i$.}
\label{Fig:quiverTrho}
\end{center}
\end{figure}

The $T[U(N)]_\rho$ theory is the 3d $\mathcal{N}=4$ quiver gauge theory with gauge group $U(N_1)\times\cdots\times U(N_L)$ and $N$ fundamental hypermultiplets for the last gauge group (see \figref{Fig:quiverTrho}).

In this appendix we shall compute the Higgs branch representation of the ellipsoid partition function for $T[U(N)]_\rho$ theory.
The partition function is given by
\begin{align}
\CZ_{U(N)}^\rho &= \int \prod_{n=1}^{L} \! \frac{dx^{(n)}}{N_n!} e^{2\pi  \sum_{n=1}^L \sum_{i=1}^{N_n}x_i^{(n)} (t_{n}-t_{n+1})}\notag\\
&\times\prod_{i<j}^{N_n}2\sinh(\pi b^\pm x_{ij}^{(n)})\prod_{i,j=1}^{N_n}\!S_b\!\left(\varepsilon+ix_{ij}^{(n)}\right) \!\prod_{i=1}^{N_{n}} \prod_{j=1}^{N_{n+1}}\! S_b\!\left(\frac{\varepsilon^*}{2} \pm i(x_i^{(n)} - x_j^{(n+1)})\right)\,,
\end{align}
with $N_{L+1}= N$ and $x_i^{(L+1)}= m_i$, the $N$ mass parameters for the flavor symmetry.

The integrand has infinite number of poles and zeros. We suppose that the physically relevant poles are only from the bifundamental hypermultiplets\footnote{This is proven for $T[U(N)]$ theory in~\secref{Sec:VortAVort}.}.
Then the contour integral can be done by taking residues from the relevant poles. The poles take the form of
\begin{equation}
	x_i^{(n)} = m_j + i\frac{(L-n+1)\varepsilon^*}{2} + ik_i^{(n)}b + i\widetilde{k}_i^{(n)}b^{-1}
\end{equation}
with $j=1,\cdots,N$ and non-negative integers $k_i^{(n)}, \widetilde{k}_i^{(n)} \ge 0$. It is convenient to define new integral variables $s_i^{(n)}$ such as
\begin{equation}
	x_i^{(n)} = s_i^{(n)}+ m_j + i\frac{(L-n+1)\varepsilon^*}{2} + ik_i^{(n)}b + i\widetilde{k}_i^{(n)}b^{-1}
\end{equation}
and perform the integral over $s_i^{(n)}$ instead of $x_i^{(n)}$.
This effectively shifts all poles to $s_i^{(n)}=0$.

With new variables the integral can be rewritten as
\begin{equation}
	\mathcal{Z}_{U(N)}^\rho(m_i,t_n) = \frac{1}{\prod_{i=1}^L\rho_i!}\sum_{\sigma \in S_{N}} \int \prod_{n=1}^{L} \prod_{i=1}^{N_n} \frac{ds^{(n)}_i}{2\pi i} \ \CZ(s,m_{\sigma(j)},t)
\end{equation}
where 
\begin{align}
	&\CZ(s,m,t) = e^{2\pi \sum_{n=1}^L\sum_{i=1}^{N_n} (t_n-t_{n+1})(s_i^{(n)}+m_i+i\frac{(L-n+1)\varepsilon^*}{2}+ik_i^{(n)}b+i\widetilde{k}_i^{(n)}b^{-1})} \\
	&\times
	\sum_{\vec{k},\vec{\widetilde{k}} \ge0}\prod_{n=1}^L\prod_{i<j}^{N_n}2\sinh \left(\pi b^\pm (s_{ij}^{(n)}\!+\!m_{ij}\!+\!ik_{ij}^{(n)}b\!+\!i\widetilde{k}_{ij}^{(n)}b^{-1})\right) \prod_{i,j=1}^{N_n}\!S\left(\varepsilon\!-\!is_{ij}^{(n)}\!-\!im_{ij}\!+\!k_{ij}^{(n)}b\!+\!\widetilde{k}_{ij}^{(n)}b^{-1}\right)  \notag\\
	& \times \prod_{i=1}^{N_{n}} \prod_{j=1}^{N_{n+1}}
	\frac{S \left(-is_i^{(n)}+is_j^{(n+1)}-im_{ij}+(k_i^{(n)}-k_j^{(n+1)})b+(\widetilde{k}_i^{(n)} - \widetilde{k}_j^{(n+1)})b^{-1}+\varepsilon^*\right)}
	{ S \left(-is_i^{(n)}+is_j^{(n+1)}-im_{ij}+(k_i^{(n)}-k_j^{(n+1)}+1)b+(\widetilde{k}_i^{(n)} - \widetilde{k}_j^{(n+1)}+1)b^{-1}\right) }  \notag
\end{align}
with $s_i^{(L+1)}=0$ and $k^{(L+1)}_i = \widetilde{k}^{(L+1)}_i=0$.
$S_{N}$ is the Weyl group of the $U(N)$ flavor symmetry and we denote collectively by $\vec{k}$ and $\vec{\widetilde{k}}$ all the (semi-)positive integer numbers $k^{(n)}_i$ and $\widetilde{k}_i^{(n)}$ respectively.
The integrand can be further simplified using the following identity
\begin{equation}\label{eq:double-sine-identity}
	S_b(x+nb+mb^{-1}) = \frac{ i^{n+m+2mn} e^{-\pi ix(nb+mb^{-1})} }{ q^{n(n-1)/4} \ \widetilde{q}^{m(m-1)/4} } (e^{2\pi ibx};q)_n(e^{2\pi ib^{-1}x};\widetilde{q})_m S_b(x)
\end{equation}
with $q\equiv e^{2\pi i b^2}, \widetilde{q} \equiv e^{2\pi i/ b^{2}}$. The  function $\CZ$ reduces to
\begin{align}
&\CZ(s,m,t) =  e^{2\pi \sum_{n=1}^L\sum_{i=1}^{N_n} (t_n-t_{n+1})(s_i^{(n)}+m_i+i\frac{(L-n+1)\varepsilon^*}{2})} \sum_{\vec{k},\vec{\widetilde{k}} \ge0} \prod_{n=1}^L\left(\frac{\tau_n}{\tau_{n+1}}\right)^{|k^{(n)}|} \, \left(\frac{\widetilde\tau_n}{\widetilde\tau_{n+1}}\right)^{\,|\widetilde{k}^{(n)}|}\nonumber  \\
&\times\! 
\prod_{i<j}^{N_n}2\sinh\pi b^\pm (s_{ij}^{(n)}+m_{ij})\prod_{i,j=1}^{N_n}S(\varepsilon+is_{ij}^{(n)}+im_{ij})
\cdot
\frac{ \Big(q\eta^{-2}\frac{\sigma_i^{(n)}}{\sigma_j^{(n)}}\frac{\mu_i}{\mu_j};q\Big)_{\!k^{(n)}_{ij}}\! }
{ \Big(\frac{\sigma_i^{(n)}}{\sigma_j^{(n)}}\frac{\mu_i}{\mu_j};q\Big)_{\!k^{(n)}_{ij}}\! }
\cdot
\frac{ \Big(\tilde{q}\widetilde\eta^{-2}\frac{\widetilde\sigma_i^{(n)}}{\widetilde\sigma_j^{(n)}}\frac{\widetilde\mu_i}{\widetilde\mu_j};\widetilde q\Big)_{\widetilde k^{(n)}_{ij}}\! }
{ \Big(\frac{\widetilde\sigma_i^{(n)}}{\widetilde\sigma_j^{(n)}}\frac{\widetilde\mu_i}{\widetilde\mu_j};\widetilde q\Big)_{\widetilde k^{(n)}_{ij}}\! }\notag \\
&\times \!\prod_{i=1}^{N_{n}} \prod_{j=1}^{N_{n+1}}\!
	\frac{ S(\varepsilon^*\!-\!is_i^{(n)}\!+\!is_j^{(n\!+\!1)}\!-\!im_{ij}) }{ S(2b_+\!-\!is_i^{(n)}\!+\!is_j^{(n\!+\!1)}\!-\!im_{ij}) }\cdot
	\frac{ \Big(\eta^{2}\frac{\sigma_i^{(n)}}{\sigma_i^{(n+1)}}\frac{\mu_i}{\mu_j} ;q \Big)_{\!k_i^{(n)}\!-\!k_j^{(n\!+\!1)}\!} }{ \Big(q\frac{\sigma_i^{(n)}}{\sigma_i^{(n+1)}}\frac{\mu_i}{\mu_j} ;q \Big)_{k_i^{(n)}-k_j^{(n\!+\!1)}} }
\cdot\frac{ \Big(\widetilde\eta^{2}\frac{\widetilde\sigma_i^{(n)}}{\widetilde\sigma_i^{(n+1)}}\frac{\widetilde\mu_i}{\widetilde\mu_j} ;\widetilde q \Big)_{\!\widetilde{k}_i^{(n)}\!-\!\widetilde{k}_j^{(n\!+\!1)}\!} }{ \Big(\widetilde q\frac{\widetilde\sigma_i^{(n)}}{\widetilde\sigma_i^{(n+1)}}\frac{\widetilde\mu_i}{\widetilde\mu_j} ;\widetilde q \Big)_{\widetilde{k}_i^{(n)}-\widetilde{k}_j^{(n\!+\!1)}} }\notag \\
&\times \! (q\eta^{-2})^{\frac{1}{2}\sum_{n=1}^L(N_{n+1}-N_{n-1})|k^{(n)}|}
	(\widetilde{q}\widetilde{\eta}^{-2})^{\frac{1}{2}\sum_{n=1}^L(N_{n+1}-N_{n-1})|\widetilde{k}^{(n)}|}\,, 
\end{align}
where
\begin{equation}
 \sigma_i^{(n)} = e^{2\pi b s_i^{(n)}}\,, \quad |k^{(n)}| = \sum_{i=1}^{N_n}k_i^{(n)} \,, \quad \widetilde\sigma_i^{(n)} = e^{2\pi b^{-1} s_i^{(n)}}\,, \quad |\widetilde{k}^{(n)}| = \sum_{i=1}^{N_n}\widetilde{k}_i^{(n)}\,.
\end{equation}
All relevant poles are simple poles located at the origin $s_i^{(n)}=0$, so the residue computation is straightforward. Finally, the residue sum provides the Higgs branch representation of the partition function of the $T[U(N)]_\rho$ theory
\begin{align}
\CZ_{U(N)}^\rho &= \frac{1}{\prod_{i=1}^L\rho_i!}\sum_{\sigma \in S_N} Z_{\rm class}(m_{\sigma(i)},t)Z_{\rm 1-loop}(m_{\sigma(i)})Z_{V}(m_{\sigma(i)})Z_{AV}(m_{\sigma(i)})\,, \notag \\
	Z_{\rm class} &=  e^{2\pi \sum_{n=1}^L\sum_{i=1}^{N_n} (t_n-t_{n+1})(m_i+i\frac{(L-n+1)\varepsilon^*}{2})}\,, \notag \\
	Z_{\rm 1-loop}&=  \prod_{n=1}^L\prod_{i<j}^{N_n}2\sinh\pi b^\pm (m_{ij})\prod_{i,j=1}^{N_n}S(im_{ij}+\varepsilon) \prod_{i=1}^{N_n} \frac{\prod_{j\neq i}^{N_{n+1}}S_b(im_{ij})}{\prod_{j=1}^{N_{n+1}}S_b(im_{ij}+\varepsilon)}\,,  \notag \\
	Z_{V} &= \sum_{\vec{k}\ge0}  \prod_{n=1}^L  \left(\frac{\tau_n}{\tau_{n+1}}\right)^{|k^{(n)}|} \left(q\eta^{-2}\right)^{\frac{1}{2}\sum_{n\!=\!1}^L(N_{n\!+\!1}\!-\!N_{n\!-\!1})|k^{(n)}|} \notag \\
	& \qquad \times \prod_{n=1}^L\prod_{i\neq j}^{N_n}\! \frac{ \Big(q\eta^{-2}\frac{\mu_i}{\mu_j};q\Big)_{\!k^{(n)}_{ij}}\! }{ \Big(\frac{\mu_i}{\mu_j};q\Big)_{\!k^{(n)}_{ij}}\!}
	\prod_{i=1}^{N_n} \prod_{j=1}^{N_{n\!+\!1}} \!\!
	\frac{ \Big(\eta^{2}\frac{\mu_i}{\mu_j} ;q \Big)_{\!k_i^{(n)}\!-\!k_j^{(n\!+\!1)}\!} }{ \Big(q\frac{\mu_i}{\mu_j} ;q \Big)_{k_i^{(n)}-k_j^{(n\!+\!1)}}}\,,  \notag \\
	Z_{AV} &= \CZ_V((z,\mu,\eta,q)\rightarrow (\widetilde{z},\widetilde\mu,\widetilde\eta,\widetilde{q}))\,.
\label{eq:PartFuncExpressionsHiggs}
\end{align}
The result consists of the classical and 1-loop determinant and non-perturbative parts and the non-perturbative contribution factorizes into the vortex and anti-vortex series.

\section{Perturbative 5d Partition Functions}\label{app:perturbative}
In this appendix, we briefly discuss perturbative partition functions and regularization issue. We will compute the 1-loop determinants of 5d theory on $S^1\times \mathbb{C}^2$ and 3d theory on $S^1\times \mathbb{C}$ where $S^1$ denotes the time circle. We will also consider the elliptic genus of 2d theory which can arise as a boundary theory of the 3d theory.

Let us start with the 1-loop determinants of 5d theories.
We can use the equivariant indices for the vector and hypermultiplets
\begin{align}
\chi_{\rm vec} &= -\frac{1+e^{-2\epsilon_+}}{2(1-e^{-\epsilon_1})(1-e^{-\epsilon_2})} \sum_\alpha e^{\alpha(a)} \,, \nonumber \\
\chi_{\rm hyper} &= \frac{e^{m-2\epsilon_+}}{(1-e^{-\epsilon_1})(1-e^{-\epsilon_2})}\sum_\rho e^{\rho(a)} \,.
\end{align}
where $\alpha$ are the roots and $\rho$ are the weights of the gauge group and $m$ is the equivariant parameter for the flavor symmetry.
The denominator factors is understood as a power series expansion in terms of $e^{\epsilon_1}$ and $e^{\epsilon_2}$.
When we compute the 1-loop determinants using these equivariant indices, there is an issue related to the boundary condition on $\partial\mathbb{C}^2$, which has not been clarified yet. We will not attempt to clarify this issue in this paper, but we will adopt the following prescription. For the multiplets in the complex representations in the gauge group, we will simply use the above equivariant indices and compute the corresponding 1-loop determinants. However, for the real multiplets, we will first take average of the equivariant index with its charge conjugation and then compute the 1-loop determinant. This prescription respects the invariance of the real multiplets under the charge conjugation.
We also remind the reader that the equivariant indices implicitly have the factor $\sum_{t\in \mathbb{Z}}e^{\frac{2\pi i}{\beta}t}$ for the Kaluza-Klein momenta along the temporal circle. In what follows we set $\beta=1$ for convenience, which can be restored by scaling other chemical potentials. 

We find the following 1-loop determinants
\begin{align}
	Z_{\rm 1-loop}^{\rm 5d,vec} &= \prod_{\alpha} \prod_{t\in\mathbb{Z}} \prod_{p,q\ge 0}^\infty \!
	\left[\left(\frac{2\pi i}{\beta}t+p\epsilon_1+q\epsilon_2 + \alpha(a)\right) \!\!\left(\frac{2\pi i}{\beta}t+(p+1)\epsilon_1+(q+1)\epsilon_2 + \alpha(a)\right)\right]^{1/2} \,, \nonumber \\
	Z_{\rm 1-loop}^{\rm 5d,hyper} &= \prod_{\rho} \prod_{t\in\mathbb{Z}} \prod_{p,q\ge 0}^\infty 
	\left(\frac{2\pi i}{\beta}t+p\epsilon_1+q\epsilon_2 + \alpha(a)+m \right)^{-1} \,.
\end{align}
where we assumed that $\rho$ is in a complex representation.

These partition functions are divergent infinite products which need to be properly regularized.
We will regularize them using Barnes' multiple gamma functions.
Barnes' gamma functions are defined as regularized infinite products \cite{Ruijsenaars2000107,Friedman2004362}
\begin{equation}
	\Gamma_N(z|w_1,\cdots,w_N) = \mathrm{exp}\left( \partial_s\zeta_N(s,z;w_1,\cdots,w_N)|_{s=0} \right) \sim \prod_{n_1,\cdots,n_N\ge0} (z+n\cdot w)^{-1} \,,
\end{equation}
where Barnes' zeta function is defined by the series
\begin{equation}
	\zeta_N(s,z|w_1,\cdots,w_N) = \sum_{n_1,\cdots,n_M \ge 0}(z+n\cdot w)^{-s} \,.
\end{equation}
When ${\rm Im}(w)>0$, the following identity holds
\begin{equation}
\Gamma_{N+1}(z|1,w)\Gamma_{N+1}(1-z|1,-w) = e^{-\pi i\zeta_{N+1}(0,z|1,w)}
\prod_{n_1,\cdots,n_N\ge0}(1-e^{2\pi i(z+n\cdot w)})^{-1} \,.
\end{equation}
The 1-loop determinants regularized using Barnes' gamma functions can be written as
\begin{align}
Z_{\rm 1-loop}^{\rm 5d,vec} &= \prod_\alpha \left[ \Gamma_3\Big( \frac{\alpha(a)}{2\pi i}\big|1,\frac{\epsilon_1}{2\pi i},\frac{\epsilon_2}{2\pi i} \Big)\Gamma_3\Big( 1-\frac{\alpha(a)}{2\pi i}\big|1,-\frac{\epsilon_1}{2\pi i},-\frac{\epsilon_2}{2\pi i} \Big) \right. \nonumber \\
&  \hspace{2cm} \times \left.\Gamma_3\Big( \frac{\alpha(a)\!+\!2\epsilon_+}{2\pi i}\big|1,\frac{\epsilon_1}{2\pi i},\frac{\epsilon_2}{2\pi i} \Big) \Gamma_3\Big( 1\!-\!\frac{\alpha(a)\!+\!2\epsilon_+}{2\pi i}\big|1,-\frac{\epsilon_1}{2\pi i},-\frac{\epsilon_2}{2\pi i} \Big) \right]^{-1/2} \,, \nonumber \\
Z_{\rm 1-loop}^{\rm 5d,hyper} &= \prod_\rho \Gamma_3\Big( \frac{\rho(a)\!+\!m}{2\pi i}\big|1,\frac{\epsilon_1}{2\pi i},\frac{\epsilon_2}{2\pi i} \Big)\Gamma_3\Big( 1\!-\!\frac{\rho(a)\!+\!m}{2\pi i}\big|1,-\frac{\epsilon_1}{2\pi i},-\frac{\epsilon_2}{2\pi i} \Big) \,.
\end{align}
Assuming $q=e^{\epsilon_1},p=e^{\epsilon_2}<1$, they can be further simplified to
\begin{align}
Z_{\rm 1-loop}^{\rm 5d,vec} &= e^{-\mathcal{F}_{\rm vec}^{5d}(a;\epsilon_1,\epsilon_2)}\prod_\alpha \prod_{n_1,n_2\ge0} \left[(1-e^{\alpha(a)}p^{n_1}q^{n_2})(1-e^{\alpha(a)}pqp^{n_1}q^{n_2})\right]^{1/2} \,,  \nonumber \\
 Z_{\rm 1-loop}^{\rm 5d,hyper} &=  e^{-\mathcal{F}_{\rm hyper}^{5d}(a,m;\epsilon_1,\epsilon_2)}\prod_\rho \prod_{n_1,n_2\ge0} (1-e^{\rho(a)+m}p^{n_1}q^{n_2})^{-1} \,,
\end{align}
where the prefactors induced after the regularization are written as
\begin{align}
\mathcal{F}_{\rm vec}^{5d} &= -\frac{\pi i}{2}\sum_\alpha \left[\zeta_3\Big(0,\frac{\alpha(a)}{2\pi i} \big|1,\frac{\epsilon_1}{2\pi i},\frac{\epsilon_2}{2\pi i}\Big)+\zeta_3\Big(0,\frac{\alpha(a)+2\epsilon_+}{2\pi i} \big|1,\frac{\epsilon_1}{2\pi i},\frac{\epsilon_2}{2\pi i}\Big)\right] \,, \nonumber \\
\mathcal{F}_{\rm hyper}^{5d} &= \pi i \sum_\rho \zeta_3\Big(0,\frac{\rho(a)+m}{2\pi i} \big|1,\frac{\epsilon_1}{2\pi i},\frac{\epsilon_2}{2\pi i}\Big) \,.
\end{align}
These prefactors encode the 1-loop corrections to the effective action of the 5d gauge theory.

We now turn to the perturbative partition functions of $\mathcal{N}=2$ field theories in three dimensions. The 3d partition function on $S^1\times D^2$ has been recently computed in \cite{Beem:2012mb} and in \cite{Yoshida:2014ssa} using localization.

The equivariant index of the chiral multiplet with R-charge $\Delta$ is
\begin{equation}
\chi_{\rm chiral}^{\rm 3d} = \frac{e^me^{-\frac{\Delta}{2}\epsilon_1}}{1-e^{-\epsilon_1}} \sum_\rho e^{\rho(a)} \,.
\end{equation}
In the 1-loop computation, we need to take into account a proper boundary condition in supersymmetric and gauge invariant fashion. For the $\mathcal{N}=2$ chiral multiplet, we can choose either the Neumann or the Dirichlet boundary condition. It appears that the boundary condition determines how to expand the denominator of the equivariant index. The computation in \cite{Yoshida:2014ssa} implies that we have to expand it in power series of $e^{-\epsilon_1}$ for the chiral multiplet with the Neumann boundary condition, while expand it in power series of $e^{\epsilon_1}$ for the chiral multiplet with the Dirichlet boundary condition. If we impose the Neumann boundary condition, the 1-loop determinant of the chiral multiplet becomes 
\begin{align}
Z_{\rm chiral,N}^{3d} &= \prod_\rho\prod_{t\in \mathbb{Z}} \prod_{n\ge0}(2\pi i t -(\Delta/2+n)\epsilon_1+ \rho(a) +m)^{-1} \nonumber \\
&= \prod_\rho \Gamma_2\Big(\frac{-\rho(a)-m+\frac{\Delta}{2}\epsilon_1}{2\pi i}\big| 1, \frac{\epsilon_1}{2\pi i} \Big) \Gamma_2\Big(1-\frac{-\rho(a)-m+\frac{\Delta}{2}\epsilon_1}{2\pi i}\big| 1, -\frac{\epsilon_1}{2\pi i} \Big) \nonumber \\
&= \prod_\rho e^{-\pi i\zeta_2(0,\frac{-\rho(a)-m+\frac{\Delta}{2}\epsilon_1}{2\pi i}|1,-\frac{\epsilon_2}{2\pi i})} \prod_{n\ge0}(1-e^{-\rho(a)-m}q^{\Delta/2}q^n)^{-1} \,.
\end{align}
We regularized the infinite products using the Barnes' gamma functions as we did in 5d theories. On the other hand, the chiral multiplet with the Dirichlet boundary condition has the following 1-loop determinant
\begin{align}
Z_{\rm chiral,D}^{3d} &= \prod_\rho\prod_{t\in \mathbb{Z}} \prod_{n\ge0}(2\pi i t +(-\Delta/2+n+1)\epsilon_1+ \rho(a) +m) \nonumber \\
&= \prod_\rho \left[\Gamma_2\Big(\frac{\rho(a)+m+(1-\frac{\Delta}{2})\epsilon_1}{2\pi i}\big| 1, \frac{\epsilon_1}{2\pi i} \Big) \Gamma_2\Big(1-\frac{\rho(a)+m+(1-\frac{\Delta}{2})\epsilon_1}{2\pi i} \big| 1, -\frac{\epsilon_1}{2\pi i} \Big)\right]^{-1} \nonumber \\
&= \prod_\rho e^{-\pi i\zeta_2(0,\frac{\rho(a)+m+(1-\frac{\Delta}{2})\epsilon_1}{2\pi i}|1,\frac{\epsilon_1}{2\pi i})} \prod_{n\ge0}(1-e^{\rho(a)+m}q^{1-\Delta/2}q^n)\,.
\end{align}
The prefactors involve the 1-loop corrections to the (mixed-) Chern-Simons terms by the matter fields.

Lastly, let us compute the elliptic genera of 2d $(0,2)$ multiplets. See \cite{Gadde:2013dda,Benini:2013xpa,Benini:2013nda} for details. The chiral multiplet with R-charge $\Delta$ contributes to the elliptic genus as
\begin{align}
Z^{2d}_{\rm chiral} &= \prod_\rho \prod_{n_1,n_2\in \mathbb{Z}} (2\pi i n_1 - (n_2 + \Delta/2)\epsilon_1+\rho(a)+m)^{-1} \nonumber \\
&= \prod_\rho \Gamma_2\Big(\frac{-\rho(a)-m+\frac{\Delta}{2}\epsilon_1}{2\pi i}\big|1,\frac{\epsilon_1}{2\pi i}\Big) \Gamma_2\Big(1-\frac{-\rho(a)-m+\frac{\Delta}{2}\epsilon_1}{2\pi i}\big|1,-\frac{\epsilon_1}{2\pi i}\Big) \nonumber \\
&\qquad \quad \times \Gamma_2\Big(\frac{\rho(a)+m+(1-\frac{\Delta}{2})\epsilon_1}{2\pi i}\big|1,\frac{\epsilon_1}{2\pi i}\Big)\Gamma_2\Big(1-\frac{\rho(a)+m+(1-\frac{\Delta}{2})\epsilon_1)\epsilon_1}{2\pi i}\big|1,-\frac{\epsilon_1}{2\pi i}\Big) \nonumber \\
&= \prod_\rho e^{-2\pi i\zeta_2(0,\frac{-\rho(a)-m+\frac{\Delta}{2}\epsilon_1}{2\pi i}|1,\frac{\epsilon_1}{2\pi i})} \ \theta(e^{-\rho(a)-m}q^{\Delta/2};q)^{-1} \, .
\end{align}
On the other hand the elliptic genus of the fermi multiplet with R-charge $\Delta$ is given by
\begin{align}
Z^{2d}_{\rm fermi} &= \prod_\rho \prod_{n_1,n_2\in \mathbb{Z}} (2\pi i n_1 - (n_2 + \Delta/2)\epsilon_1+\rho(a)+m) \nonumber \\
&= \prod_\rho e^{2\pi i\zeta_2(0,\frac{-\rho(a)-m+\frac{\Delta}{2}\epsilon_1}{2\pi i}|1,\frac{\epsilon_1}{2\pi i})} \ \theta(e^{-\rho(a)-m}q^{\Delta/2};q)\,.
\end{align}

\bibliography{cpn1}
\bibliographystyle{JHEP}

\end{document}